\documentclass[longauth]{aa}

\usepackage{amsmath}
\usepackage{amssymb}
\usepackage[utf8]{inputenc}
\usepackage{graphicx}
\usepackage{txfonts}
\usepackage{subfigure}
\usepackage{natbib}
\usepackage{appendix}
\usepackage[switch, modulo]{lineno}

\usepackage[normalem]{ulem}

\usepackage{euclid}

\usepackage{hyperref} \hypersetup{ colorlinks=true, linkcolor=blue, filecolor=magenta, urlcolor=blue, citecolor=blue}
\usepackage[utf8]{inputenc}
\usepackage{graphicx}

\usepackage{chngcntr}
\usepackage{appendix}
\usepackage{etoolbox}
\usepackage{xcolor}
\usepackage{diagbox}
\usepackage{multicol}
\usepackage{comment}
\usepackage{siunitx}
\usepackage{hyperref}

\usepackage{multirow}
\usepackage{tabulary}
\usepackage[para]{threeparttable}
\usepackage{array,booktabs,longtable,tabularx}
\newcolumntype{L}{>{\raggedright\arraybackslash}X}
\usepackage{ltablex}

\usepackage{etoolbox}
\usepackage{csquotes}
\usepackage{soul}

\newcommand{\orcid}[1]{\unskip\protect\href{https://orcid.org/#1}{\protect\includegraphics[width=8pt,clip]{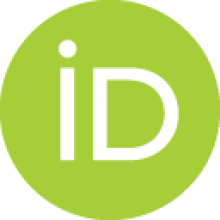}}}

\newcommand{\mstar}{\mbox{$M_\ast$}\xspace}
\newcommand{\msunh}{\mbox{$h^{-1}\, M_\odot$}\xspace}
\newcommand{\msun}{\mbox{$M_\odot$}\xspace}
\newcommand{\disperse}{\mbox{{\tt DisPerSE}}\xspace}

\newcommand{\deep}{\mbox{$\mathcal{D}$}\xspace}
\newcommand{\deepmass}{\mbox{$\mathcal{D}_{M_\ast}$}\xspace}
\newcommand{\deepnoise}{\mbox{$\mathcal{D}_{\rm noise}$}\xspace}
\newcommand{\deepsamp}{\mbox{$\mathcal{D}_{\rm 60\%}$}\xspace}
\newcommand{\deepnoisesamp}{\mbox{$\mathcal{D}_{\rm noise,60\%}$}\xspace}
\newcommand{\deepwf}{\mbox{$\mathcal{D}^{\rm w FoG}$}\xspace}
\newcommand{\deepwof}{\mbox{$\mathcal{D}^{\rm FoG,corr}$}\xspace}
\newcommand{\deepnoisesampwf}{\mbox{$\mathcal{D}^{\rm w FoG}_{\rm noise,60\%}$}\xspace}
\newcommand{\deepnoisesampwof}{\mbox{$\mathcal{D}^{\rm FoG,corr}_{\rm noise,60\%}$}\xspace}

\newcommand{\deepmasssamp}{\mbox{$\mathcal{D}_{M_\ast,\rm 60\%}$}\xspace}
\newcommand{\deepmassnoisesamp}{\mbox{$\mathcal{D}_{M_\ast,\rm noise,\rm 60\%}$}\xspace}
\newcommand{\deepmassnoisesampwf}{\mbox{$\mathcal{D}^{\rm w FoG}_{M_\star,\rm noise,60\%}$}\xspace}
\newcommand{\deepmassnoisesampwof}{\mbox{$\mathcal{D}^{\rm FoG,corr}_{M_\star,\rm noise,60\%}$}\xspace}

\newcommand{\deepmasswf}{\mbox{$\mathcal{D}^{\rm w FoG}_{M_\ast}$}\xspace}
\newcommand{\deepmasswof}{\mbox{$\mathcal{D}^{\rm FoG,corr}_{M_\ast}$}\xspace}

\newcommand{\flag}{\mbox{Flagship}\xspace}
\newcommand{\gaea}{\mbox{GAEA}\xspace}

\definecolor{dodgerblue}{rgb}{0.12, 0.56, 1.0}
\definecolor{applegreen}{rgb}{0.55, 0.71, 0.0}
\definecolor{pink}{rgb}{0.858, 0.688, 0.688}

\begin{document}

\title{ \Euclid  
preparation} \subtitle{3D reconstruction of the cosmic web with simulated Euclid Deep spectroscopic samples}

\author{Euclid Collaboration: K.~Kraljic\orcid{0000-0001-6180-0245}\thanks{\email{katarina.kraljic@astro.unistra.fr}}\inst{\ref{aff1}}
\and C.~Laigle\orcid{0009-0008-5926-818X}\inst{\ref{aff2}}
\and M.~Balogh\orcid{0000-0003-4849-9536}\inst{\ref{aff3},\ref{aff4}}
\and P.~Jablonka\orcid{0000-0002-9655-1063}\inst{\ref{aff5}}
\and U.~Kuchner\orcid{0000-0002-0035-5202}\inst{\ref{aff6}}
\and N.~Malavasi\orcid{0000-0001-9033-7958}\inst{\ref{aff7}}
\and F.~Sarron\orcid{0000-0001-8376-0360}\inst{\ref{aff8},\ref{aff9}}
\and C.~Pichon\orcid{0000-0003-0695-6735}\inst{\ref{aff2},\ref{aff10}}
\and G.~De~Lucia\orcid{0000-0002-6220-9104}\inst{\ref{aff11}}
\and M.~Bethermin\orcid{0000-0002-3915-2015}\inst{\ref{aff1}}
\and F.~Durret\orcid{0000-0002-6991-4578}\inst{\ref{aff12}}
\and M.~Fumagalli\orcid{0000-0001-6676-3842}\inst{\ref{aff11},\ref{aff13}}
\and C.~Gouin\orcid{0000-0002-8837-9953}\inst{\ref{aff2}}
\and M.~Magliocchetti\orcid{0000-0001-9158-4838}\inst{\ref{aff14}}
\and J.~G.~Sorce\orcid{0000-0002-2307-2432}\inst{\ref{aff15},\ref{aff16}}
\and O.~Cucciati\orcid{0000-0002-9336-7551}\inst{\ref{aff17}}
\and F.~Fontanot\orcid{0000-0003-4744-0188}\inst{\ref{aff11},\ref{aff18}}
\and M.~Hirschmann\orcid{0000-0002-3301-3321}\inst{\ref{aff19}}
\and Y.~Kang\orcid{0009-0000-8588-7250}\inst{\ref{aff20}}
\and M.~Spinelli\orcid{0000-0003-0148-3254}\inst{\ref{aff21},\ref{aff11},\ref{aff22}}
\and N.~Aghanim\orcid{0000-0002-6688-8992}\inst{\ref{aff16}}
\and A.~Amara\inst{\ref{aff23}}
\and S.~Andreon\orcid{0000-0002-2041-8784}\inst{\ref{aff24}}
\and N.~Auricchio\orcid{0000-0003-4444-8651}\inst{\ref{aff17}}
\and C.~Baccigalupi\orcid{0000-0002-8211-1630}\inst{\ref{aff18},\ref{aff11},\ref{aff25},\ref{aff26}}
\and M.~Baldi\orcid{0000-0003-4145-1943}\inst{\ref{aff27},\ref{aff17},\ref{aff28}}
\and S.~Bardelli\orcid{0000-0002-8900-0298}\inst{\ref{aff17}}
\and A.~Biviano\orcid{0000-0002-0857-0732}\inst{\ref{aff11},\ref{aff18}}
\and E.~Branchini\orcid{0000-0002-0808-6908}\inst{\ref{aff29},\ref{aff30},\ref{aff24}}
\and M.~Brescia\orcid{0000-0001-9506-5680}\inst{\ref{aff31},\ref{aff32}}
\and J.~Brinchmann\orcid{0000-0003-4359-8797}\inst{\ref{aff33},\ref{aff34},\ref{aff35}}
\and S.~Camera\orcid{0000-0003-3399-3574}\inst{\ref{aff36},\ref{aff37},\ref{aff38}}
\and G.~Ca\~nas-Herrera\orcid{0000-0003-2796-2149}\inst{\ref{aff39},\ref{aff40},\ref{aff41}}
\and V.~Capobianco\orcid{0000-0002-3309-7692}\inst{\ref{aff38}}
\and C.~Carbone\orcid{0000-0003-0125-3563}\inst{\ref{aff42}}
\and J.~Carretero\orcid{0000-0002-3130-0204}\inst{\ref{aff43},\ref{aff44}}
\and R.~Casas\orcid{0000-0002-8165-5601}\inst{\ref{aff45},\ref{aff46}}
\and S.~Casas\orcid{0000-0002-4751-5138}\inst{\ref{aff47}}
\and F.~J.~Castander\orcid{0000-0001-7316-4573}\inst{\ref{aff46},\ref{aff45}}
\and M.~Castellano\orcid{0000-0001-9875-8263}\inst{\ref{aff48}}
\and G.~Castignani\orcid{0000-0001-6831-0687}\inst{\ref{aff17}}
\and S.~Cavuoti\orcid{0000-0002-3787-4196}\inst{\ref{aff32},\ref{aff49}}
\and K.~C.~Chambers\orcid{0000-0001-6965-7789}\inst{\ref{aff50}}
\and A.~Cimatti\inst{\ref{aff51}}
\and C.~Colodro-Conde\inst{\ref{aff52}}
\and G.~Congedo\orcid{0000-0003-2508-0046}\inst{\ref{aff53}}
\and C.~J.~Conselice\orcid{0000-0003-1949-7638}\inst{\ref{aff54}}
\and L.~Conversi\orcid{0000-0002-6710-8476}\inst{\ref{aff55},\ref{aff56}}
\and Y.~Copin\orcid{0000-0002-5317-7518}\inst{\ref{aff57}}
\and F.~Courbin\orcid{0000-0003-0758-6510}\inst{\ref{aff58},\ref{aff59}}
\and H.~M.~Courtois\orcid{0000-0003-0509-1776}\inst{\ref{aff60}}
\and A.~Da~Silva\orcid{0000-0002-6385-1609}\inst{\ref{aff61},\ref{aff62}}
\and H.~Degaudenzi\orcid{0000-0002-5887-6799}\inst{\ref{aff20}}
\and S.~de~la~Torre\inst{\ref{aff63}}
\and H.~Dole\orcid{0000-0002-9767-3839}\inst{\ref{aff16}}
\and M.~Douspis\orcid{0000-0003-4203-3954}\inst{\ref{aff16}}
\and F.~Dubath\orcid{0000-0002-6533-2810}\inst{\ref{aff20}}
\and C.~A.~J.~Duncan\orcid{0009-0003-3573-0791}\inst{\ref{aff53},\ref{aff54}}
\and X.~Dupac\inst{\ref{aff56}}
\and S.~Dusini\orcid{0000-0002-1128-0664}\inst{\ref{aff64}}
\and S.~Escoffier\orcid{0000-0002-2847-7498}\inst{\ref{aff65}}
\and M.~Farina\orcid{0000-0002-3089-7846}\inst{\ref{aff14}}
\and R.~Farinelli\inst{\ref{aff17}}
\and S.~Ferriol\inst{\ref{aff57}}
\and F.~Finelli\orcid{0000-0002-6694-3269}\inst{\ref{aff17},\ref{aff66}}
\and P.~Fosalba\orcid{0000-0002-1510-5214}\inst{\ref{aff45},\ref{aff46}}
\and N.~Fourmanoit\orcid{0009-0005-6816-6925}\inst{\ref{aff65}}
\and M.~Frailis\orcid{0000-0002-7400-2135}\inst{\ref{aff11}}
\and E.~Franceschi\orcid{0000-0002-0585-6591}\inst{\ref{aff17}}
\and M.~Fumana\orcid{0000-0001-6787-5950}\inst{\ref{aff42}}
\and S.~Galeotta\orcid{0000-0002-3748-5115}\inst{\ref{aff11}}
\and K.~George\orcid{0000-0002-1734-8455}\inst{\ref{aff67}}
\and W.~Gillard\orcid{0000-0003-4744-9748}\inst{\ref{aff65}}
\and B.~Gillis\orcid{0000-0002-4478-1270}\inst{\ref{aff53}}
\and C.~Giocoli\orcid{0000-0002-9590-7961}\inst{\ref{aff17},\ref{aff28}}
\and J.~Gracia-Carpio\inst{\ref{aff7}}
\and A.~Grazian\orcid{0000-0002-5688-0663}\inst{\ref{aff68}}
\and F.~Grupp\inst{\ref{aff7},\ref{aff67}}
\and S.~V.~H.~Haugan\orcid{0000-0001-9648-7260}\inst{\ref{aff69}}
\and W.~Holmes\inst{\ref{aff70}}
\and F.~Hormuth\inst{\ref{aff71}}
\and A.~Hornstrup\orcid{0000-0002-3363-0936}\inst{\ref{aff72},\ref{aff73}}
\and K.~Jahnke\orcid{0000-0003-3804-2137}\inst{\ref{aff74}}
\and M.~Jhabvala\inst{\ref{aff75}}
\and B.~Joachimi\orcid{0000-0001-7494-1303}\inst{\ref{aff76}}
\and E.~Keih\"anen\orcid{0000-0003-1804-7715}\inst{\ref{aff77}}
\and S.~Kermiche\orcid{0000-0002-0302-5735}\inst{\ref{aff65}}
\and A.~Kiessling\orcid{0000-0002-2590-1273}\inst{\ref{aff70}}
\and M.~Kilbinger\orcid{0000-0001-9513-7138}\inst{\ref{aff78}}
\and B.~Kubik\orcid{0009-0006-5823-4880}\inst{\ref{aff57}}
\and M.~K\"ummel\orcid{0000-0003-2791-2117}\inst{\ref{aff67}}
\and M.~Kunz\orcid{0000-0002-3052-7394}\inst{\ref{aff79}}
\and H.~Kurki-Suonio\orcid{0000-0002-4618-3063}\inst{\ref{aff80},\ref{aff81}}
\and A.~M.~C.~Le~Brun\orcid{0000-0002-0936-4594}\inst{\ref{aff82}}
\and S.~Ligori\orcid{0000-0003-4172-4606}\inst{\ref{aff38}}
\and P.~B.~Lilje\orcid{0000-0003-4324-7794}\inst{\ref{aff69}}
\and V.~Lindholm\orcid{0000-0003-2317-5471}\inst{\ref{aff80},\ref{aff81}}
\and I.~Lloro\orcid{0000-0001-5966-1434}\inst{\ref{aff83}}
\and G.~Mainetti\orcid{0000-0003-2384-2377}\inst{\ref{aff84}}
\and D.~Maino\inst{\ref{aff85},\ref{aff42},\ref{aff86}}
\and E.~Maiorano\orcid{0000-0003-2593-4355}\inst{\ref{aff17}}
\and O.~Mansutti\orcid{0000-0001-5758-4658}\inst{\ref{aff11}}
\and S.~Marcin\inst{\ref{aff87}}
\and O.~Marggraf\orcid{0000-0001-7242-3852}\inst{\ref{aff88}}
\and M.~Martinelli\orcid{0000-0002-6943-7732}\inst{\ref{aff48},\ref{aff89}}
\and N.~Martinet\orcid{0000-0003-2786-7790}\inst{\ref{aff63}}
\and F.~Marulli\orcid{0000-0002-8850-0303}\inst{\ref{aff90},\ref{aff17},\ref{aff28}}
\and R.~Massey\orcid{0000-0002-6085-3780}\inst{\ref{aff91}}
\and S.~Maurogordato\inst{\ref{aff21}}
\and E.~Medinaceli\orcid{0000-0002-4040-7783}\inst{\ref{aff17}}
\and S.~Mei\orcid{0000-0002-2849-559X}\inst{\ref{aff92},\ref{aff93}}
\and Y.~Mellier\inst{\ref{aff12},\ref{aff2}}
\and M.~Meneghetti\orcid{0000-0003-1225-7084}\inst{\ref{aff17},\ref{aff28}}
\and E.~Merlin\orcid{0000-0001-6870-8900}\inst{\ref{aff48}}
\and G.~Meylan\inst{\ref{aff5}}
\and A.~Mora\orcid{0000-0002-1922-8529}\inst{\ref{aff94}}
\and M.~Moresco\orcid{0000-0002-7616-7136}\inst{\ref{aff90},\ref{aff17}}
\and L.~Moscardini\orcid{0000-0002-3473-6716}\inst{\ref{aff90},\ref{aff17},\ref{aff28}}
\and R.~Nakajima\orcid{0009-0009-1213-7040}\inst{\ref{aff88}}
\and C.~Neissner\orcid{0000-0001-8524-4968}\inst{\ref{aff95},\ref{aff44}}
\and S.-M.~Niemi\orcid{0009-0005-0247-0086}\inst{\ref{aff39}}
\and C.~Padilla\orcid{0000-0001-7951-0166}\inst{\ref{aff95}}
\and S.~Paltani\orcid{0000-0002-8108-9179}\inst{\ref{aff20}}
\and F.~Pasian\orcid{0000-0002-4869-3227}\inst{\ref{aff11}}
\and K.~Pedersen\inst{\ref{aff96}}
\and W.~J.~Percival\orcid{0000-0002-0644-5727}\inst{\ref{aff4},\ref{aff3},\ref{aff97}}
\and V.~Pettorino\inst{\ref{aff39}}
\and S.~Pires\orcid{0000-0002-0249-2104}\inst{\ref{aff78}}
\and G.~Polenta\orcid{0000-0003-4067-9196}\inst{\ref{aff98}}
\and M.~Poncet\inst{\ref{aff99}}
\and L.~A.~Popa\inst{\ref{aff100}}
\and L.~Pozzetti\orcid{0000-0001-7085-0412}\inst{\ref{aff17}}
\and F.~Raison\orcid{0000-0002-7819-6918}\inst{\ref{aff7}}
\and R.~Rebolo\orcid{0000-0003-3767-7085}\inst{\ref{aff52},\ref{aff101},\ref{aff102}}
\and A.~Renzi\orcid{0000-0001-9856-1970}\inst{\ref{aff103},\ref{aff64}}
\and J.~Rhodes\orcid{0000-0002-4485-8549}\inst{\ref{aff70}}
\and G.~Riccio\inst{\ref{aff32}}
\and E.~Romelli\orcid{0000-0003-3069-9222}\inst{\ref{aff11}}
\and M.~Roncarelli\orcid{0000-0001-9587-7822}\inst{\ref{aff17}}
\and C.~Rosset\orcid{0000-0003-0286-2192}\inst{\ref{aff92}}
\and E.~Rossetti\orcid{0000-0003-0238-4047}\inst{\ref{aff27}}
\and R.~Saglia\orcid{0000-0003-0378-7032}\inst{\ref{aff67},\ref{aff7}}
\and Z.~Sakr\orcid{0000-0002-4823-3757}\inst{\ref{aff104},\ref{aff105},\ref{aff106}}
\and A.~G.~S\'anchez\orcid{0000-0003-1198-831X}\inst{\ref{aff7}}
\and D.~Sapone\orcid{0000-0001-7089-4503}\inst{\ref{aff107}}
\and B.~Sartoris\orcid{0000-0003-1337-5269}\inst{\ref{aff67},\ref{aff11}}
\and P.~Schneider\orcid{0000-0001-8561-2679}\inst{\ref{aff88}}
\and T.~Schrabback\orcid{0000-0002-6987-7834}\inst{\ref{aff108}}
\and M.~Scodeggio\inst{\ref{aff42}}
\and A.~Secroun\orcid{0000-0003-0505-3710}\inst{\ref{aff65}}
\and E.~Sefusatti\orcid{0000-0003-0473-1567}\inst{\ref{aff11},\ref{aff18},\ref{aff25}}
\and G.~Seidel\orcid{0000-0003-2907-353X}\inst{\ref{aff74}}
\and M.~Seiffert\orcid{0000-0002-7536-9393}\inst{\ref{aff70}}
\and S.~Serrano\orcid{0000-0002-0211-2861}\inst{\ref{aff45},\ref{aff109},\ref{aff46}}
\and P.~Simon\inst{\ref{aff88}}
\and C.~Sirignano\orcid{0000-0002-0995-7146}\inst{\ref{aff103},\ref{aff64}}
\and G.~Sirri\orcid{0000-0003-2626-2853}\inst{\ref{aff28}}
\and L.~Stanco\orcid{0000-0002-9706-5104}\inst{\ref{aff64}}
\and J.~Steinwagner\orcid{0000-0001-7443-1047}\inst{\ref{aff7}}
\and P.~Tallada-Cresp\'{i}\orcid{0000-0002-1336-8328}\inst{\ref{aff43},\ref{aff44}}
\and A.~N.~Taylor\inst{\ref{aff53}}
\and H.~I.~Teplitz\orcid{0000-0002-7064-5424}\inst{\ref{aff110}}
\and I.~Tereno\orcid{0000-0002-4537-6218}\inst{\ref{aff61},\ref{aff111}}
\and N.~Tessore\orcid{0000-0002-9696-7931}\inst{\ref{aff76}}
\and S.~Toft\orcid{0000-0003-3631-7176}\inst{\ref{aff112},\ref{aff113}}
\and R.~Toledo-Moreo\orcid{0000-0002-2997-4859}\inst{\ref{aff114}}
\and F.~Torradeflot\orcid{0000-0003-1160-1517}\inst{\ref{aff44},\ref{aff43}}
\and I.~Tutusaus\orcid{0000-0002-3199-0399}\inst{\ref{aff105}}
\and L.~Valenziano\orcid{0000-0002-1170-0104}\inst{\ref{aff17},\ref{aff66}}
\and J.~Valiviita\orcid{0000-0001-6225-3693}\inst{\ref{aff80},\ref{aff81}}
\and T.~Vassallo\orcid{0000-0001-6512-6358}\inst{\ref{aff67},\ref{aff11}}
\and G.~Verdoes~Kleijn\orcid{0000-0001-5803-2580}\inst{\ref{aff115}}
\and A.~Veropalumbo\orcid{0000-0003-2387-1194}\inst{\ref{aff24},\ref{aff30},\ref{aff29}}
\and D.~Vibert\orcid{0009-0008-0607-631X}\inst{\ref{aff63}}
\and Y.~Wang\orcid{0000-0002-4749-2984}\inst{\ref{aff110}}
\and J.~Weller\orcid{0000-0002-8282-2010}\inst{\ref{aff67},\ref{aff7}}
\and A.~Zacchei\orcid{0000-0003-0396-1192}\inst{\ref{aff11},\ref{aff18}}
\and G.~Zamorani\orcid{0000-0002-2318-301X}\inst{\ref{aff17}}
\and E.~Zucca\orcid{0000-0002-5845-8132}\inst{\ref{aff17}}
\and V.~Allevato\orcid{0000-0001-7232-5152}\inst{\ref{aff32}}
\and M.~Ballardini\orcid{0000-0003-4481-3559}\inst{\ref{aff116},\ref{aff117},\ref{aff17}}
\and M.~Bolzonella\orcid{0000-0003-3278-4607}\inst{\ref{aff17}}
\and E.~Bozzo\orcid{0000-0002-8201-1525}\inst{\ref{aff20}}
\and C.~Burigana\orcid{0000-0002-3005-5796}\inst{\ref{aff118},\ref{aff66}}
\and R.~Cabanac\orcid{0000-0001-6679-2600}\inst{\ref{aff105}}
\and M.~Calabrese\orcid{0000-0002-2637-2422}\inst{\ref{aff119},\ref{aff42}}
\and A.~Cappi\inst{\ref{aff17},\ref{aff21}}
\and D.~Di~Ferdinando\inst{\ref{aff28}}
\and J.~A.~Escartin~Vigo\inst{\ref{aff7}}
\and L.~Gabarra\orcid{0000-0002-8486-8856}\inst{\ref{aff120}}
\and W.~G.~Hartley\inst{\ref{aff20}}
\and J.~Mart\'{i}n-Fleitas\orcid{0000-0002-8594-569X}\inst{\ref{aff121}}
\and S.~Matthew\orcid{0000-0001-8448-1697}\inst{\ref{aff53}}
\and N.~Mauri\orcid{0000-0001-8196-1548}\inst{\ref{aff51},\ref{aff28}}
\and R.~B.~Metcalf\orcid{0000-0003-3167-2574}\inst{\ref{aff90},\ref{aff17}}
\and A.~A.~Nucita\inst{\ref{aff122},\ref{aff123},\ref{aff124}}
\and A.~Pezzotta\orcid{0000-0003-0726-2268}\inst{\ref{aff125},\ref{aff7}}
\and M.~P\"ontinen\orcid{0000-0001-5442-2530}\inst{\ref{aff80}}
\and C.~Porciani\orcid{0000-0002-7797-2508}\inst{\ref{aff88}}
\and I.~Risso\orcid{0000-0003-2525-7761}\inst{\ref{aff126}}
\and V.~Scottez\orcid{0009-0008-3864-940X}\inst{\ref{aff12},\ref{aff127}}
\and M.~Sereno\orcid{0000-0003-0302-0325}\inst{\ref{aff17},\ref{aff28}}
\and M.~Tenti\orcid{0000-0002-4254-5901}\inst{\ref{aff28}}
\and M.~Viel\orcid{0000-0002-2642-5707}\inst{\ref{aff18},\ref{aff11},\ref{aff26},\ref{aff25},\ref{aff128}}
\and M.~Wiesmann\orcid{0009-0000-8199-5860}\inst{\ref{aff69}}
\and Y.~Akrami\orcid{0000-0002-2407-7956}\inst{\ref{aff129},\ref{aff130}}
\and S.~Alvi\orcid{0000-0001-5779-8568}\inst{\ref{aff116}}
\and I.~T.~Andika\orcid{0000-0001-6102-9526}\inst{\ref{aff131},\ref{aff132}}
\and S.~Anselmi\orcid{0000-0002-3579-9583}\inst{\ref{aff64},\ref{aff103},\ref{aff133}}
\and M.~Archidiacono\orcid{0000-0003-4952-9012}\inst{\ref{aff85},\ref{aff86}}
\and F.~Atrio-Barandela\orcid{0000-0002-2130-2513}\inst{\ref{aff134}}
\and A.~Balaguera-Antolinez\orcid{0000-0001-5028-3035}\inst{\ref{aff52},\ref{aff135}}
\and P.~Bergamini\orcid{0000-0003-1383-9414}\inst{\ref{aff85},\ref{aff17}}
\and D.~Bertacca\orcid{0000-0002-2490-7139}\inst{\ref{aff103},\ref{aff68},\ref{aff64}}
\and A.~Blanchard\orcid{0000-0001-8555-9003}\inst{\ref{aff105}}
\and L.~Blot\orcid{0000-0002-9622-7167}\inst{\ref{aff136},\ref{aff82}}
\and H.~B\"ohringer\orcid{0000-0001-8241-4204}\inst{\ref{aff7},\ref{aff137},\ref{aff138}}
\and S.~Borgani\orcid{0000-0001-6151-6439}\inst{\ref{aff139},\ref{aff18},\ref{aff11},\ref{aff25},\ref{aff128}}
\and M.~L.~Brown\orcid{0000-0002-0370-8077}\inst{\ref{aff54}}
\and S.~Bruton\orcid{0000-0002-6503-5218}\inst{\ref{aff140}}
\and A.~Calabro\orcid{0000-0003-2536-1614}\inst{\ref{aff48}}
\and B.~Camacho~Quevedo\orcid{0000-0002-8789-4232}\inst{\ref{aff18},\ref{aff26},\ref{aff11},\ref{aff45},\ref{aff46}}
\and F.~Caro\inst{\ref{aff48}}
\and C.~S.~Carvalho\inst{\ref{aff111}}
\and T.~Castro\orcid{0000-0002-6292-3228}\inst{\ref{aff11},\ref{aff25},\ref{aff18},\ref{aff128}}
\and R.~Chary\orcid{0000-0001-7583-0621}\inst{\ref{aff110},\ref{aff141}}
\and F.~Cogato\orcid{0000-0003-4632-6113}\inst{\ref{aff90},\ref{aff17}}
\and S.~Conseil\orcid{0000-0002-3657-4191}\inst{\ref{aff57}}
\and T.~Contini\orcid{0000-0003-0275-938X}\inst{\ref{aff105}}
\and A.~R.~Cooray\orcid{0000-0002-3892-0190}\inst{\ref{aff142}}
\and S.~Davini\orcid{0000-0003-3269-1718}\inst{\ref{aff30}}
\and F.~De~Paolis\orcid{0000-0001-6460-7563}\inst{\ref{aff122},\ref{aff123},\ref{aff124}}
\and G.~Desprez\orcid{0000-0001-8325-1742}\inst{\ref{aff115}}
\and A.~D\'iaz-S\'anchez\orcid{0000-0003-0748-4768}\inst{\ref{aff143}}
\and J.~J.~Diaz\orcid{0000-0003-2101-1078}\inst{\ref{aff52}}
\and S.~Di~Domizio\orcid{0000-0003-2863-5895}\inst{\ref{aff29},\ref{aff30}}
\and J.~M.~Diego\orcid{0000-0001-9065-3926}\inst{\ref{aff144}}
\and P.~Dimauro\orcid{0000-0001-7399-2854}\inst{\ref{aff145},\ref{aff48}}
\and P.-A.~Duc\orcid{0000-0003-3343-6284}\inst{\ref{aff1}}
\and A.~Enia\orcid{0000-0002-0200-2857}\inst{\ref{aff27},\ref{aff17}}
\and Y.~Fang\inst{\ref{aff67}}
\and A.~G.~Ferrari\orcid{0009-0005-5266-4110}\inst{\ref{aff28}}
\and A.~Finoguenov\orcid{0000-0002-4606-5403}\inst{\ref{aff80}}
\and A.~Fontana\orcid{0000-0003-3820-2823}\inst{\ref{aff48}}
\and A.~Franco\orcid{0000-0002-4761-366X}\inst{\ref{aff123},\ref{aff122},\ref{aff124}}
\and K.~Ganga\orcid{0000-0001-8159-8208}\inst{\ref{aff92}}
\and J.~Garc\'ia-Bellido\orcid{0000-0002-9370-8360}\inst{\ref{aff129}}
\and T.~Gasparetto\orcid{0000-0002-7913-4866}\inst{\ref{aff11}}
\and R.~Gavazzi\orcid{0000-0002-5540-6935}\inst{\ref{aff63},\ref{aff2}}
\and E.~Gaztanaga\orcid{0000-0001-9632-0815}\inst{\ref{aff46},\ref{aff45},\ref{aff146}}
\and F.~Giacomini\orcid{0000-0002-3129-2814}\inst{\ref{aff28}}
\and F.~Gianotti\orcid{0000-0003-4666-119X}\inst{\ref{aff17}}
\and G.~Gozaliasl\orcid{0000-0002-0236-919X}\inst{\ref{aff147},\ref{aff80}}
\and M.~Guidi\orcid{0000-0001-9408-1101}\inst{\ref{aff27},\ref{aff17}}
\and C.~M.~Gutierrez\orcid{0000-0001-7854-783X}\inst{\ref{aff148}}
\and A.~Hall\orcid{0000-0002-3139-8651}\inst{\ref{aff53}}
\and H.~Hildebrandt\orcid{0000-0002-9814-3338}\inst{\ref{aff149}}
\and J.~Hjorth\orcid{0000-0002-4571-2306}\inst{\ref{aff96}}
\and S.~Joudaki\orcid{0000-0001-8820-673X}\inst{\ref{aff43}}
\and J.~J.~E.~Kajava\orcid{0000-0002-3010-8333}\inst{\ref{aff150},\ref{aff151}}
\and V.~Kansal\orcid{0000-0002-4008-6078}\inst{\ref{aff152},\ref{aff153}}
\and D.~Karagiannis\orcid{0000-0002-4927-0816}\inst{\ref{aff116},\ref{aff22}}
\and K.~Kiiveri\inst{\ref{aff77}}
\and C.~C.~Kirkpatrick\inst{\ref{aff77}}
\and S.~Kruk\orcid{0000-0001-8010-8879}\inst{\ref{aff56}}
\and M.~Lattanzi\orcid{0000-0003-1059-2532}\inst{\ref{aff117}}
\and V.~Le~Brun\orcid{0000-0002-5027-1939}\inst{\ref{aff63}}
\and J.~Le~Graet\orcid{0000-0001-6523-7971}\inst{\ref{aff65}}
\and L.~Legrand\orcid{0000-0003-0610-5252}\inst{\ref{aff154},\ref{aff155}}
\and M.~Lembo\orcid{0000-0002-5271-5070}\inst{\ref{aff2}}
\and F.~Lepori\orcid{0009-0000-5061-7138}\inst{\ref{aff156}}
\and G.~Leroy\orcid{0009-0004-2523-4425}\inst{\ref{aff157},\ref{aff91}}
\and G.~F.~Lesci\orcid{0000-0002-4607-2830}\inst{\ref{aff90},\ref{aff17}}
\and J.~Lesgourgues\orcid{0000-0001-7627-353X}\inst{\ref{aff47}}
\and L.~Leuzzi\orcid{0009-0006-4479-7017}\inst{\ref{aff17}}
\and T.~I.~Liaudat\orcid{0000-0002-9104-314X}\inst{\ref{aff158}}
\and S.~J.~Liu\orcid{0000-0001-7680-2139}\inst{\ref{aff14}}
\and A.~Loureiro\orcid{0000-0002-4371-0876}\inst{\ref{aff159},\ref{aff160}}
\and J.~Macias-Perez\orcid{0000-0002-5385-2763}\inst{\ref{aff161}}
\and G.~Maggio\orcid{0000-0003-4020-4836}\inst{\ref{aff11}}
\and E.~A.~Magnier\orcid{0000-0002-7965-2815}\inst{\ref{aff50}}
\and F.~Mannucci\orcid{0000-0002-4803-2381}\inst{\ref{aff162}}
\and R.~Maoli\orcid{0000-0002-6065-3025}\inst{\ref{aff163},\ref{aff48}}
\and C.~J.~A.~P.~Martins\orcid{0000-0002-4886-9261}\inst{\ref{aff164},\ref{aff33}}
\and L.~Maurin\orcid{0000-0002-8406-0857}\inst{\ref{aff16}}
\and M.~Miluzio\inst{\ref{aff56},\ref{aff165}}
\and P.~Monaco\orcid{0000-0003-2083-7564}\inst{\ref{aff139},\ref{aff11},\ref{aff25},\ref{aff18}}
\and C.~Moretti\orcid{0000-0003-3314-8936}\inst{\ref{aff26},\ref{aff128},\ref{aff11},\ref{aff18},\ref{aff25}}
\and G.~Morgante\inst{\ref{aff17}}
\and S.~Nadathur\orcid{0000-0001-9070-3102}\inst{\ref{aff146}}
\and K.~Naidoo\orcid{0000-0002-9182-1802}\inst{\ref{aff146}}
\and A.~Navarro-Alsina\orcid{0000-0002-3173-2592}\inst{\ref{aff88}}
\and S.~Nesseris\orcid{0000-0002-0567-0324}\inst{\ref{aff129}}
\and L.~Pagano\orcid{0000-0003-1820-5998}\inst{\ref{aff116},\ref{aff117}}
\and F.~Passalacqua\orcid{0000-0002-8606-4093}\inst{\ref{aff103},\ref{aff64}}
\and K.~Paterson\orcid{0000-0001-8340-3486}\inst{\ref{aff74}}
\and L.~Patrizii\inst{\ref{aff28}}
\and A.~Pisani\orcid{0000-0002-6146-4437}\inst{\ref{aff65}}
\and D.~Potter\orcid{0000-0002-0757-5195}\inst{\ref{aff156}}
\and S.~Quai\orcid{0000-0002-0449-8163}\inst{\ref{aff90},\ref{aff17}}
\and M.~Radovich\orcid{0000-0002-3585-866X}\inst{\ref{aff68}}
\and P.-F.~Rocci\inst{\ref{aff16}}
\and G.~Rodighiero\orcid{0000-0002-9415-2296}\inst{\ref{aff103},\ref{aff68}}
\and S.~Sacquegna\orcid{0000-0002-8433-6630}\inst{\ref{aff122},\ref{aff123},\ref{aff124}}
\and M.~Sahl\'en\orcid{0000-0003-0973-4804}\inst{\ref{aff166}}
\and D.~B.~Sanders\orcid{0000-0002-1233-9998}\inst{\ref{aff50}}
\and A.~Schneider\orcid{0000-0001-7055-8104}\inst{\ref{aff156}}
\and D.~Sciotti\orcid{0009-0008-4519-2620}\inst{\ref{aff48},\ref{aff89}}
\and E.~Sellentin\inst{\ref{aff167},\ref{aff41}}
\and L.~C.~Smith\orcid{0000-0002-3259-2771}\inst{\ref{aff168}}
\and K.~Tanidis\orcid{0000-0001-9843-5130}\inst{\ref{aff120}}
\and C.~Tao\orcid{0000-0001-7961-8177}\inst{\ref{aff65}}
\and G.~Testera\inst{\ref{aff30}}
\and R.~Teyssier\orcid{0000-0001-7689-0933}\inst{\ref{aff169}}
\and S.~Tosi\orcid{0000-0002-7275-9193}\inst{\ref{aff29},\ref{aff30},\ref{aff24}}
\and A.~Troja\orcid{0000-0003-0239-4595}\inst{\ref{aff103},\ref{aff64}}
\and M.~Tucci\inst{\ref{aff20}}
\and C.~Valieri\inst{\ref{aff28}}
\and A.~Venhola\orcid{0000-0001-6071-4564}\inst{\ref{aff170}}
\and D.~Vergani\orcid{0000-0003-0898-2216}\inst{\ref{aff17}}
\and G.~Verza\orcid{0000-0002-1886-8348}\inst{\ref{aff171}}
\and P.~Vielzeuf\orcid{0000-0003-2035-9339}\inst{\ref{aff65}}
\and N.~A.~Walton\orcid{0000-0003-3983-8778}\inst{\ref{aff168}}}
										   
\institute{Universit\'e de Strasbourg, CNRS, Observatoire astronomique de Strasbourg, UMR 7550, 67000 Strasbourg, France\label{aff1}
\and
Institut d'Astrophysique de Paris, UMR 7095, CNRS, and Sorbonne Universit\'e, 98 bis boulevard Arago, 75014 Paris, France\label{aff2}
\and
Department of Physics and Astronomy, University of Waterloo, Waterloo, Ontario N2L 3G1, Canada\label{aff3}
\and
Waterloo Centre for Astrophysics, University of Waterloo, Waterloo, Ontario N2L 3G1, Canada\label{aff4}
\and
Institute of Physics, Laboratory of Astrophysics, Ecole Polytechnique F\'ed\'erale de Lausanne (EPFL), Observatoire de Sauverny, 1290 Versoix, Switzerland\label{aff5}
\and
School of Physics and Astronomy, University of Nottingham, University Park, Nottingham NG7 2RD, UK\label{aff6}
\and
Max Planck Institute for Extraterrestrial Physics, Giessenbachstr. 1, 85748 Garching, Germany\label{aff7}
\and
Institut de Recherche en Informatique de Toulouse (IRIT), Universit\'e de Toulouse, CNRS, Toulouse INP, UT3, 31062 Toulouse, France\label{aff8}
\and
Laboratoire MCD, Centre de Biologie Int\'egrative (CBI), Universit\'e de Toulouse, CNRS, UT3, 31062 Toulouse, France\label{aff9}
\and
Kyung Hee University, Dept. of Astronomy \& Space Science, Yongin-shi, Gyeonggi-do 17104, Republic of Korea\label{aff10}
\and
INAF-Osservatorio Astronomico di Trieste, Via G. B. Tiepolo 11, 34143 Trieste, Italy\label{aff11}
\and
Institut d'Astrophysique de Paris, 98bis Boulevard Arago, 75014, Paris, France\label{aff12}
\and
Dipartimento di Fisica ``G. Occhialini", Universit\`a degli Studi di Milano Bicocca, Piazza della Scienza 3, 20126 Milano, Italy\label{aff13}
\and
INAF-Istituto di Astrofisica e Planetologia Spaziali, via del Fosso del Cavaliere, 100, 00100 Roma, Italy\label{aff14}
\and
Univ. Lille, CNRS, Centrale Lille, UMR 9189 CRIStAL, 59000 Lille, France\label{aff15}
\and
Universit\'e Paris-Saclay, CNRS, Institut d'astrophysique spatiale, 91405, Orsay, France\label{aff16}
\and
INAF-Osservatorio di Astrofisica e Scienza dello Spazio di Bologna, Via Piero Gobetti 93/3, 40129 Bologna, Italy\label{aff17}
\and
IFPU, Institute for Fundamental Physics of the Universe, via Beirut 2, 34151 Trieste, Italy\label{aff18}
\and
Institute of Physics, Laboratory for Galaxy Evolution, Ecole Polytechnique F\'ed\'erale de Lausanne, Observatoire de Sauverny, CH-1290 Versoix, Switzerland\label{aff19}
\and
Department of Astronomy, University of Geneva, ch. d'Ecogia 16, 1290 Versoix, Switzerland\label{aff20}
\and
Universit\'e C\^{o}te d'Azur, Observatoire de la C\^{o}te d'Azur, CNRS, Laboratoire Lagrange, Bd de l'Observatoire, CS 34229, 06304 Nice cedex 4, France\label{aff21}
\and
Department of Physics and Astronomy, University of the Western Cape, Bellville, Cape Town, 7535, South Africa\label{aff22}
\and
School of Mathematics and Physics, University of Surrey, Guildford, Surrey, GU2 7XH, UK\label{aff23}
\and
INAF-Osservatorio Astronomico di Brera, Via Brera 28, 20122 Milano, Italy\label{aff24}
\and
INFN, Sezione di Trieste, Via Valerio 2, 34127 Trieste TS, Italy\label{aff25}
\and
SISSA, International School for Advanced Studies, Via Bonomea 265, 34136 Trieste TS, Italy\label{aff26}
\and
Dipartimento di Fisica e Astronomia, Universit\`a di Bologna, Via Gobetti 93/2, 40129 Bologna, Italy\label{aff27}
\and
INFN-Sezione di Bologna, Viale Berti Pichat 6/2, 40127 Bologna, Italy\label{aff28}
\and
Dipartimento di Fisica, Universit\`a di Genova, Via Dodecaneso 33, 16146, Genova, Italy\label{aff29}
\and
INFN-Sezione di Genova, Via Dodecaneso 33, 16146, Genova, Italy\label{aff30}
\and
Department of Physics "E. Pancini", University Federico II, Via Cinthia 6, 80126, Napoli, Italy\label{aff31}
\and
INAF-Osservatorio Astronomico di Capodimonte, Via Moiariello 16, 80131 Napoli, Italy\label{aff32}
\and
Instituto de Astrof\'isica e Ci\^encias do Espa\c{c}o, Universidade do Porto, CAUP, Rua das Estrelas, PT4150-762 Porto, Portugal\label{aff33}
\and
Faculdade de Ci\^encias da Universidade do Porto, Rua do Campo de Alegre, 4150-007 Porto, Portugal\label{aff34}
\and
European Southern Observatory, Karl-Schwarzschild-Str.~2, 85748 Garching, Germany\label{aff35}
\and
Dipartimento di Fisica, Universit\`a degli Studi di Torino, Via P. Giuria 1, 10125 Torino, Italy\label{aff36}
\and
INFN-Sezione di Torino, Via P. Giuria 1, 10125 Torino, Italy\label{aff37}
\and
INAF-Osservatorio Astrofisico di Torino, Via Osservatorio 20, 10025 Pino Torinese (TO), Italy\label{aff38}
\and
European Space Agency/ESTEC, Keplerlaan 1, 2201 AZ Noordwijk, The Netherlands\label{aff39}
\and
Institute Lorentz, Leiden University, Niels Bohrweg 2, 2333 CA Leiden, The Netherlands\label{aff40}
\and
Leiden Observatory, Leiden University, Einsteinweg 55, 2333 CC Leiden, The Netherlands\label{aff41}
\and
INAF-IASF Milano, Via Alfonso Corti 12, 20133 Milano, Italy\label{aff42}
\and
Centro de Investigaciones Energ\'eticas, Medioambientales y Tecnol\'ogicas (CIEMAT), Avenida Complutense 40, 28040 Madrid, Spain\label{aff43}
\and
Port d'Informaci\'{o} Cient\'{i}fica, Campus UAB, C. Albareda s/n, 08193 Bellaterra (Barcelona), Spain\label{aff44}
\and
Institut d'Estudis Espacials de Catalunya (IEEC),  Edifici RDIT, Campus UPC, 08860 Castelldefels, Barcelona, Spain\label{aff45}
\and
Institute of Space Sciences (ICE, CSIC), Campus UAB, Carrer de Can Magrans, s/n, 08193 Barcelona, Spain\label{aff46}
\and
Institute for Theoretical Particle Physics and Cosmology (TTK), RWTH Aachen University, 52056 Aachen, Germany\label{aff47}
\and
INAF-Osservatorio Astronomico di Roma, Via Frascati 33, 00078 Monteporzio Catone, Italy\label{aff48}
\and
INFN section of Naples, Via Cinthia 6, 80126, Napoli, Italy\label{aff49}
\and
Institute for Astronomy, University of Hawaii, 2680 Woodlawn Drive, Honolulu, HI 96822, USA\label{aff50}
\and
Dipartimento di Fisica e Astronomia "Augusto Righi" - Alma Mater Studiorum Universit\`a di Bologna, Viale Berti Pichat 6/2, 40127 Bologna, Italy\label{aff51}
\and
Instituto de Astrof\'{\i}sica de Canarias, V\'{\i}a L\'actea, 38205 La Laguna, Tenerife, Spain\label{aff52}
\and
Institute for Astronomy, University of Edinburgh, Royal Observatory, Blackford Hill, Edinburgh EH9 3HJ, UK\label{aff53}
\and
Jodrell Bank Centre for Astrophysics, Department of Physics and Astronomy, University of Manchester, Oxford Road, Manchester M13 9PL, UK\label{aff54}
\and
European Space Agency/ESRIN, Largo Galileo Galilei 1, 00044 Frascati, Roma, Italy\label{aff55}
\and
ESAC/ESA, Camino Bajo del Castillo, s/n., Urb. Villafranca del Castillo, 28692 Villanueva de la Ca\~nada, Madrid, Spain\label{aff56}
\and
Universit\'e Claude Bernard Lyon 1, CNRS/IN2P3, IP2I Lyon, UMR 5822, Villeurbanne, F-69100, France\label{aff57}
\and
Institut de Ci\`{e}ncies del Cosmos (ICCUB), Universitat de Barcelona (IEEC-UB), Mart\'{i} i Franqu\`{e}s 1, 08028 Barcelona, Spain\label{aff58}
\and
Instituci\'o Catalana de Recerca i Estudis Avan\c{c}ats (ICREA), Passeig de Llu\'{\i}s Companys 23, 08010 Barcelona, Spain\label{aff59}
\and
UCB Lyon 1, CNRS/IN2P3, IUF, IP2I Lyon, 4 rue Enrico Fermi, 69622 Villeurbanne, France\label{aff60}
\and
Departamento de F\'isica, Faculdade de Ci\^encias, Universidade de Lisboa, Edif\'icio C8, Campo Grande, PT1749-016 Lisboa, Portugal\label{aff61}
\and
Instituto de Astrof\'isica e Ci\^encias do Espa\c{c}o, Faculdade de Ci\^encias, Universidade de Lisboa, Campo Grande, 1749-016 Lisboa, Portugal\label{aff62}
\and
Aix-Marseille Universit\'e, CNRS, CNES, LAM, Marseille, France\label{aff63}
\and
INFN-Padova, Via Marzolo 8, 35131 Padova, Italy\label{aff64}
\and
Aix-Marseille Universit\'e, CNRS/IN2P3, CPPM, Marseille, France\label{aff65}
\and
INFN-Bologna, Via Irnerio 46, 40126 Bologna, Italy\label{aff66}
\and
Universit\"ats-Sternwarte M\"unchen, Fakult\"at f\"ur Physik, Ludwig-Maximilians-Universit\"at M\"unchen, Scheinerstrasse 1, 81679 M\"unchen, Germany\label{aff67}
\and
INAF-Osservatorio Astronomico di Padova, Via dell'Osservatorio 5, 35122 Padova, Italy\label{aff68}
\and
Institute of Theoretical Astrophysics, University of Oslo, P.O. Box 1029 Blindern, 0315 Oslo, Norway\label{aff69}
\and
Jet Propulsion Laboratory, California Institute of Technology, 4800 Oak Grove Drive, Pasadena, CA, 91109, USA\label{aff70}
\and
Felix Hormuth Engineering, Goethestr. 17, 69181 Leimen, Germany\label{aff71}
\and
Technical University of Denmark, Elektrovej 327, 2800 Kgs. Lyngby, Denmark\label{aff72}
\and
Cosmic Dawn Center (DAWN), Denmark\label{aff73}
\and
Max-Planck-Institut f\"ur Astronomie, K\"onigstuhl 17, 69117 Heidelberg, Germany\label{aff74}
\and
NASA Goddard Space Flight Center, Greenbelt, MD 20771, USA\label{aff75}
\and
Department of Physics and Astronomy, University College London, Gower Street, London WC1E 6BT, UK\label{aff76}
\and
Department of Physics and Helsinki Institute of Physics, Gustaf H\"allstr\"omin katu 2, University of Helsinki, 00014 Helsinki, Finland\label{aff77}
\and
Universit\'e Paris-Saclay, Universit\'e Paris Cit\'e, CEA, CNRS, AIM, 91191, Gif-sur-Yvette, France\label{aff78}
\and
Universit\'e de Gen\`eve, D\'epartement de Physique Th\'eorique and Centre for Astroparticle Physics, 24 quai Ernest-Ansermet, CH-1211 Gen\`eve 4, Switzerland\label{aff79}
\and
Department of Physics, P.O. Box 64, University of Helsinki, 00014 Helsinki, Finland\label{aff80}
\and
Helsinki Institute of Physics, Gustaf H{\"a}llstr{\"o}min katu 2, University of Helsinki, 00014 Helsinki, Finland\label{aff81}
\and
Laboratoire d'etude de l'Univers et des phenomenes eXtremes, Observatoire de Paris, Universit\'e PSL, Sorbonne Universit\'e, CNRS, 92190 Meudon, France\label{aff82}
\and
SKAO, Jodrell Bank, Lower Withington, Macclesfield SK11 9FT, United Kingdom\label{aff83}
\and
Centre de Calcul de l'IN2P3/CNRS, 21 avenue Pierre de Coubertin 69627 Villeurbanne Cedex, France\label{aff84}
\and
Dipartimento di Fisica "Aldo Pontremoli", Universit\`a degli Studi di Milano, Via Celoria 16, 20133 Milano, Italy\label{aff85}
\and
INFN-Sezione di Milano, Via Celoria 16, 20133 Milano, Italy\label{aff86}
\and
University of Applied Sciences and Arts of Northwestern Switzerland, School of Computer Science, 5210 Windisch, Switzerland\label{aff87}
\and
Universit\"at Bonn, Argelander-Institut f\"ur Astronomie, Auf dem H\"ugel 71, 53121 Bonn, Germany\label{aff88}
\and
INFN-Sezione di Roma, Piazzale Aldo Moro, 2 - c/o Dipartimento di Fisica, Edificio G. Marconi, 00185 Roma, Italy\label{aff89}
\and
Dipartimento di Fisica e Astronomia "Augusto Righi" - Alma Mater Studiorum Universit\`a di Bologna, via Piero Gobetti 93/2, 40129 Bologna, Italy\label{aff90}
\and
Department of Physics, Institute for Computational Cosmology, Durham University, South Road, Durham, DH1 3LE, UK\label{aff91}
\and
Universit\'e Paris Cit\'e, CNRS, Astroparticule et Cosmologie, 75013 Paris, France\label{aff92}
\and
CNRS-UCB International Research Laboratory, Centre Pierre Bin\'etruy, IRL2007, CPB-IN2P3, Berkeley, USA\label{aff93}
\and
Telespazio UK S.L. for European Space Agency (ESA), Camino bajo del Castillo, s/n, Urbanizacion Villafranca del Castillo, Villanueva de la Ca\~nada, 28692 Madrid, Spain\label{aff94}
\and
Institut de F\'{i}sica d'Altes Energies (IFAE), The Barcelona Institute of Science and Technology, Campus UAB, 08193 Bellaterra (Barcelona), Spain\label{aff95}
\and
DARK, Niels Bohr Institute, University of Copenhagen, Jagtvej 155, 2200 Copenhagen, Denmark\label{aff96}
\and
Perimeter Institute for Theoretical Physics, Waterloo, Ontario N2L 2Y5, Canada\label{aff97}
\and
Space Science Data Center, Italian Space Agency, via del Politecnico snc, 00133 Roma, Italy\label{aff98}
\and
Centre National d'Etudes Spatiales -- Centre spatial de Toulouse, 18 avenue Edouard Belin, 31401 Toulouse Cedex 9, France\label{aff99}
\and
Institute of Space Science, Str. Atomistilor, nr. 409 M\u{a}gurele, Ilfov, 077125, Romania\label{aff100}
\and
Consejo Superior de Investigaciones Cientificas, Calle Serrano 117, 28006 Madrid, Spain\label{aff101}
\and
Universidad de La Laguna, Departamento de Astrof\'{\i}sica, 38206 La Laguna, Tenerife, Spain\label{aff102}
\and
Dipartimento di Fisica e Astronomia "G. Galilei", Universit\`a di Padova, Via Marzolo 8, 35131 Padova, Italy\label{aff103}
\and
Institut f\"ur Theoretische Physik, University of Heidelberg, Philosophenweg 16, 69120 Heidelberg, Germany\label{aff104}
\and
Institut de Recherche en Astrophysique et Plan\'etologie (IRAP), Universit\'e de Toulouse, CNRS, UPS, CNES, 14 Av. Edouard Belin, 31400 Toulouse, France\label{aff105}
\and
Universit\'e St Joseph; Faculty of Sciences, Beirut, Lebanon\label{aff106}
\and
Departamento de F\'isica, FCFM, Universidad de Chile, Blanco Encalada 2008, Santiago, Chile\label{aff107}
\and
Universit\"at Innsbruck, Institut f\"ur Astro- und Teilchenphysik, Technikerstr. 25/8, 6020 Innsbruck, Austria\label{aff108}
\and
Satlantis, University Science Park, Sede Bld 48940, Leioa-Bilbao, Spain\label{aff109}
\and
Infrared Processing and Analysis Center, California Institute of Technology, Pasadena, CA 91125, USA\label{aff110}
\and
Instituto de Astrof\'isica e Ci\^encias do Espa\c{c}o, Faculdade de Ci\^encias, Universidade de Lisboa, Tapada da Ajuda, 1349-018 Lisboa, Portugal\label{aff111}
\and
Cosmic Dawn Center (DAWN)\label{aff112}
\and
Niels Bohr Institute, University of Copenhagen, Jagtvej 128, 2200 Copenhagen, Denmark\label{aff113}
\and
Universidad Polit\'ecnica de Cartagena, Departamento de Electr\'onica y Tecnolog\'ia de Computadoras,  Plaza del Hospital 1, 30202 Cartagena, Spain\label{aff114}
\and
Kapteyn Astronomical Institute, University of Groningen, PO Box 800, 9700 AV Groningen, The Netherlands\label{aff115}
\and
Dipartimento di Fisica e Scienze della Terra, Universit\`a degli Studi di Ferrara, Via Giuseppe Saragat 1, 44122 Ferrara, Italy\label{aff116}
\and
Istituto Nazionale di Fisica Nucleare, Sezione di Ferrara, Via Giuseppe Saragat 1, 44122 Ferrara, Italy\label{aff117}
\and
INAF, Istituto di Radioastronomia, Via Piero Gobetti 101, 40129 Bologna, Italy\label{aff118}
\and
Astronomical Observatory of the Autonomous Region of the Aosta Valley (OAVdA), Loc. Lignan 39, I-11020, Nus (Aosta Valley), Italy\label{aff119}
\and
Department of Physics, Oxford University, Keble Road, Oxford OX1 3RH, UK\label{aff120}
\and
Aurora Technology for European Space Agency (ESA), Camino bajo del Castillo, s/n, Urbanizacion Villafranca del Castillo, Villanueva de la Ca\~nada, 28692 Madrid, Spain\label{aff121}
\and
Department of Mathematics and Physics E. De Giorgi, University of Salento, Via per Arnesano, CP-I93, 73100, Lecce, Italy\label{aff122}
\and
INFN, Sezione di Lecce, Via per Arnesano, CP-193, 73100, Lecce, Italy\label{aff123}
\and
INAF-Sezione di Lecce, c/o Dipartimento Matematica e Fisica, Via per Arnesano, 73100, Lecce, Italy\label{aff124}
\and
INAF - Osservatorio Astronomico di Brera, via Emilio Bianchi 46, 23807 Merate, Italy\label{aff125}
\and
INAF-Osservatorio Astronomico di Brera, Via Brera 28, 20122 Milano, Italy, and INFN-Sezione di Genova, Via Dodecaneso 33, 16146, Genova, Italy\label{aff126}
\and
ICL, Junia, Universit\'e Catholique de Lille, LITL, 59000 Lille, France\label{aff127}
\and
ICSC - Centro Nazionale di Ricerca in High Performance Computing, Big Data e Quantum Computing, Via Magnanelli 2, Bologna, Italy\label{aff128}
\and
Instituto de F\'isica Te\'orica UAM-CSIC, Campus de Cantoblanco, 28049 Madrid, Spain\label{aff129}
\and
CERCA/ISO, Department of Physics, Case Western Reserve University, 10900 Euclid Avenue, Cleveland, OH 44106, USA\label{aff130}
\and
Technical University of Munich, TUM School of Natural Sciences, Physics Department, James-Franck-Str.~1, 85748 Garching, Germany\label{aff131}
\and
Max-Planck-Institut f\"ur Astrophysik, Karl-Schwarzschild-Str.~1, 85748 Garching, Germany\label{aff132}
\and
Laboratoire Univers et Th\'eorie, Observatoire de Paris, Universit\'e PSL, Universit\'e Paris Cit\'e, CNRS, 92190 Meudon, France\label{aff133}
\and
Departamento de F{\'\i}sica Fundamental. Universidad de Salamanca. Plaza de la Merced s/n. 37008 Salamanca, Spain\label{aff134}
\and
Instituto de Astrof\'isica de Canarias (IAC); Departamento de Astrof\'isica, Universidad de La Laguna (ULL), 38200, La Laguna, Tenerife, Spain\label{aff135}
\and
Center for Data-Driven Discovery, Kavli IPMU (WPI), UTIAS, The University of Tokyo, Kashiwa, Chiba 277-8583, Japan\label{aff136}
\and
Ludwig-Maximilians-University, Schellingstrasse 4, 80799 Munich, Germany\label{aff137}
\and
Max-Planck-Institut f\"ur Physik, Boltzmannstr. 8, 85748 Garching, Germany\label{aff138}
\and
Dipartimento di Fisica - Sezione di Astronomia, Universit\`a di Trieste, Via Tiepolo 11, 34131 Trieste, Italy\label{aff139}
\and
California Institute of Technology, 1200 E California Blvd, Pasadena, CA 91125, USA\label{aff140}
\and
University of California, Los Angeles, CA 90095-1562, USA\label{aff141}
\and
Department of Physics \& Astronomy, University of California Irvine, Irvine CA 92697, USA\label{aff142}
\and
Departamento F\'isica Aplicada, Universidad Polit\'ecnica de Cartagena, Campus Muralla del Mar, 30202 Cartagena, Murcia, Spain\label{aff143}
\and
Instituto de F\'isica de Cantabria, Edificio Juan Jord\'a, Avenida de los Castros, 39005 Santander, Spain\label{aff144}
\and
Observatorio Nacional, Rua General Jose Cristino, 77-Bairro Imperial de Sao Cristovao, Rio de Janeiro, 20921-400, Brazil\label{aff145}
\and
Institute of Cosmology and Gravitation, University of Portsmouth, Portsmouth PO1 3FX, UK\label{aff146}
\and
Department of Computer Science, Aalto University, PO Box 15400, Espoo, FI-00 076, Finland\label{aff147}
\and
Instituto de Astrof\'\i sica de Canarias, c/ Via Lactea s/n, La Laguna 38200, Spain. Departamento de Astrof\'\i sica de la Universidad de La Laguna, Avda. Francisco Sanchez, La Laguna, 38200, Spain\label{aff148}
\and
Ruhr University Bochum, Faculty of Physics and Astronomy, Astronomical Institute (AIRUB), German Centre for Cosmological Lensing (GCCL), 44780 Bochum, Germany\label{aff149}
\and
Department of Physics and Astronomy, Vesilinnantie 5, University of Turku, 20014 Turku, Finland\label{aff150}
\and
Serco for European Space Agency (ESA), Camino bajo del Castillo, s/n, Urbanizacion Villafranca del Castillo, Villanueva de la Ca\~nada, 28692 Madrid, Spain\label{aff151}
\and
ARC Centre of Excellence for Dark Matter Particle Physics, Melbourne, Australia\label{aff152}
\and
Centre for Astrophysics \& Supercomputing, Swinburne University of Technology,  Hawthorn, Victoria 3122, Australia\label{aff153}
\and
DAMTP, Centre for Mathematical Sciences, Wilberforce Road, Cambridge CB3 0WA, UK\label{aff154}
\and
Kavli Institute for Cosmology Cambridge, Madingley Road, Cambridge, CB3 0HA, UK\label{aff155}
\and
Department of Astrophysics, University of Zurich, Winterthurerstrasse 190, 8057 Zurich, Switzerland\label{aff156}
\and
Department of Physics, Centre for Extragalactic Astronomy, Durham University, South Road, Durham, DH1 3LE, UK\label{aff157}
\and
IRFU, CEA, Universit\'e Paris-Saclay 91191 Gif-sur-Yvette Cedex, France\label{aff158}
\and
Oskar Klein Centre for Cosmoparticle Physics, Department of Physics, Stockholm University, Stockholm, SE-106 91, Sweden\label{aff159}
\and
Astrophysics Group, Blackett Laboratory, Imperial College London, London SW7 2AZ, UK\label{aff160}
\and
Univ. Grenoble Alpes, CNRS, Grenoble INP, LPSC-IN2P3, 53, Avenue des Martyrs, 38000, Grenoble, France\label{aff161}
\and
INAF-Osservatorio Astrofisico di Arcetri, Largo E. Fermi 5, 50125, Firenze, Italy\label{aff162}
\and
Dipartimento di Fisica, Sapienza Universit\`a di Roma, Piazzale Aldo Moro 2, 00185 Roma, Italy\label{aff163}
\and
Centro de Astrof\'{\i}sica da Universidade do Porto, Rua das Estrelas, 4150-762 Porto, Portugal\label{aff164}
\and
HE Space for European Space Agency (ESA), Camino bajo del Castillo, s/n, Urbanizacion Villafranca del Castillo, Villanueva de la Ca\~nada, 28692 Madrid, Spain\label{aff165}
\and
Theoretical astrophysics, Department of Physics and Astronomy, Uppsala University, Box 516, 751 37 Uppsala, Sweden\label{aff166}
\and
Mathematical Institute, University of Leiden, Einsteinweg 55, 2333 CA Leiden, The Netherlands\label{aff167}
\and
Institute of Astronomy, University of Cambridge, Madingley Road, Cambridge CB3 0HA, UK\label{aff168}
\and
Department of Astrophysical Sciences, Peyton Hall, Princeton University, Princeton, NJ 08544, USA\label{aff169}
\and
Space physics and astronomy research unit, University of Oulu, Pentti Kaiteran katu 1, FI-90014 Oulu, Finland\label{aff170}
\and
Center for Computational Astrophysics, Flatiron Institute, 162 5th Avenue, 10010, New York, NY, USA\label{aff171}}    

\date{}

\titlerunning{\Euclid 3D cosmic web reconstruction}
   \authorrunning{Euclid Consortium: K. Kraljic et al.}

 \abstract{
The ongoing \Euclid mission aims to measure spectroscopic redshifts for approximately two million galaxies using the \ha\ line emission detected in near-infrared slitless spectroscopic data from the Euclid Deep Fields, leveraging both the red and blue grisms. These measurements will reach a flux limit of $5\times 10^{-17}\,{\rm erg}\,{\rm cm}^{-2}\,{\rm s}^{-1}$
in the redshift range $0.4<z<1.8$, opening the door to numerous scientific investigations involving galaxy evolution, extending well beyond the mission's core objectives. The achieved \ha\ luminosity depth will lead to a sufficiently high sampling, enabling the reconstruction of the large-scale galaxy environment.
We assess the quality of the reconstruction of the galaxy cosmic web environment with the expected spectroscopic dataset in Euclid Deep Fields. The analysis is carried out on the Flagship and \gaea galaxy mock catalogues. The quality of the reconstruction is first evaluated using simple geometrical and topological statistics measured on the cosmic web network, namely the length of filaments, the area of walls, the volume of voids, and its connectivity and multiplicity. We then quantify how accurately gradients in galaxy properties with distance from filaments can be recovered. As expected, the small-scale redshift-space distortions, such as Fingers-of-God, have a strong impact on filament lengths and connectivity, but can be mitigated by compressing galaxy groups, identified with an anisotropic group finder, before skeleton extraction. The cosmic web reconstruction is biased when relying solely on \ha\ emitters. This limitation can be mitigated by applying stellar mass weighting during the cosmic web reconstruction. However, this approach introduces non-trivial biases that need to be accounted for when comparing to theoretical predictions. Redshift uncertainties pose the greatest challenge in recovering the expected dependence of galaxy properties, though the well-established stellar mass transverse gradients towards filaments can still be observed, albeit with reduced significance.

}
\maketitle

\section{Introduction}
\label{sec:Intro}

Since the first observations in the late 1970s, revealing the existence of coherent patterns on scales larger than those of galaxy clusters, mapping the large-scale structure of the Universe has become possible thanks to large galaxy redshift surveys. 

Early observations of the nearby Universe uncovering complex structures of interconnected superclusters \citep[e.g.,][]{Davis1982}, and enabling the unexpected discoveries of the first large cosmic voids \citep[e.g.,][]{Kirshner1981}, provided an initial hint that the spatial distribution of galaxies is highly inhomogeneous. It was quickly confirmed that this is a general feature of the large-scale distribution of galaxies, once observations from surveys covering wider areas on the sky started to become available. 
Beginning with the first redshift slices of the Center of Astrophysics redshift survey \citep[CfA,][]{deLapparent1986}, the
progressively increasing depth and coverage offered by the next generations of surveys such as the Las Campanas Redshift Survey \citep[LCRS,][]{Shectman1996}, the 2dF Galaxy Redshift Survey \citep[2dFGRS,][]{Colless2001}, the Sloan Digital Sky Survey \citep[SDSS,][for DR7]{SDSS,SDSS_DR72009}, the 6dF Galaxy Survey \citep[6dFGS,][]{Jones2004,Jones2009}, and the Galaxy and Mass Assembly \citep[GAMA,][]{GAMA2011} survey, allow us today to map the large-scale structure of the nearby Universe ($z \lesssim 0.3$) in unprecedented detail. These surveys revealed a cosmic landscape where galaxies were distributed within high-density peaks, intermediate-density filaments, and walls, which enclose low-density, nearly empty voids.
This view has been extended up to $z \simeq 1$ by the VIMOS Public Extragalactic Redshift Survey \citep[VIPERS,][]{Vipers2014} that encompasses a volume and galaxy sampling density comparable to those of spectroscopic surveys of the local Universe. Further improvement, in terms of significantly increased galaxy number density and depth compared to any of these surveys in the comparable redshift range, will be achieved by the ongoing Dark Energy Spectroscopic Instrument \citep[DESI;][]{DESI2016} collaboration, which already collected high-confidence spectroscopic redshifts \citep[][]{McCullough2024} for more
than ten million galaxies \citep{DESI_DR1_2025}.

Mapping the large-scale structure in three dimensions at even higher redshifts ($z \gtrsim 1$) is currently impossible with existing spectroscopic surveys, due to their rapidly decreasing completeness and sampling number density. For the time being, the density field at high redshifts  is observationally accessible only through the tomographic reconstruction using the Lyman-$\alpha$ forest absorption of light from bright background sources, such as quasars, typically at $z\sim2.5-3$ \citep[e.g.,][]{Lee2016,Ravoux2020}. The redshift range $1 \lesssim z \lesssim 2$, near the peak epoch of star formation \citep[e.g.,][]{MadauDickinson2014} -- the period when star formation in the Universe was at its highest -- remains largely uncharted territory in understanding the co-evolution of galaxies and large-scale structures.

The web-like pattern observed in the distribution of galaxies, spanning scales from a few to over a hundred megaparsecs and revealed by large galaxy redshift surveys, is now understood within the framework of the so-called cosmic web \citep[e.g.,][]{Klypin1983,Klypin1993,Bond1996}. This structure connects observed galaxy clusters through a network of filaments, which arise from initial fluctuations in the primordial density field and are amplified by anisotropic gravitational collapse \citep{lynden-bell64,zeldovich70} during later cosmic times. One of the most important features of this network is that it naturally sets the large-scale environment within which galaxies form and evolve. Since more than a decade now, interest has been shifting from extensively studied high-density regions, such as galaxy groups and clusters \citep[e.g.,][and references therein]{davis1976,Dressler1980,Dressler1997,Goto2003,Blanton2003,Baldry2006,Bamford2009,Cucciati2010,Burton2013,Cucciati2017}, toward intermediate-density filaments and walls. These anisotropic large-scale environments seem to play a role in shaping at least some of the galaxy properties. 
Indeed, observational studies of the local and higher-$z$ Universe ($z \lesssim 0.9$) have demonstrated that more massive and/or passive galaxies tend to reside closer to large-scale filaments compared to their lower-mass and/or star-forming counterparts \citep[e.g.,][]{Chen2017,Kuutma2017,Malavasi2017,Kraljic2018,Laigle2018,Winkel2021}. This trend aligns qualitatively with results from large hydrodynamical simulations \citep[][]{Kraljic2018,Laigle2018,Hasan2023,Bulichi2024} and with theoretical expectations \citep{Musso2018}, 
since the assembly history of galaxies encoded in a conditional excursion set is biased by the eigenvalues and eigenvectors of  anisotropic tides.   

The role of the cosmic web in modulating other galaxy properties, i.e. beyond stellar mass and star-formation activity, could also be modelled in this framework, but has so far only been explored at low redshifts ($z\lesssim 0.2$). In particular, it was found that after controlling for stellar mass, halo mass, or density, a clear signature of the impact of the cosmic filaments, walls, and nodes can be found for galaxy age, stellar metallicity, and element abundance ratio [$\alpha$/Fe] \citep[][]{Winkel2021}, gas-phase metallicity \citep[][]{Donnan2022}, or \ion{H}{i} fraction \citep[][]{Kleiner2017,Crone_Odekon2018}. Another property of galaxies that shows a dependence on their large-scale environment, as expected by the tidal torque theory (for a review, see \citeauthor{Schafer2009} \citeyear{Schafer2009} and \citeauthor{Codis2015} \citeyear{Codis2015} for the corresponding theory of constrained tidal torques near filaments), is their angular momentum (or spin) orientation, as identified in low-$z$ observations \citep[e.g.][]{LeeErdogdu2007,Tempel2013,TempelLibeskind2013,Zhang2013,Zhang2015,Pahwa2016,Krolewski2019,Kraljic2021,Barsanti2022}

An alternative approach to studying the impact of the cosmic network on galaxy properties involves analysing its connectivity, i.e., the number of filaments connected to a given node of the cosmic web. This serves as a probe for the geometry of accretion at halo and galaxy scales \citep[][]{Codis2018b}. When applied to SDSS data \citep[see also, e.g.,][for measurements on galaxy groups and cluster scales]{DarraghFord2019,Sarron2019,Einasto2020,Einasto2021,Smith2023}, more massive galaxies were found to exhibit higher connectivity \citep[][]{Kraljic2020a}, a result consistent with theoretical predictions \citep[][]{Codis2018b}. At fixed stellar mass, galaxy properties such as star-formation activity and morphology also show some dependence on connectivity: less star-forming and less rotation-supported galaxies tend to be more connected. This trend is qualitatively consistent with the findings of hydrodynamical simulations \citep[][]{Kraljic2020a}.

Our understanding of how the anisotropic large-scale environment shapes galaxy properties remains observationally constrained to the low-redshift Universe ($z \lesssim 0.9$), with the majority of studies focusing on the nearby Universe ($z \lesssim 0.2$). However, ongoing and upcoming surveys, such as \Euclid \citep{laureijs2011}, the PFS (Prime Focus Spectrograph) Galaxy Evolution survey \citep[][]{PFS_GE2022} at the Subaru Telescope, MOONRISE \citep[the main GTO MOONS extra-galactic survey,][]{MOONRISE2020} at the Very Large Telescope (VLT), and Nancy Grace Roman Space Telescope \citep[Roman;][]{Akeson_2019}, will enable us to extend these analyses to redshifts between 1 and 2. This epoch is critical for unraveling the details of gas accretion onto galaxies, its conversion into stars, and the physical processes responsible for the quenching of star formation.

The ongoing Euclid survey \citep{EuclidSkyOverview} will primarily focus on characterising the nature of dark energy and understanding the distribution of dark matter in the Universe. However, while \Euclid has been specifically prepared to meet these core science objectives, it will also address a wide range of other scientific questions. This ``by-product" research will be notably made possible by an extensive database of approximately two million galaxies observed over a $53\, {\rm deg}^2$ area (the so-called Euclid Deep Fields, EDFs hereafter). This will, for the first time, enable detailed tracing of the large-scale environment of galaxies between redshifts 1 and 2, providing new insights into its connection to galaxy growth and properties.

In this paper, we examine to what extent the cosmic web environment of galaxies can be reconstructed using the upcoming spectroscopic dataset in EDFs. We use Euclid Deep mock galaxy catalogues to evaluate the quality of cosmic web reconstruction. Specifically, we investigate how factors such as the selection function, redshift-space distortions, and the anticipated uncertainties in redshift measurements affect the reconstruction quality. This analysis is conducted using various geometrical and topological properties of the cosmic web.
We also examine how accurately stellar-mass gradients toward cosmic web filaments can be recovered. A key objective of this study is to provide practical guidelines for cosmic web reconstruction using the EDF dataset. While this paper focuses on the three-dimensional reconstruction of the cosmic web, its two-dimensional counterpart is discussed in \cite{Malavasi_2025}, and the reconstruction of cluster-scale filaments is addressed in Sarron et al. (in prep.).
In the present approach, the skeleton is used as an effective summary statistics. An alternative is simulation-based inference via forward modelling, which aims to match the full data set to mocks without constructing specific estimators \citep[e.g.,][]{SBI_Cranmer2020}. These methods have recently gained some traction given the available computing power \citep[e.g.,][]{Angulo_2021,Kobayashi_2022,Hou_2024}. 
Similarly, alternative approaches exist for the reconstruction of the large-scale environment \citep[see][for a detailed comparison of different estimators]{Libeskind_2018}.

The paper is organised as follows.
Section~\ref{sec:data} presents the Euclid Deep Survey and simulated galaxy catalogues, together with methods used to create mock data and extract the cosmic web. Section~\ref{sec:results} quantifies the quality of the cosmic web reconstruction and its ability to recover the stellar-mass gradients with respect to cosmic filaments. Section~\ref{sec:discussion} provides guidelines for cosmic web reconstruction with the Euclid Deep dataset, outlines possible science cases, and discusses possible synergies with other surveys. Section~\ref{sec:conclusion} summarises the key results and concludes. Finally, Appendix~\ref{appendix:stats} investigates the impact of the selection function of galaxies and reduced sampling on the distribution of filament lengths, while Appendix~\ref{appendix:connect} focuses on their impact on connectivity and multiplicity. Appendix~\ref{appendix:grads} complements the analysis of stellar mass gradients.

\section{Data}
\label{sec:data}

\subsection{The Euclid Deep Survey (EDS)}
\label{sec:EDS}
The EDS is described in \cite{EuclidSkyOverview}, but we summarise its main characteristics below. In brief, it includes three non-contiguous fields: the EDF North (EDF-N, $20\, {\rm deg}^2$), the EDF Fornax (EDF-F, $10\, {\rm deg}^2$) and the EDF South (EDF-S, $23\, {\rm deg}^2$). All together, they cover $53\, {\rm deg}^2$ and will reach a $5\, \sigma$ point source depth of at least two magnitudes deeper than the Euclid Wide Survey \citep[EWS,][]{Scaramella-EP1}, i.e. $\sim 28.2$~AB mag in $\IE$ and 26.4~AB mag in $\YE$, $\JE$, and $\HE$. These fields will be complemented by deep photometry from the Cosmic DAWN survey in the UV, optical, and IR that will be very valuable for deriving reliable masses and star-formation histories for the observed galaxies \citep{EP-McPartland,EP-Zalesky}. In addition, the Euclid Auxiliary Fields (EAF), designed to serve the calibration of the VIS and NISP instruments, will reach an almost similar depth to the EDF, with the special case of the self-calibration field which will be ultra-deep (29.4~AB mag in $\IE$ and 27.7~AB mag in $\YE$, $\JE$, and $\HE$ in $2.5\, {\rm deg}^2$). 
The EAF will cover a total of $9\, {\rm deg}^2$ distributed over seven fields, all of them already having deep multi-wavelength coverage in a large number of bands covering the whole electromagnetic spectrum. The forecasts presented in this paper are relevant for both the EDF and at least the three largest EAF (COSMOS, SXDS, and the self-calibration field).

On the spectroscopic side, the red grisms will allow the detection of \ha\ emitters over $0.84<z<1.88$ 
and will operate over the full survey (EWS, EDS, and EAF). The blue grism will operate only on the EDF and EAF, and will allow the \ha\ detection down to $z=0.41$. Since the forecasts presented in the present work concern the EDS, we adopt the redshift range $0.4<z<1.8$ throughout this paper. The expected flux limit in the EDFs is $5\times 10^{-17}\,{\rm erg}\,{\rm cm}^{-2}\,{\rm s}^{-1}$ for sources with ${\rm S/N}>3.5$. These objects virtually all meet the expected photometric magnitude limits of the EDS.

\subsection{Performances of the EDS for cosmic web mapping}
\label{sec:perf}

\paragraph{Redshift accuracy:} The red grisms have a resolving power of ${\cal{R}_{\rm RG}} \geq 480$ for sources with 0\arcsecf5 diameter, and the blue grism has a resolving power of ${\cal{R}_{\rm BG}} \geq 400$. The expected redshift accuracy is $\sigma(z)<0.001(1+z)$, confirmed by the analysis using the first \Euclid Quick Data Release (Q1) \citep[][]{Q1-TP007}.

\paragraph{Completeness and purity:} Slitless spectroscopy is essentially dispersed imaging. The features of the various overlapping spectra must be separated from each other by taking advantage of the fact that the same field is observed with several dispersion angles. A complex extraction and decontamination process produces 1D spectra, performed with the OU-SIR unit \citep{Q1-TP006}.The probability distribution function of the redshift is then determined from these spectra using a spectral template fitting algorithm, performed by the OU-SPE unit \citep{Q1-TP007}. The algorithm returns the most probable redshift as the first solution, which is then used for cosmic web reconstruction. In addition, a reliability score is assigned to each galaxy and a threshold is determined to obtain a good compromise between the completeness and purity of the galaxy sample\footnote{Completeness is defined as the fraction of objects recovered with a correct redshift from a full sample above a certain intrinsic flux limit and in a certain redshift range, while purity is the fraction of correctly measured redshifts in the recovered sample.}.

This threshold will exclude a significant fraction of galaxies above the flux limit, leading to incompleteness and a higher flux cut. For the cosmic web reconstruction, we will need high purity, and thus a very high OSU-SPE reliability cut. Similarly to \cite{2022A&A...658A..20H}, we assume that it will lead to a completeness of 60\% and a higher flux limit (see Sect.~\ref{sec:mocking_EDS}). This high purity will in turn allow us to neglect the existence of catastrophic redshift failures in our current analysis.

\subsection{Mock catalogues}
In order to make forecasts about the quality of the cosmic web reconstruction with the upcoming Euclid data, we rely on two simulated sets, the Euclid Flagship 2 and the GAEA simulations.

\subsubsection{The Flagship galaxy mock catalogues}

The Euclid Flagship Simulation (\flag, hereafter) is described in great detail in \cite{EuclidSkyFlagship}.  Here, we only summarise its main features.  

The Flagship lightcone, produced on the fly out to $z=3$, is based on an $N$-body dark matter simulation of four trillion dark matter particles in a periodic box of $3600\, \hMpc$ on a side, leading to the particle mass resolution of $\sim 10^9\, \msunh$. The simulation was performed using the code {\tt PKDGRAV3} \citep{Potter_Stadel2016}, 
with cosmological parameters $h=0.67$, $\Omm=0.319$, $\Omb=0.049$, $n_{\rm s}=0.96$, $A_{\rm s}=2.1\times10^{-9}$.

Dark matter halos, identified with the {\tt ROCKSTAR} halo finder \citep{2013ApJ...762..109B}, were populated by galaxies using a combination of halo occupation distribution (HOD) and abundance matching (AM) techniques, satisfying some of the observed relations between galaxy properties. Following the HOD prescription, each halo was assigned a central galaxy and a number of satellites depending on the halo mass, reproducing observational constraints of galaxy clustering in the local Universe \citep{2011ApJ...736...59Z}. Luminosities were assigned to galaxies by performing abundance matching between the halo mass function and the galaxy luminosity function, calibrated on the local observations \citep{2003ApJ...592..819B, 2005ApJ...631..208B} in a way to match the observed clustering of galaxies as a function of color \citep{2011ApJ...736...59Z}.

The star-formation rate (SFR) was computed from the rest-frame ultraviolet luminosity, following the \cite{Kennicutt1998} relation for a Chabrier initial mass function \citep[IMF;][]{Chabrier2003}. The stellar mass was computed from the galaxy luminosity and the stellar mass-to-luminosity ratio. The \ha\ line flux was computed from the SFR using the \cite{Kennicutt1998} relation adapted to the \cite{Chabrier2003} IMF, using the un-extincted ultraviolet absolute magnitude. The dust extinguished \ha\ flux was then computed following \cite{Calzetti2000} and \cite{Saito2020}.
Finally, the resulting \ha\ flux distribution was calibrated to the empirical models of \cite{2016A&A...590A...3P}, namely model~1 and model~3 (named m1 and m3, hereafter).

Validation of the \flag galaxy catalogue by comparing it with observations is presented in \cite{EuclidSkyFlagship}, showing good agreement for many galaxy properties, distributions, and relations. Among them, the stellar mass function and the SFR-stellar mass relation show good consistency when compared to observational data up to $z\sim3$.

The full galaxy catalogue covers one octant of the sky ($\sim 5157 \,{\rm deg}^2$) centred at approximately the North Galactic Pole ($145\, {\rm deg} < {\rm RA} < 235 \, {\rm deg}$, $0\, {\rm deg} < {\rm Dec} < 90\, {\rm deg}$), and samples a redshift range $0 < z < 3$. However, only a limited area ($150\, {\rm deg} < {\rm RA} < 155 \, {\rm deg}$, $5\, {\rm deg} <{\rm Dec}< 10\, {\rm deg}$), with no magnitude or line flux cut, can be used to simulate the EDF (see Sect.~\ref{sec:mocking_EDS}). \flag catalogues were accessed through CosmoHub \citep[][]{Carretero2017,Tallada2020}.

\subsubsection{The GAEA lightcone}

GAEA (GAlaxy Evolution and Assembly)\footnote{\href{https://sites.google.com/inaf.it/gaea/}{https://sites.google.com/inaf.it/gaea/}} is a semi-analytic model \citep[see e.g.,][]{DeLucia2014,Hirschmann2016,Fontanot2017} run on the Millennium simulation \citep{Springel2005} containing $2160^3$ dark matter particles in a periodic box of $500 \, \hMpc$ on one side, leading to a particle mass resolution of $8.6 \times 10^8 \msunh$. The Millennium simulation was performed using the code {\tt GADGET} \citep{Springel2001} with cosmological parameters   
$h=0.73$, $\Omm=0.25$, $\Omb=0.045$, $n_{\rm s}=1$, $\sigma_8=0.9$. From this simulation, a lightcone was created following \cite{Zoldan2017}. 

The GAEA semi-analytic model traces the evolution of galaxy populations inside DM halos by self-consistently treating gas, metal and energy recycling, as well as chemical enrichment, using physically or observationally motivated prescriptions.
Two versions of this model have been implemented. These are described in \cite{Hirschmann2016} and \cite{Fontanot2020}, and named respectively ECLH and ECLQ in the following. One of the major differences between these two models is an improved prescription for AGN feedback in ECLQ. While both ECLH and ECLQ models include radio-mode AGN feedback, ECLQ also includes a quasar-driven wind component, following an improved modelling of cold-gas accretion onto the supermassive black hole, based on both analytic approaches and high-resolution simulations.

To compute the dust-attenuated \ha\ flux, the non-attenuated \ha\ luminosity was derived from the SFR following the \cite{Kennicutt1994} relation (rescaled to follow the \citealt{Chabrier2003} IMF). The obtained \ha\ flux has then been attenuated by dust following the dust attenuation curve of \cite{Calzetti2000}.

GAEA was shown to well reproduce a number of observations, such as the evolution of the galaxy stellar mass function and the cosmic SFR density up to $z \sim 7$ \citep{Fontanot2017}, or the evolution of the stellar mass-gas metallicity relation up to $z \sim 2$ \citep{Hirschmann2016}.

The full galaxy catalogue covers a region on the plane of the sky with a diameter of $5.27 \deg$ ($\sim 22 \deg^2$), samples a redshift range $0 < z < 4$, and includes all galaxies with magnitude $H \leq 25$ (i.e. deep enough to allow for the construction of EDF mocks; see Sect.~\ref{sec:mocking_EDS}).

\subsubsection{Mocking the EDS}
\label{sec:mocking_EDS}

To mimic the EDFs, we select from the full Flagship volume the $\sim 25\, {\rm deg}^2$ region that has no magnitude or line flux cut, in the redshift range $0.4 < z < 1.8$. The same redshift selection is applied to the \gaea galaxy catalogues, having comparable sky coverage.

\paragraph{Flux limit:}
In this work, we select all galaxies with an \ha\ flux limit of
$6\times 10^{-17}\,{\rm erg}\,{\rm cm}^{-2}\,{\rm s}^{-1}$, which is a more conservative threshold compared to the predicted limit (see Sect.~\ref{sec:EDS}).
This selection ensures that, as expected, all objects satisfy the EDS photometric magnitude limits. To assess the impact of galaxy selection on the quality of cosmic web reconstruction, we also consider a stellar mass-limited sample of galaxies for each mock. In order to keep the same total number density, we select from each mock, ordered in decreasing stellar mass, the same number of galaxies as in its flux-limited counterpart. The resulting stellar mass limits are $10^{9.5}\, \msun$, $10^{9.8}\, \msun$, for the Flagship m1 and m3 models, respectively, and $10^{9.8}\, \msun$ for both GAEA models. 
The \mstar- and \ha\ flux-limited samples therefore contain by construction different populations of galaxies. 
\ha\ flux-limited samples have fewer satellites (and therefore more centrals), and miss an important fraction of quiescent massive galaxies compared to \mstar-limited catalogues.

In the following, flux-limited catalogues are denoted \deep, while mass-limited catalogues are denoted with `\mstar' in the subscript, i.e. \deepmass.

\paragraph{Completeness:}
To mimic the expected sample completeness of $60\, \%$ due to observational effects (see Sect.~\ref{sec:perf}), we randomly discard $40 \, \%$ of our galaxies \citep[as done in e.g.,][]{2022A&A...658A..20H}. We note that the incompleteness will not be truly random. However, since the incompleteness has not been characterised in detail against parameters, such as the local projected density, a more precise model remains currently unavailable.
In the following, the corresponding catalogues are denoted with `$60 \, \%$' in the subscript, for example, \deepsamp.

\paragraph{Redshift uncertainties:}
To model the redshift uncertainties associated with Euclid measurement, we perturb galaxy redshifts, as taken from the mocks, using a Gaussian redshift-dependent error with an RMS of $\sigma_z = 0.001(1+z)$. 
For each model, we make five realisations and the corresponding catalogues are denoted with `noise' in the subscript in the following, for example, \deepnoise.

\paragraph{Fiducial realisation:} Applying all these constraints to the selection of galaxies, i.e. \ha\ flux limit of $6\times 10^{-17}{\rm erg}\,{\rm cm}^{-2}\,{\rm s}^{-1}$, redshift uncertainty $\sigma_z =0.001(1+z)$, and completeness of 60\%, leads to fiducial realisations mimicking the EDFs, denoted in the following \deepnoisesamp.

\subsubsection{Qualitative assessment of the mocks}
\label{sec:quality_mocks}

\begin{figure}
\centering\includegraphics[width=\columnwidth]{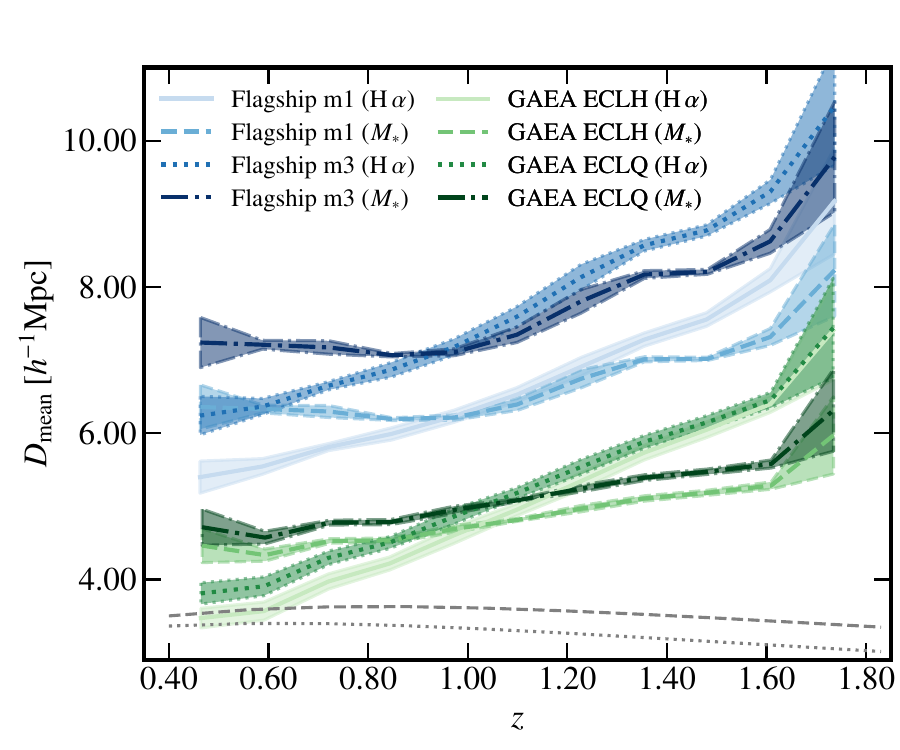}
\caption{Mean intergalactic separation in \flag (blue colours) and \gaea (green colours) simulations for all considered models (m1 and m3 for \flag, ECLH and ECLQ for \gaea). For each model, the fiducial galaxy selection based on the \ha\ flux (\ha) is compared to the stellar mass-based selection (\mstar). Shaded regions correspond to the standard deviation across five mocks for each model. Grey dotted and dashed lines represent the corresponding redshift uncertainties converted into distances for \flag and \gaea simulations, respectively.
\gaea models show smaller differences between each other and compared to the \flag simulation models, in particular for the \ha\ selection of galaxies. Above $z\sim0.9$, the \ha\ selection follows more closely the selection based on stellar mass for the \flag simulation, with respect to \gaea.
}
\label{fig:mean_dist}
\end{figure}

\begin{figure*}
\centering\includegraphics[width=0.9\textwidth]{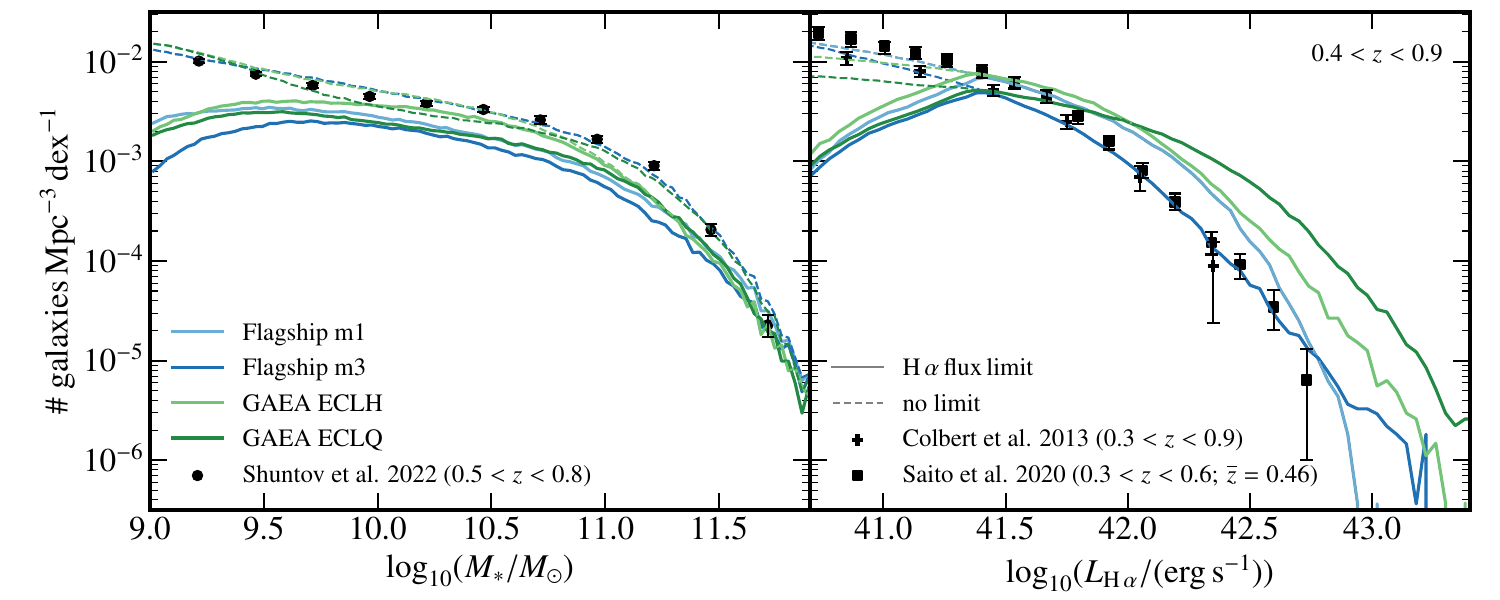}\\[-5pt]
\centering\includegraphics[width=0.9\textwidth]{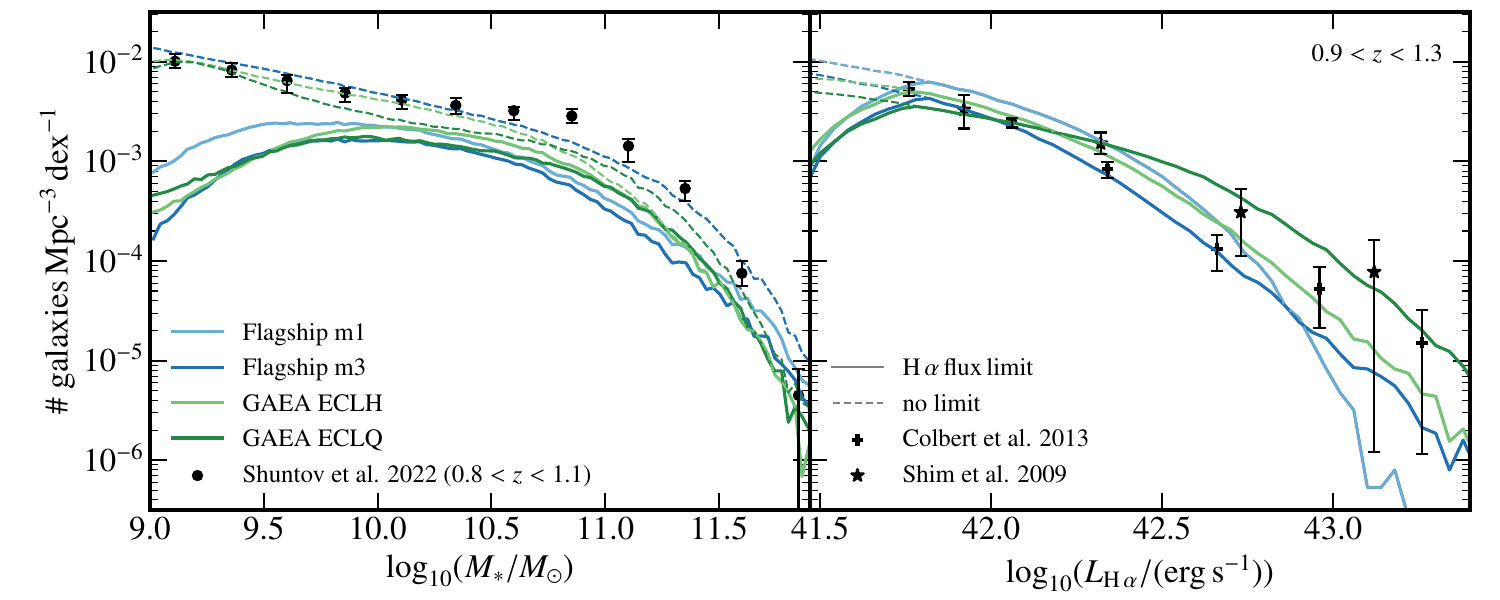}\\[-5pt]
\centering\includegraphics[width=0.9\textwidth]{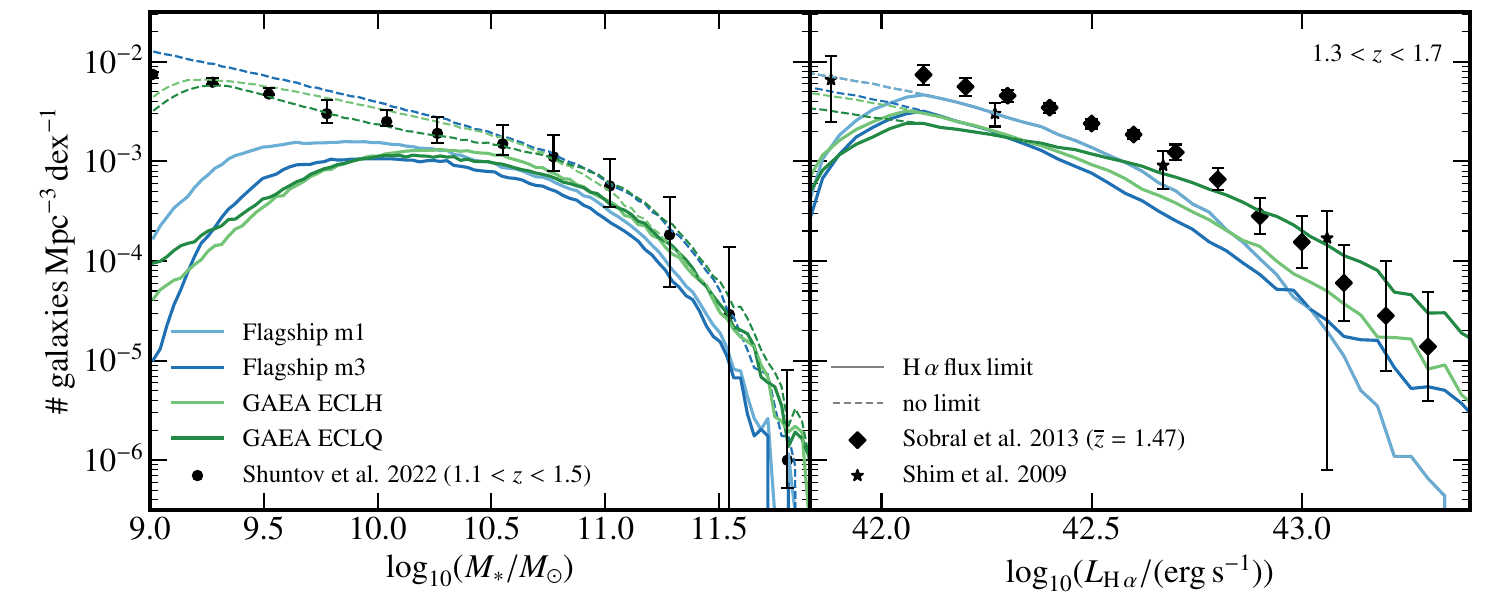}
\caption{Stellar mass (left) and \ha\ luminosity  functions (right) in three redshift bins, $0.4<z<0.9$ (top), $0.9<z<1.3$ (middle), and $1.3<z<1.7$ (bottom), in all models considered in this work for \ha\ flux limited samples (coloured solid lines) and for samples without any limit (coloured dashed lines). Black symbols correspond to observational data at these redshifts, COSMOS2020 \citep[][]{Shuntov2022} for stellar mass functions and the Emission Line COSMOS catalogue \citep[][]{Saito2020}, HST-NICMOS \citep{Shim2009}, HST WISP \citep{Colbert2013}, and HiZELS \citep{Sobral2013} for \ha\ luminosity functions.
}
\label{fig:MF_LF}
\end{figure*}

\begin{figure*}
\centering\includegraphics[width=0.497\textwidth]{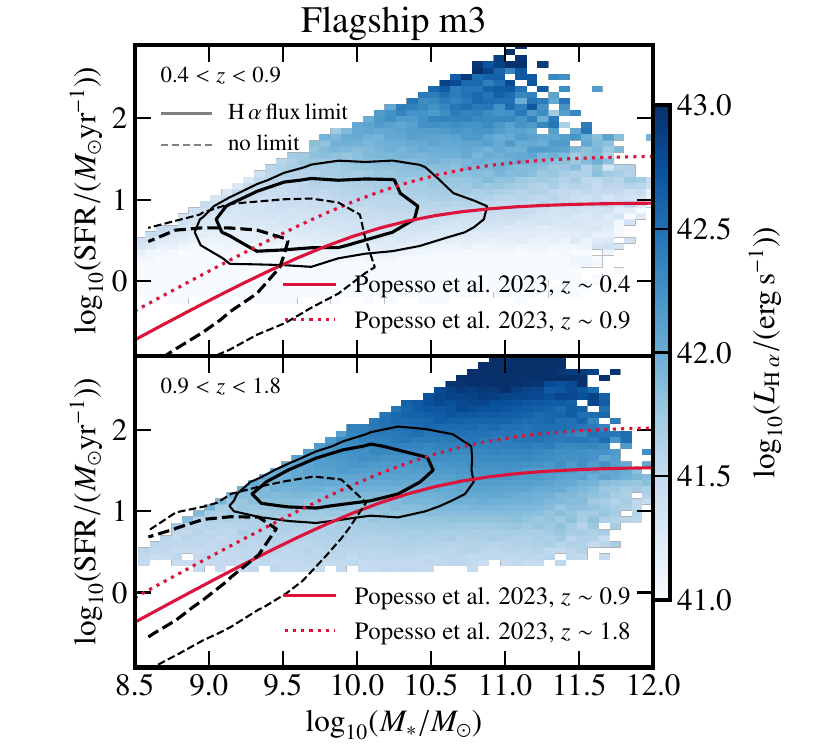}
\centering\includegraphics[width=0.497\textwidth]{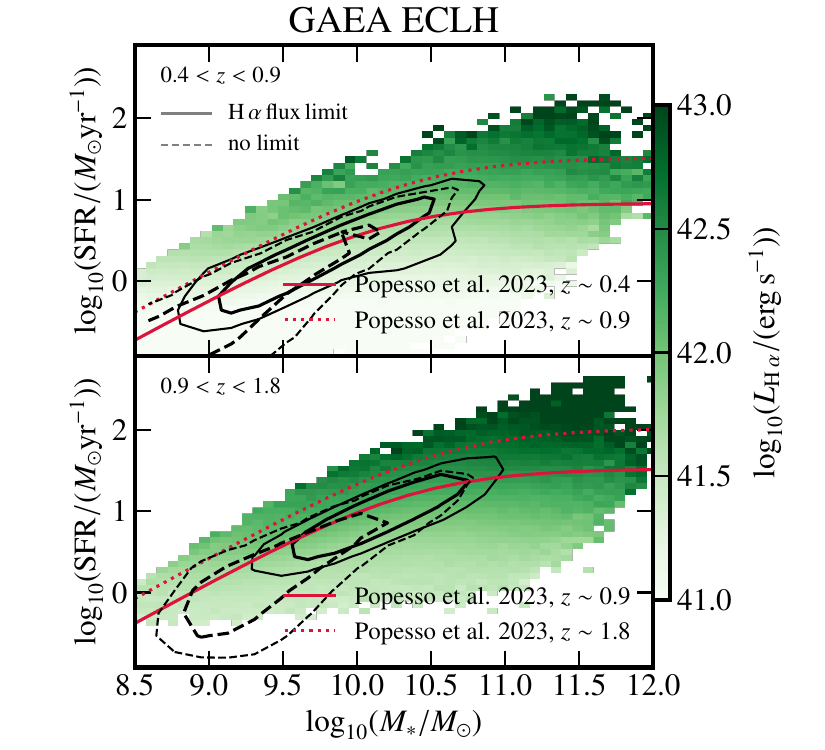}
\caption{The star-forming main sequence in the \flag (left) and \gaea (right) simulations (models m3 and ECLH, respectively) at $0.4<z<0.9$ (top) and $0.9<z<1.8$ (bottom), color-coded by the \ha\ luminosity, which essentially correlates with SFR. The red lines correspond to the compilation of observational data presented in \cite{Popesso2023}. The contours encompass 50\% and 75\% of the galaxy distribution for \ha\ flux-limited sample (solid lines) and for the sample without any limit (dashed lines). 
}
\label{fig:main_sequence}
\end{figure*}

\begin{figure*}
\centering\includegraphics[width=0.497\textwidth]{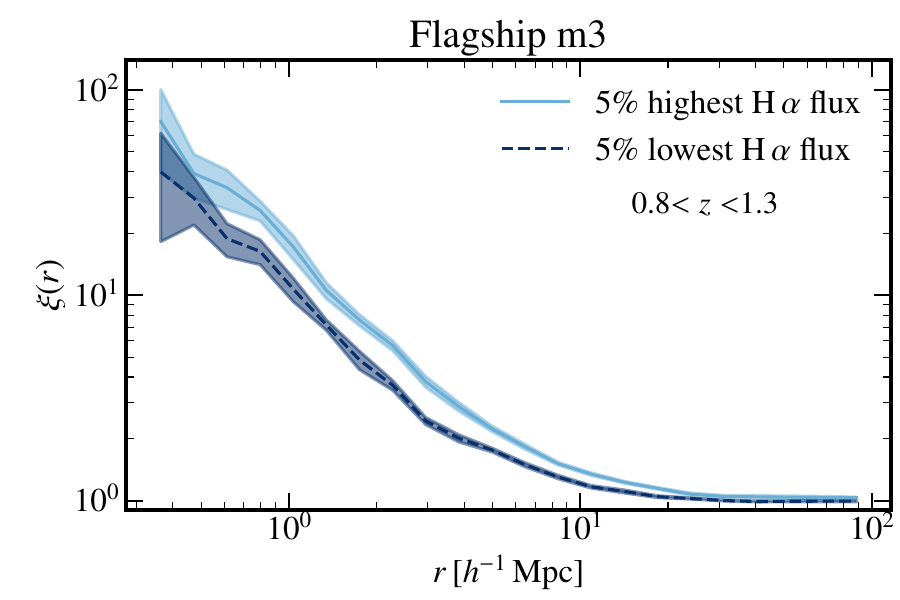}
\centering\includegraphics[width=0.497\textwidth]{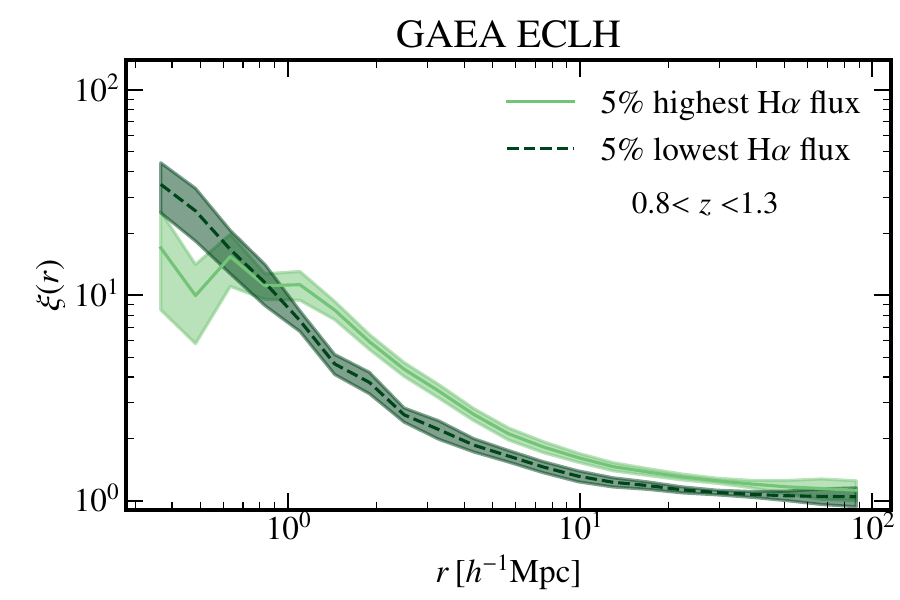}
\caption{Two-point correlation functions of galaxies in the redshift range $0.8<z<1.3$ for galaxies with the $5 \, \%$ highest and lowest \ha\ flux in the \flag m3 (left) and \gaea ECLH (right) mocks. 
Shaded regions correspond to jackknife error bars.
At small separations ($\lesssim 1 \, h^{-1}\, {\rm Mpc}$), brightest galaxies show enhanced clustering in \flag compared to their low \ha\ flux counterparts, while brightest galaxies in \gaea show comparable (or reduced) clustering to their lower brightness counterparts. At larger separations ($\gtrsim 1 \,h^{-1}\, {\rm Mpc}$), brightest galaxies are more clustered compared to their low \ha\ flux counterparts in all mocks.
}
\label{fig:2pcf}
\end{figure*}

Being based on different $N$-body simulations and using different methods to assign galaxies to dark matter halos and to compute their properties, the \flag and \gaea simulations are expected to provide different predictions in terms of galaxy distribution, galaxy clustering and their dependence on stellar mass or \ha\ flux. In this section, we provide a qualitative assessment of the mock catalogues without additional observational biases, i.e. using true redshifts and a full sampling, by focusing on their mean intergalactic separation, stellar mass and \ha\ luminosity functions, main sequence and \ha-dependent galaxy clustering.

\paragraph{Mean intergalactic separation:} 
The mean number density of \ha-selected galaxies across the entire redshift range is higher in \gaea compared to \flag,  
$1.2\times10^{-2}\, (\hMpc)^{-3}$ and $9.7\times10^{-3}\, (\hMpc)^{-3}$ for ECLH and ECLQ \gaea models, respectively, and $3.9\times10^{-3}\, (\hMpc)^{-3}$ and $2.5\times10^{-3}\, (\hMpc)^{-3}$ for the two \flag models m1 and m3, respectively. This translates into the smaller mean intergalactic separations for \gaea compared to \flag, shown in Fig.~\ref{fig:mean_dist}.
The two \gaea models show very little difference in mean intergalactic separations in particular when galaxy selection is based on the \ha\ flux and above $z \sim 0.9$. The two \flag models show larger differences in the mean intergalactic separation compared to \gaea, regardless of galaxy selection, which is roughly constant across the entire redshift range considered in this work.
For comparison, the mean galaxy separations are shown also for the \mstar-limited samples for all mocks. As expected, they show a flatter redshift dependence compared to \ha\ galaxy selection for all the mocks, especially at redshifts below $z \sim 1.4$. We anticipate that the expected large redshift uncertainties (see the dotted lines in Fig.~\ref{fig:mean_dist}) are going to play a major role in reducing the quality of the cosmic web reconstruction.

\paragraph{Stellar mass and \ha\ luminosity functions:} The stellar mass (SMF) and \ha\ luminosity functions (LF) for all \flag and \gaea models are displayed in Fig.~\ref{fig:MF_LF} in three different redshift bins. To highlight the impact of \ha\ flux selection on these observables, we also show the results for samples without any \ha\ flux limit. The SMFs of \flag and \gaea models are compared with the COSMOS2020 observational dataset \citep{Shuntov2022}, while for the LFs the models are compared to data from the Emission Line COSMOS catalogue \citep[][]{Saito2020}, HST-NICMOS \citep{Shim2009}, HST WISP \citep{Colbert2013}, and HiZELS \citep{Sobral2013}. 

Overall, the SMFs of \flag and \gaea galaxies without the \ha\ flux limit (dashed lines) agree well with the observational measurements in all redshift bins.\footnote{We note that we have not transformed the stellar masses of mock galaxies to units of the Hubble constant used in the observed galaxy catalogues.} This is not surprising, particularly for \gaea, since the SMF was used to calibrate their models. 

The application of an \ha\ flux limit translates non-trivially to the change in the SMF in a model-dependent way.   
The SMFs for the \ha\ flux-limited samples show a reduced amplitude at all stellar masses for all models except \gaea ECLH for which the SMFs overlap at the high-mass end (the stellar mass at which the deviation occurs depends on the redshift). Overall, the SMFs follow a qualitatively similar trend for all models. 
The steep increase at the high-mass end is followed by a shallower slope at intermediate masses and a downturn of the SMF at low masses. The stellar mass at which the SMFs start to decrease at the low-mass end increases with increasing redshift and depends on the model. For \flag it corresponds to $\sim10^{9.5}\, \msun$, $10^{9.5}\, \msun$, and $10^{9.65}\, \msun$ in the three increasing redshift bins, while for \gaea the corresponding masses are slightly higher, particularly in the two highest redshift bins ($10^{9.75}\, \msun$ and $10^{10.15}\, \msun$).

The \ha\ LF of observed galaxies at low redshifts \citep[$0.4 < z < 0.9$;][]{Colbert2013,Saito2020} is well reproduced by \flag, the model m3 in particular. This is expected, given that the LF of galaxies was among the observables used to calibrate the \flag mocks. \gaea mocks reproduce reasonably well the LF of galaxies at the faint end, whereas they over-predict the number of observed galaxies at the bright end. At intermediate redshifts ($0.9 < z < 1.3$), the \ha\ LF of the observed galaxies \citep[][]{Shim2009,Colbert2013} lies well within the range spanned by the four galaxy mocks, making them a fairly good representation of galaxies
at luminosities $L_{{\rm H}\, \alpha} \gtrsim 10^{41.8}\, {\rm erg} \, \rm{s}^{-1}$, corresponding to the \ha\ flux limit of EDS at $z=1.3$.  At higher redshifts ($1.3<z<1.7$), the \ha\ LFs of all mocks under-predict observations \citep[][]{Colbert2013,Saito2020} below $L_{{\rm H}\, \alpha} \sim 10^{42.8}\, {\rm erg} \, \rm{s}^{-1}$, while at higher luminosities, most models agree well with observations. \gaea in particular reproduces well the observed LF at the bright end in spite of the fact that this observable has not been used to calibrate the models.   

We note that the drop (down turn) of the LF at the faint end for the \ha\ flux-limited sample is a consequence of the relatively large width of the redshift bins. The luminosity at which the two LFs (for full and \ha\ flux-limited samples) start to deviate corresponds to the \ha\ flux limit of the upper bound of each redshift bin. 
This corresponds to the \ha\ luminosity of $10^{40.51}\, {\rm erg} \, \rm{s}^{-1}$, 
$10^{41.37}\, {\rm erg} \, \rm{s}^{-1}$, $10^{41.77}\, {\rm erg} \, \rm{s}^{-1}$, and $10^{42.07}\, {\rm erg} \, \rm{s}^{-1}$, at redshifts $0.4$, $0.9$, $1.3$, and $1.7$, respectively.

\paragraph{Main sequence:} 
Figure~\ref{fig:main_sequence} shows the relation between the stellar mass and star-formation rate of galaxies in the \flag and \gaea simulations in two redshift bins, $0.4<z<0.9$ (top panels) and $0.9<z<1.8$ (bottom panels). We only show models m3 and ECLH, but the results are similar for m1 and ECLQ.  All galaxy mocks recover the expected correlation between the SFR and \ha\ luminosity. Comparison with the compilation of observational data from \cite{Popesso2023} confirms that, in general, there is good agreement between observations and simulations. The shape of the star-forming main sequence for \ha\ flux-limited sample (solid black lines) is better reproduced in \gaea compared to \flag across a wide range of stellar masses in both redshift bins. However, both models show a shift compared to the observed main sequence. In \gaea, galaxies lie below, whereas in \flag, they tend to be above the observed relation.

\paragraph{Galaxy clustering:}
Figure~\ref{fig:2pcf} shows the two-point correlation functions of galaxies, relying on the Landy--Szalay estimator \citep{LandySzalay1993}, in the redshift range $0.8<z<1.3$ for the $5 \, \%$ \ha\ flux-brightest and the $5 \, \%$ least bright galaxies in the \flag m3 and \gaea ECLH mocks. Similar results are found in \flag m1 and \gaea ECLQ. In both simulations, the clustering is generally higher for the $5 \, \%$ brightest galaxies compared to their $5 \, \%$ lowest \ha\ flux counterparts at large separations ($\gtrsim 1 \, \hMpc$). At separations $\lesssim 1$ \hMpc, the $5 \, \%$ brightest galaxies in \gaea  show reduced clustering with respect to their lower-brightness counterparts, whereas in \flag, the clustering of the brightest galaxies continues to be higher, presumably better tracing the substructures. 
We note that galaxy clustering was among the constraints used during the construction of \flag mocks to set the number of satellites and assign colour types to galaxies, while this information was not considered at all for the \gaea models. We note also that the considered redshift range bin is quite large. However, the median (and also mean) redshifts of the $5 \, \%$ highest and lowest \ha\ flux galaxies are comparable ($\bar{z} \sim 1.0$ for \flag m3 and \gaea ECLH; $\bar{z} \sim 1.0$ for the $5 \, \%$ highest and $\bar{z} \sim 1.1$ for the $5 \, \%$ lowest \ha\ flux sample in \flag m1 and \gaea ECLQ).
Therefore, the observed differences in the clustering of these populations are unlikely to be driven by their redshifts alone. We finally note that qualitatively similar results are found when considering $10 \, \%$ cut in the \ha\ flux, and/or for higher redshift range, e.g. $1.3<z<1.8$.

\subsection{Cosmic web reconstruction}

\begin{table}
\centering
\begin{threeparttable}
\caption{Optimal linking lengths (in units of the mean intergalactic separation).} 
\label{tab:linking_lengths}
\begin{tabular*}{0.95\columnwidth}{@{\extracolsep{\fill}}l|cc|cc}
\hline
& & & & \\[-9pt]
 &  \multicolumn{2}{c|}{\flag} & \multicolumn{2}{c}{\gaea} \\
 \cline{2-5}
 & & & & \\[-9pt]
 &  m3 & m1 & ECLH& ECLQ\\
 \hline
 & & & & \\[-8pt]
\deep & (0.07,14) & (0.07,15) & (0.09,19) & (0.09,20)\\
\deepmass & (0.07,12) & (0.07,14) & (0.09,16) & (0.09,17)\\
\tnote{a}\deepnoisesamp & (0.07,25) & (0.07,28) & (0.09,29) & (0.09,29)\\
\hline
\end{tabular*}
\tablefoot{
Values in parenthesis ($b_{\perp}$, $R$) indicate the linking length perpendicular to the line-of-sight $b_{\perp}$ and the radial expansion factor $R~=~b_\parallel/b_\perp$, relating the projected ($b_\perp$) and line-of-sight ($b_\parallel$) linking lengths.\\
\tablefoottext{a}{Same values for linking lengths are applied for \mstar and \ha-limited samples.}
}
\end{threeparttable}   
\end{table}

\begin{figure*}
\centering\includegraphics[width=\textwidth]{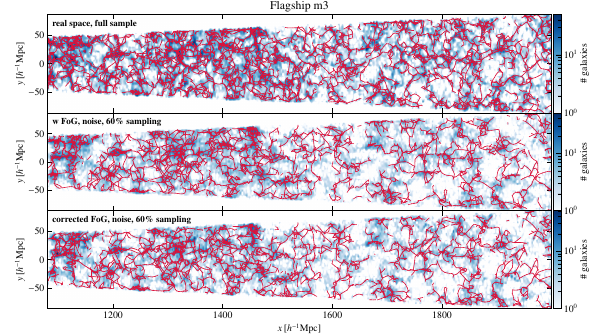}
\caption{Visualisation of a $\sim 40\, h^{-1}\, {\rm Mpc}$ thick slice of the galaxy distribution from the \flag m3 mock and the corresponding cosmic web skeleton reconstructed without weighting Delaunay tessellation for the reference catalogue without the FoG effect, without added noise and with $100\, \%$ completeness (top; \deep), with the FoG, added redshift error and $60\, \%$ sampling (middle; \deepnoisesampwf) and after correcting for the FoG effect (bottom; \deepnoisesampwof). 
For the sake of clarity, only $0.4<z<0.85$ range of the lightcone is shown.
}
\label{fig:cw_flag}
\end{figure*}

\begin{figure*}
\centering\includegraphics[width=\textwidth]{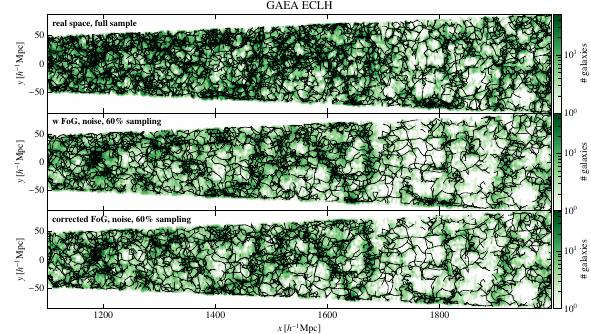}
\caption{As in Fig.~\ref{fig:cw_flag}, but for the \gaea ECLH mock. The higher galaxy number density of the \gaea ECLH mock compared to \flag m3 (Fig.~\ref{fig:cw_flag}) can be clearly seen.
}
\label{fig:cw_gaea}
\end{figure*}

To extract the cosmic web network, we rely on the publicly available and widely used structure finder \disperse \citep[][]{Sousbie2011a,Sousbie2011}. To account for the redshift-space distortions that affect any three-dimensional galaxy distribution relying on redshift-based distance measurements, we follow the method outlined in \cite{Kraljic2018}. This approach has been previously adopted for the cosmic web reconstruction in three-dimensional space using spectroscopic surveys, such as GAMA or SDSS \citep[][]{Kraljic2018,Kraljic2020a}. Here we only briefly describe its main steps. 

To minimise the impact of redshift-space distortions induced by the random motions of galaxies within virialized haloes, the so-called Fingers-of-God (FoG) effect \citep[e.g.,][]{Jackson1972}, we first identify the galaxy groups. This is done using an anisotropic Friends-of-Friends (FoF) algorithm that operates on the projected perpendicular and parallel separations of galaxies \citep[see][for details on the group finder algorithm]{Treyer2018} calibrated on the \flag and \gaea mock catalogues. Table~\ref{tab:linking_lengths} shows the resulting optimal linking lengths, in units of the mean intergalactic separation, for all models used in this work. 

The next step consists of the radial compression of the groups such that the dispersions of their member galaxies in transverse and radial directions are
equal \citep[see also e.g.,][]{Tegmark2004}. The resulting isotropic galaxy distribution within the groups about their centres minimises the impact of elongated structures along the line-of-sight (the FoG effect) that could be misidentified as filaments of the cosmic web.

Finally, \disperse is used to coherently identify all the components of the cosmic web, i.e., voids, walls, filaments, and nodes, directly from the inhomogeneous distribution of galaxies, relying on discrete Morse theory \citep{Forman2001}. To deal with such a discrete data set, \disperse builds on the Delaunay tessellation allowing one to provide a scale-free Delaunay Tessellation Field Estimator \citep[DTFE;][]{Schaap2000} density and reconstruct the local topology. In this work we will consider the cosmic web reconstruction relying on both the non-weighted and stellar mass-weighted Delaunay tessellation.
To deal with noisy data, such as galaxy catalogues, \disperse implements the concept of the topological persistence allowing to effectively filter out the topologically less robust features, i.e. features that would disappear or change after resampling of or adding a noise to the underlying field of the galaxy distribution. The level of filtering is controlled by the persistence threshold $N_\sigma$, such that higher values of $N_\sigma$ select structures that are topologically more robust with respect to noise. The fiducial value used throughout this work is $N_\sigma=5$. This higher value, compared to the more commonly used $N_\sigma=3$, allows us to better highlight challenges of reconstructing cosmic web structures when working
with \ha\ flux-limited, rather than stellar mass-limited, noisy data (see Sect.~\ref{sec:results}). Lastly, the cosmic web skeleton was smoothed in post-processing three times.

For illustration, Figs.~\ref{fig:cw_flag} and~\ref{fig:cw_gaea} show a $\sim 40\, \hMpc$ thick slice of the distribution of galaxies within the redshift range $0.4 < z < 0.85$ from the \flag and \gaea mock catalogues (models m3 and ECLH, respectively) together with the corresponding network of filaments, reconstructed using unweighted Delaunay tessellation, for the reference catalogue, i.e. without FoG effect and with 100\% completeness (top), after adding redshift error, $60\, \%$  sampling and FoG effect (middle), and after correcting for the FoG effect (bottom). 
This visual inspection allows us to already identify some of the key factors impacting the quality of the cosmic web reconstruction, namely, the FoG effect, redshift uncertainty, and the incompleteness of the underlying galaxy sample.

In the following, catalogues including redshift-space distortions are denoted with `wFoG' in the superscript, for example \deepwf, while catalogues with applied correction for the FoG effect have `FoG,corr' in the superscript, such as \deepwof.

\section{Results} 
\label{sec:results}

To assess the quality of the cosmic web reconstruction expected for the EDFs, we consider three different measures. These involve geometrical and topological properties of the cosmic web and transverse stellar-mass gradients of galaxies with respect to filaments of the cosmic web, i.e. the observed and theoretically expected trend of increasing galaxies' stellar mass with their decreasing distance from filaments.

\subsection{The geometrical cosmic web measures}
\label{sec:stats}

\begin{figure*}
\centering\includegraphics[width=0.45\textwidth]{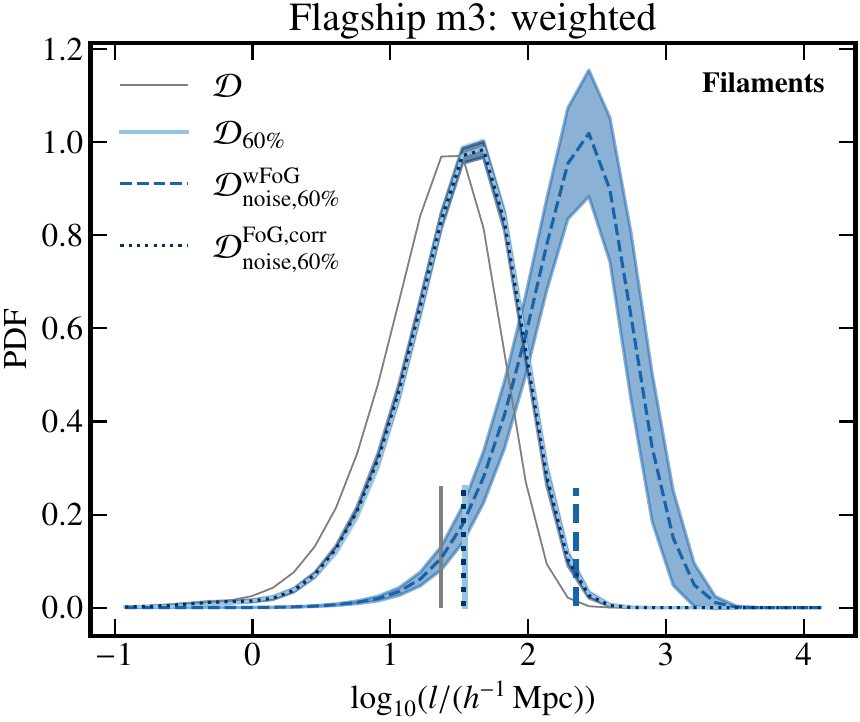}
\centering\includegraphics[width=0.45\textwidth]{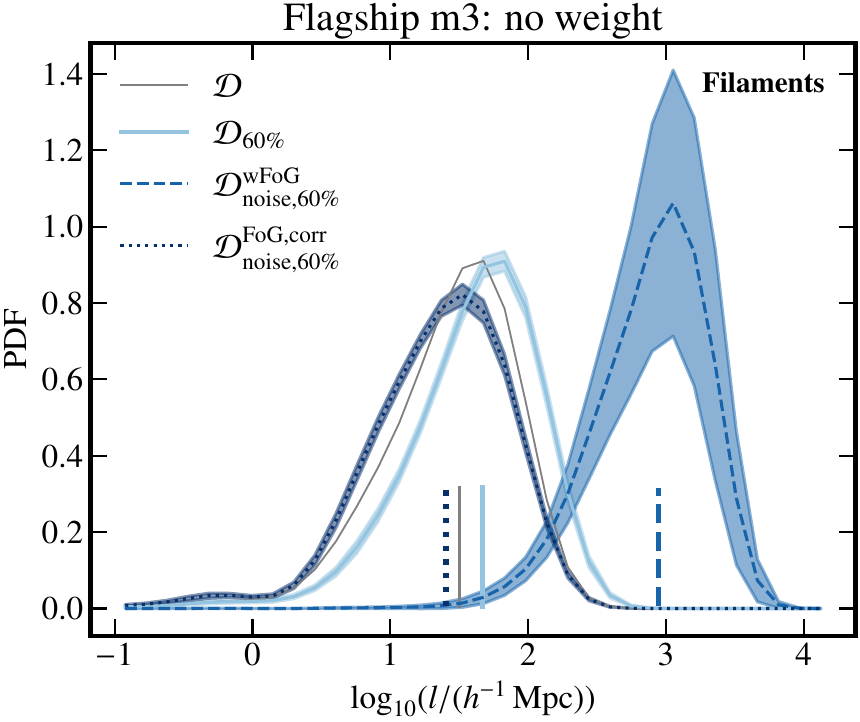}
\\
\centering\includegraphics[width=0.45\textwidth]{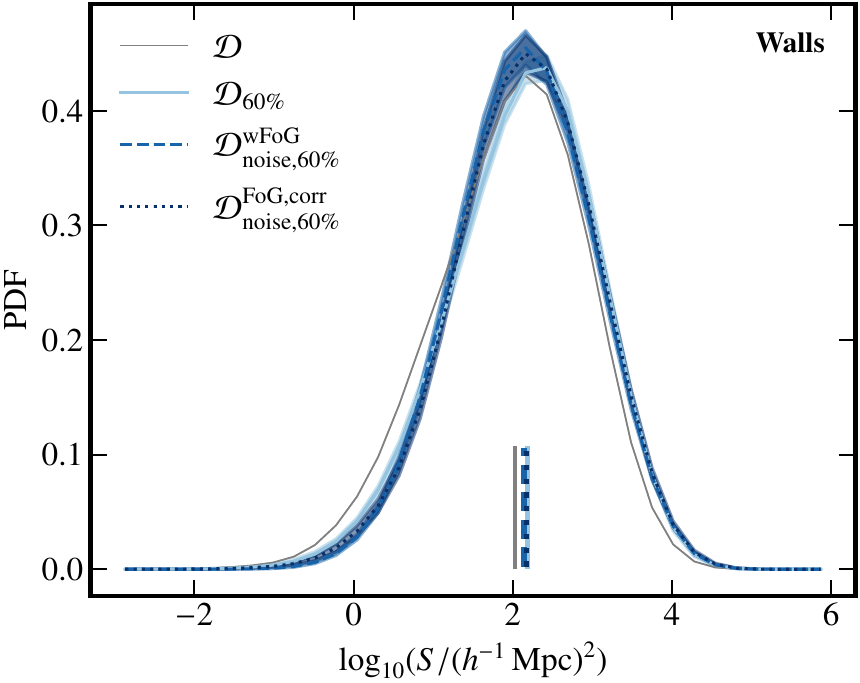}
\centering\includegraphics[width=0.45\textwidth]{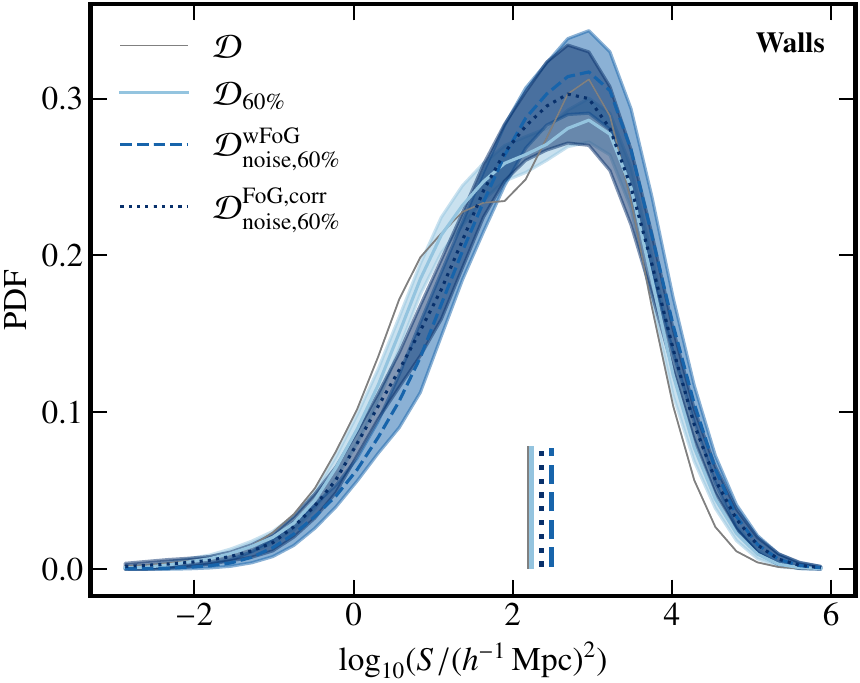}
\\
\centering\includegraphics[width=0.45\textwidth]{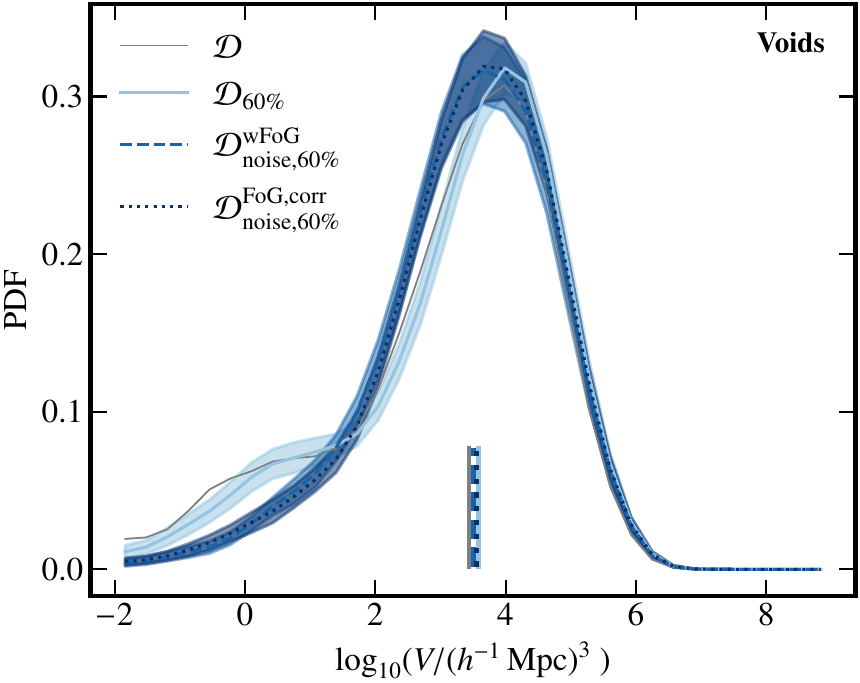}
\centering\includegraphics[width=0.45\textwidth]{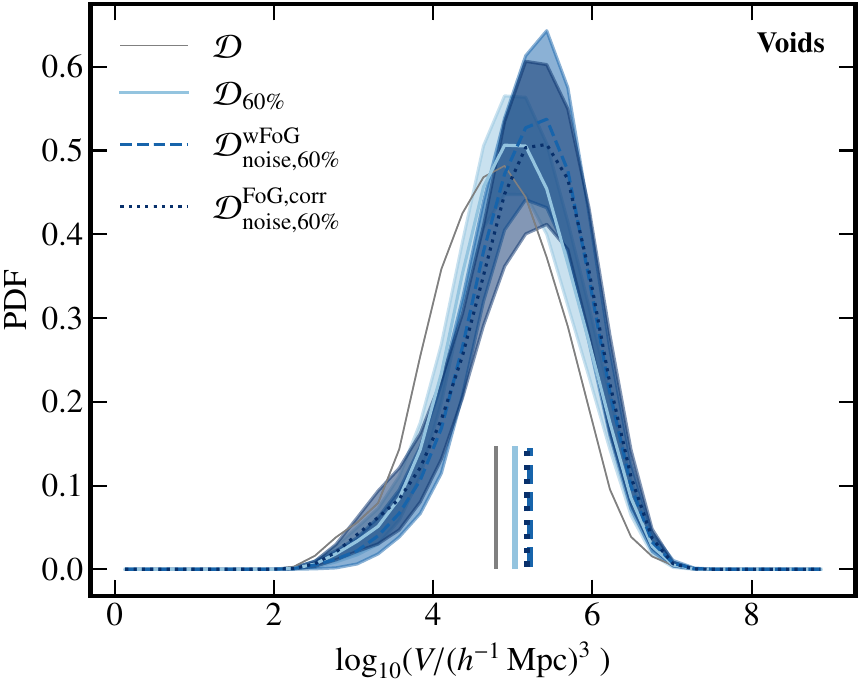}
\caption{PDF of filament lengths (top), wall areas (middle) and void volumes (bottom) for EDS mocks without redshift-space distortions for full (\deep; solid grey lines), and 60\% (\deepsamp; solid coloured lines) sampling, including the FoG effect (\deepnoisesampwf; dashed coloured lines) and after the compression of FoG (\deepnoisesampwof; coloured dotted lines) for the \flag (model m3) simulation.
PDFs obtained using the stellar mass-weighted Delaunay tessellation for the cosmic web reconstruction (left) are compared with the reconstruction without weighting (right). Vertical lines indicate the medians of the distributions. Shaded regions correspond to the standard deviation across five mocks. The FoG mainly impacts the length of filaments which tend to be longer. After the compression of groups, the PDFs are in a good agreement with the reference distribution. Weighting the tessellation helps bringing in better agreement the distributions of the reference catalogue without the FoG effect (\deepsamp) and after the correction for the FoG effect (\deepnoisesampwof). 
For the PDFs of void volumes obtained for the reconstruction without weighted tessellation, to avoid spurious border effects, only voids with volume larger than the volume corresponding to the mean intergalactic separation are considered.
}
\label{fig:PDF_stats_flag_all}
\end{figure*}

\begin{figure*}
\centering\includegraphics[width=0.45\textwidth]{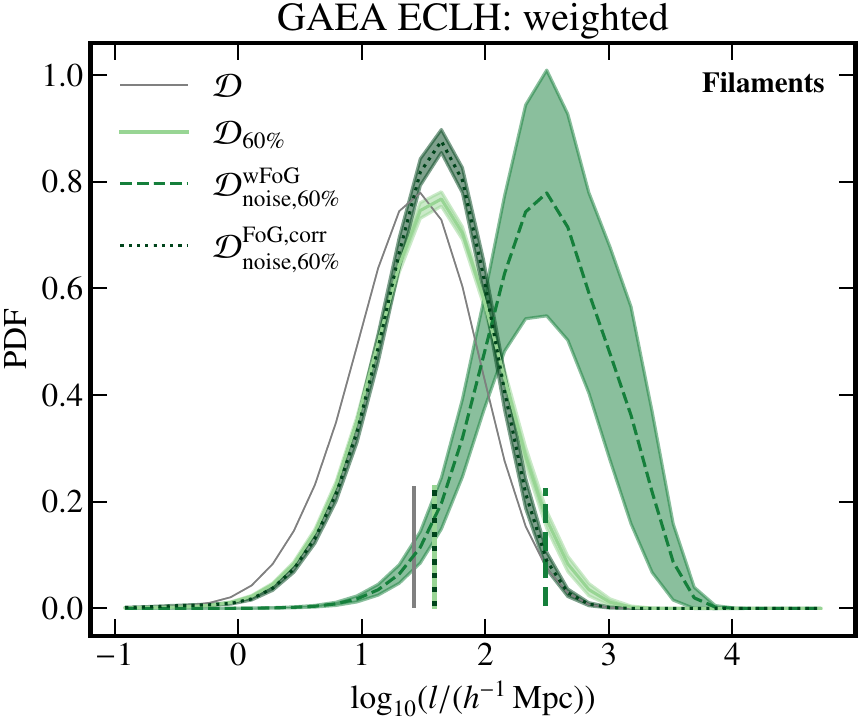}
\centering\includegraphics[width=0.45\textwidth]{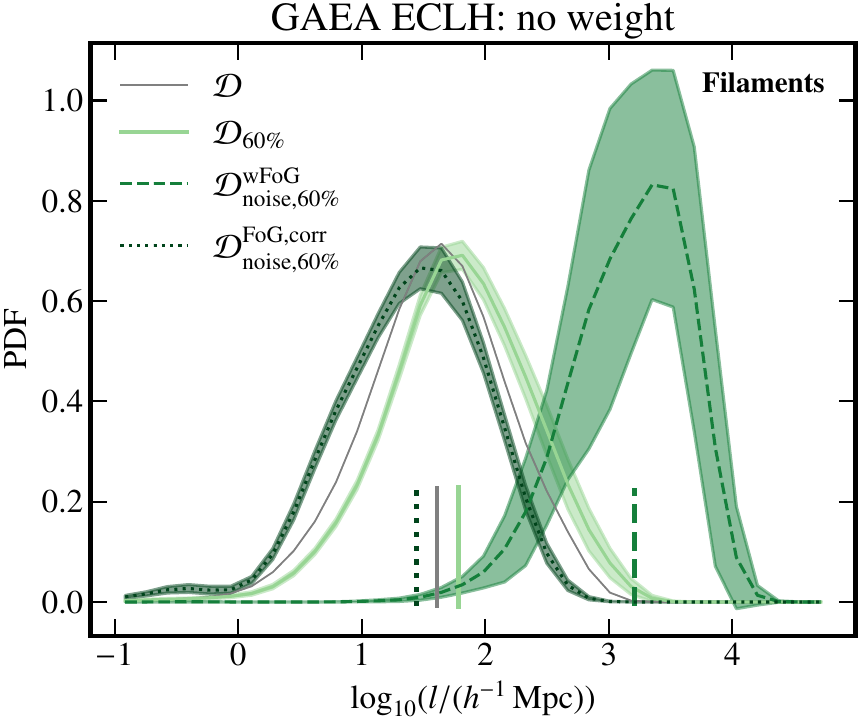}
\caption{As in Fig.~\ref{fig:PDF_stats_flag_all} but for \gaea mocks (model ECLH) and filament lengths alone.
}
\label{fig:PDF_stats_gaea_all}
\end{figure*}

\begin{figure*}
\centering\includegraphics[width=0.49\textwidth]{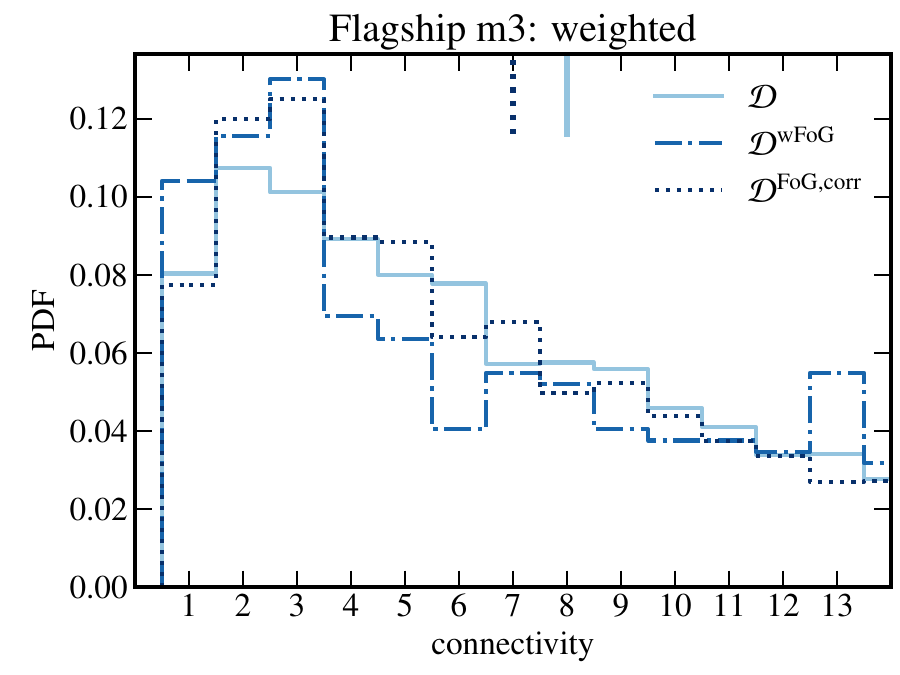}
\centering\includegraphics[width=0.49\textwidth]{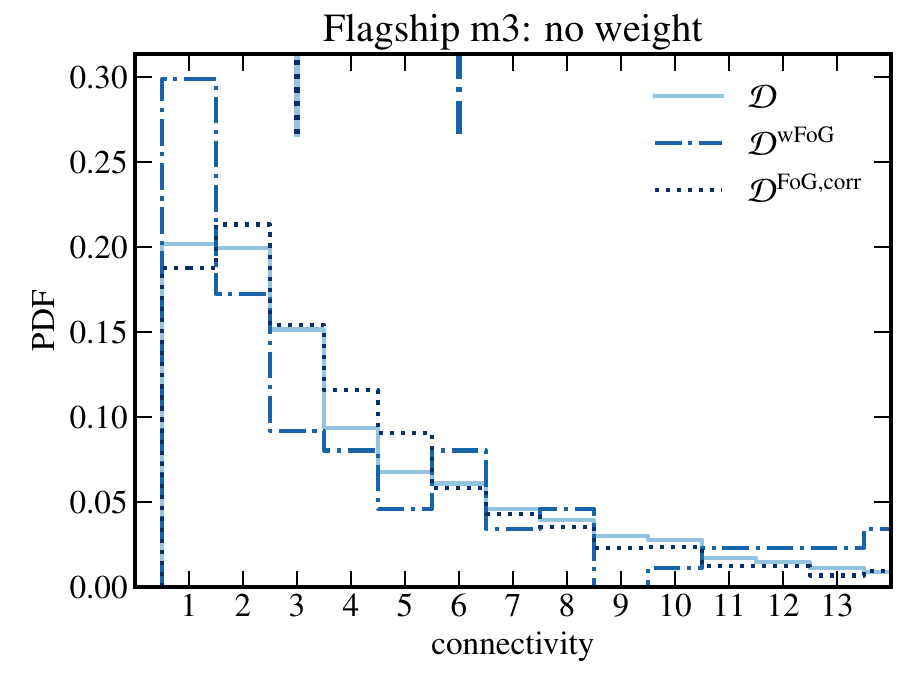}\\
\centering\includegraphics[width=0.49\textwidth]{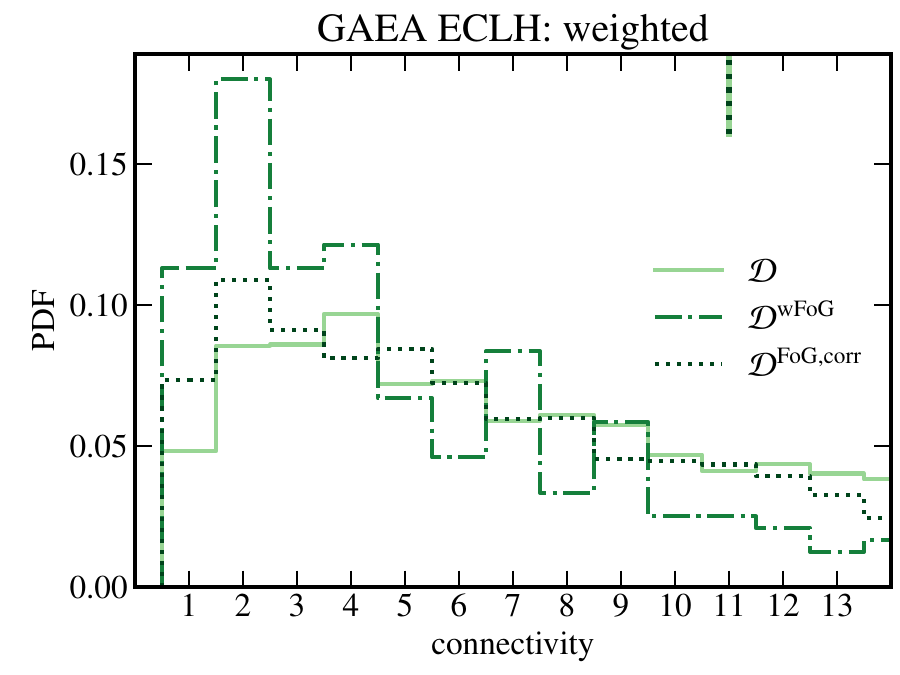}
\centering\includegraphics[width=0.49\textwidth]{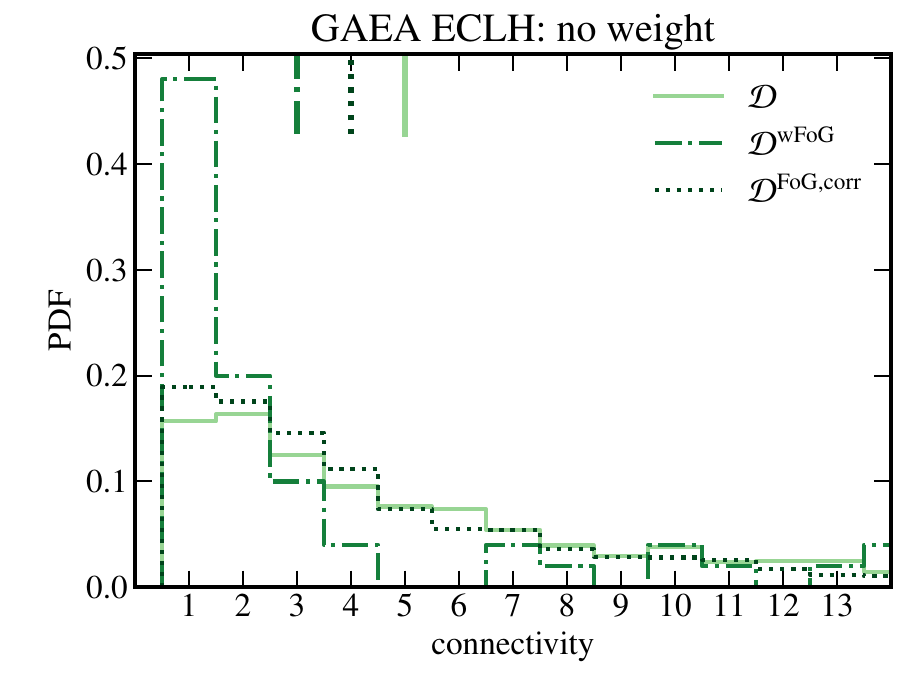}
\caption{PDFs of the connectivity of central galaxies for the \ha-limited selection of galaxies in the full sample without redshift error (\deep) for \flag (model m3, top panels) and \gaea (model ECLH, bottom panels),
with stellar mass weighting of the skeleton (left) and without weighting (right). In each panel, vertical lines indicate the medians of distributions. For a better visibility, the histograms are limited to the connectivity values below 14. Mean and median values for all distributions are reported in Tables~\ref{tab:con_flag} and \ref{tab:con_gaea}. The redshift-space distortions have a strong impact on the connectivity of galaxies, but this effect that can be reasonably well corrected for.
}
\label{fig:PDF_connect}
\end{figure*}

Geometrical measures of individual cosmic web components, such as the length of filaments, the area of walls, and the volume of voids, provide a straightforward way to assess the impact of different parameters on the quality of the cosmic web reconstruction. We explore in particular the impact of the FoG effect, redshift errors, and incompleteness of the galaxy sample. 

Figure~\ref{fig:PDF_stats_flag_all} shows the probability distribution function (PDF) of the lengths of filaments\footnote{In practice, half length of filaments, measured as the curvilinear distance between filament-type saddle points and maxima, the nodes of the cosmic web.}, areas of walls, and volumes of voids for \flag mock, m3 model (m1 model leads to the same conclusions) for the cosmic web reconstruction with (left) and without (right) the stellar mass-weighted Delaunay tessellation. Similarly, Fig.~\ref{fig:PDF_stats_gaea_all} presents the PDF of the filaments' lengths for \gaea mock, model ECLH (the same conclusions apply to the ECLQ model). For walls and voids (not shown), qualitatively similar results and conclusions to those of \flag are obtained.    

As expected, the FoG effect has a strong impact on the filaments of the cosmic web, and much less so on its other components. When this effect is not corrected for, the reconstructed filaments tend to be too long as manifested by the shift of the distribution of filaments' lengths toward larger values (\deepnoisesampwf) compared to the reference sample without redshift-space distortions (\deepsamp). 
After correcting for the FoG effect (\deepnoisesampwof), the distributions of filaments' lengths are in excellent agreement, both in terms of medians and overall shape for \flag. For \gaea, the correction is not perfect, showing some residual deficit of long and excess of intermediate length filaments. We note, however, that the medians of the distributions are comparable.

Weighting the Delaunay tessellation by the stellar mass of galaxies turns out to be important for recovering a better agreement between the distributions before and after correcting for the FoG effect when the galaxy sample is not stellar mass-limited, as is the case for the EDS. 
As can be seen, the agreement between the distributions of filaments' lengths does improve once the correction of the FoG effect is applied, but it is not as good as in the case of weighting.  
This is a direct consequence of the galaxy sample being \ha\ flux- rather than stellar mass-limited 
(see Fig.~\ref{fig:PDF_stats_mass_vs_halpha}). 
For stellar mass-limited galaxy samples the method used to deal with the FoG is efficient even without weighting the tessellation. The  sample selection based on the \ha\ flux is also responsible for our failure to correct completely for the FoG effect seen for \gaea. The underlying reason is the inability to properly reconstruct galaxy groups, virialized structures responsible for small-scale redshift-space distortions, when the sample is not stellar mass-limited. As discussed in Sect.~\ref{sec:quality_mocks}, \flag and \gaea mocks show different clustering on small scales, 
with brightest galaxies tracing presumably more closely substructures in \flag models, therefore 
resembling more the stellar mass selection.

The incompleteness of the galaxy sample manifests, as expected, by a shift of the distributions toward higher values (compare \deepsamp vs. \deep in Figs. ~\ref{fig:PDF_stats_flag_all} and \ref{fig:PDF_stats_gaea_all}), in particular for the filaments' lengths. Walls and voids are impacted to a much lesser degree, especially when weighting is applied. 

Redshift error, on top of the sample selection (whether it is stellar mass or \ha\ flux selected), impacts strongly our ability to correct for the FoG effect (see Fig.~\ref{fig:PDF_stats_mass_noise}). However, the correction works better for the stellar mass-limited sample.

In summary, the small-scale redshift-space distortions strongly impact the cosmic web reconstruction, filamentary network in particular, regardless of the sample selection and regardless of the weighting of the tessellation. The applied correction for the FoG effect works better for stellar mass-limited samples. On top of the sample selection, the redshift uncertainties hinder our ability to correct for the FoG effect.

\subsection{Connectivity and multiplicity}
\label{sec:connect}

\begin{figure*}
\centering\includegraphics[width=0.49\textwidth]{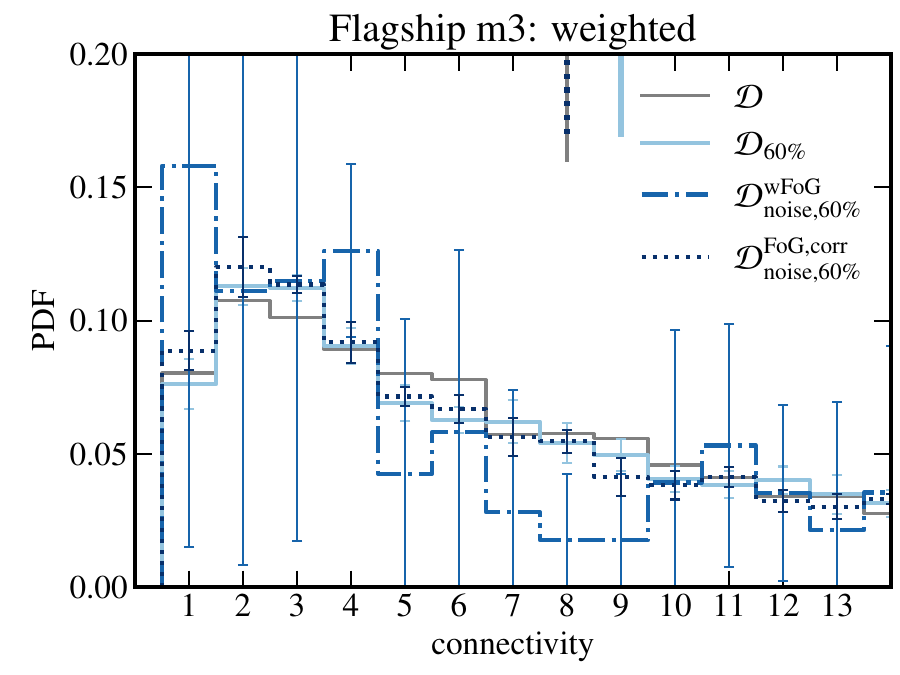}
\centering\includegraphics[width=0.49\textwidth]{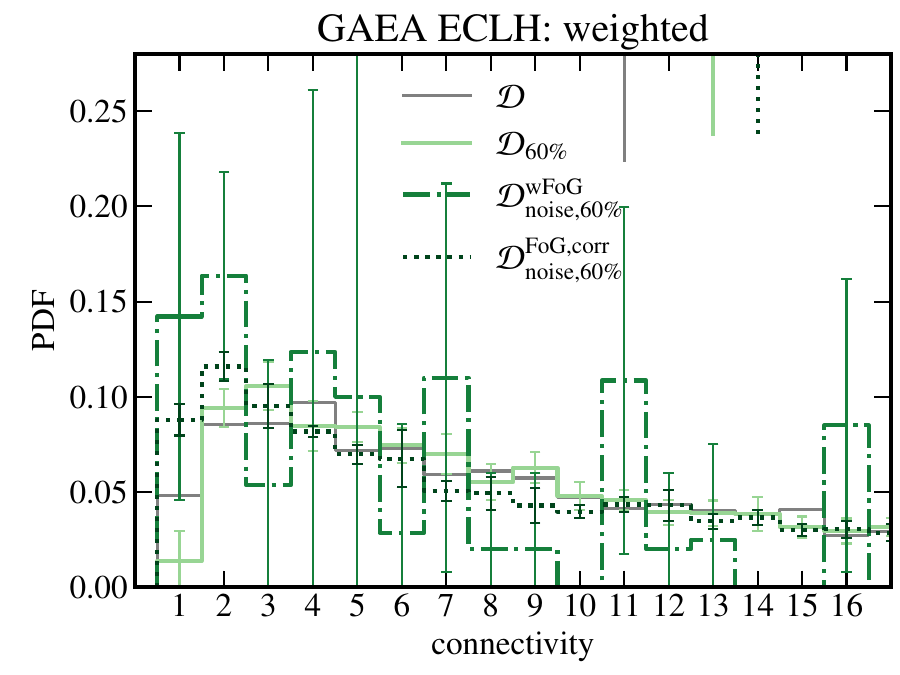}
\caption{PDFs of the connectivity of central galaxies for the \ha-limited selection of galaxies in the fiducial sample (\deepnoisesamp; coloured lines) for \flag (model m3, left panel) and \gaea (model ECLH, right panel), with stellar mass weighting of the skeleton. The full sample without noise (\deep) is shown for comparison (grey lines). In each panel, vertical lines indicate the medians of distributions and error bars correspond to the standard deviation across five mocks. For a better visibility, the histograms are limited to the connectivity values below 14. Mean and median values for all distributions are reported in Tables~\ref{tab:con_flag} and \ref{tab:con_gaea}. Redshift errors and incompleteness of the sample decrease the quality of the cosmic web reconstruction, which can be improved by stellar mass weighting.
}
\label{fig:PDF_connect_noise}
\end{figure*}

\begin{figure*}
\centering\includegraphics[width=0.49\textwidth]{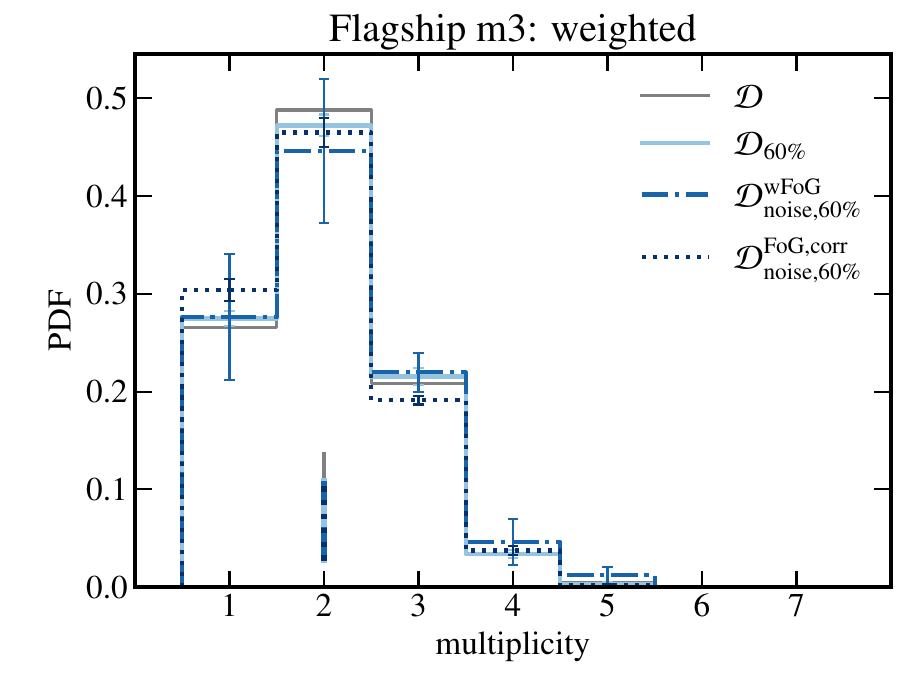}
\centering\includegraphics[width=0.49\textwidth]{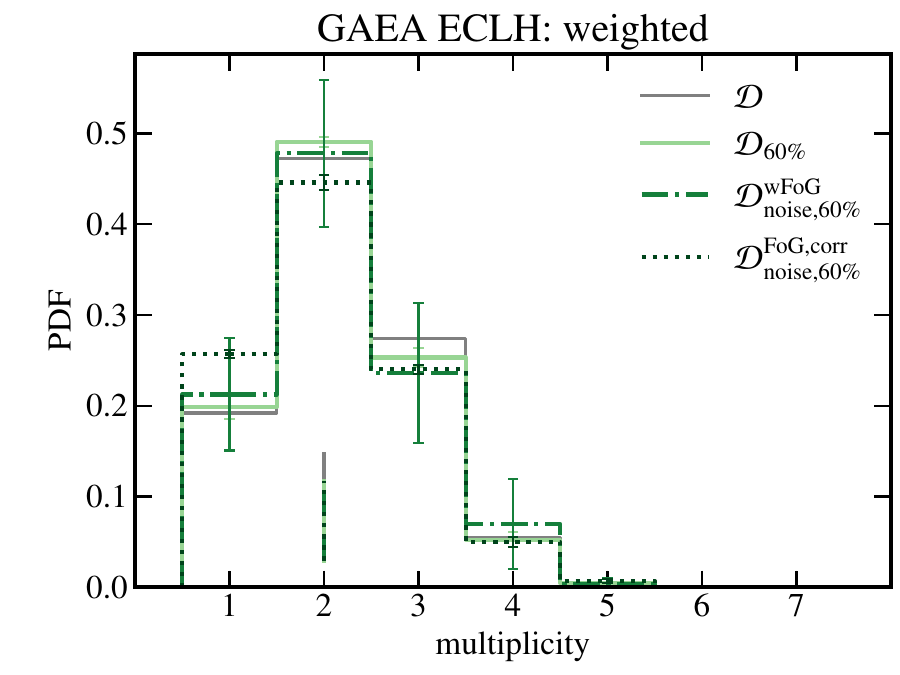}
\caption{PDFs of the multiplicity of central galaxies for an \ha-limited selection of galaxies in the fiducial sample (\deepnoisesamp; coloured lines) for \flag (model m3, left panel) and \gaea (model ECLH, right panel), with stellar mass weighting of the skeleton. The full sample without noise (\deep) is shown for a comparison (grey lines). In each panel, vertical lines indicate the medians of distributions and error bars correspond to the standard deviation across five mocks. The mean and median values for all distributions are reported in Tables~\ref{tab:multi_flag} and \ref{tab:multi_gaea}. The multiplicity of the cosmic web shows only a weak sensitivity to the FoG effect, redshift error, and sample completeness. 
}
\label{fig:PDF_multi}
\end{figure*}

\begin{figure*}
\centering\includegraphics[width=0.49\textwidth]{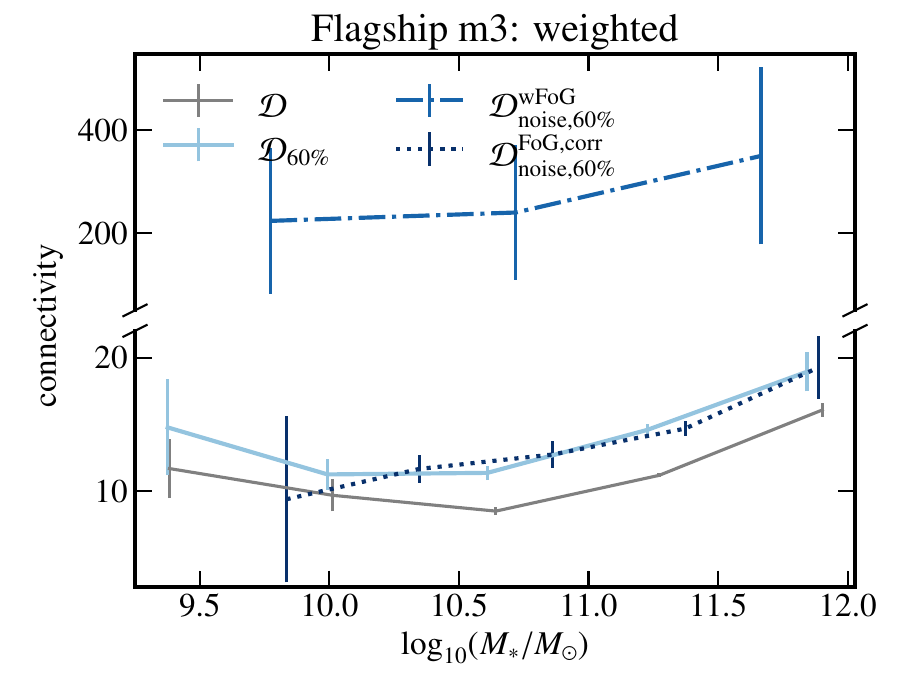}
\centering\includegraphics[width=0.49\textwidth]{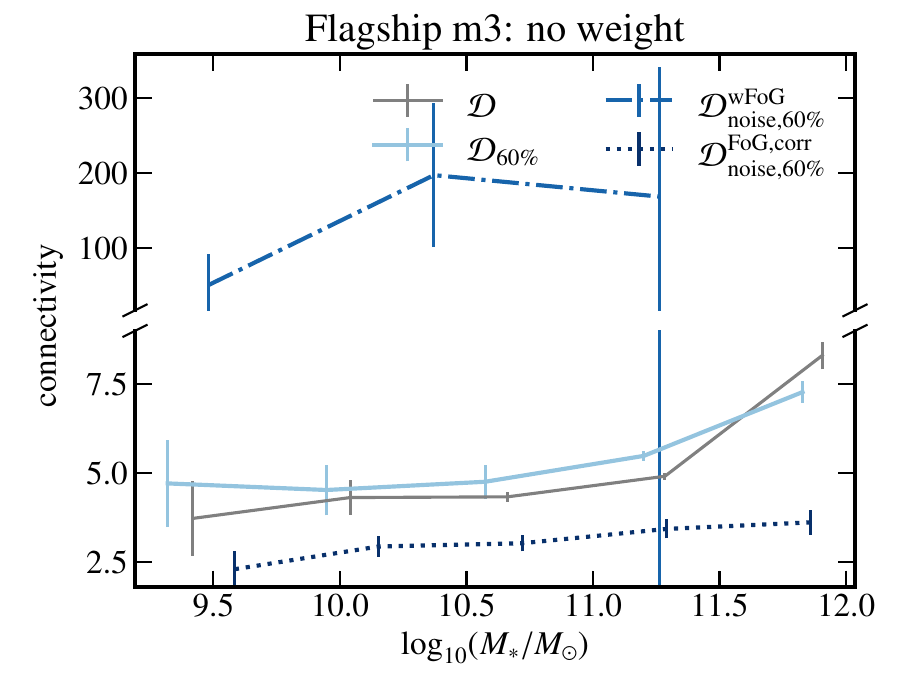}\\
\centering\includegraphics[width=0.49\textwidth]{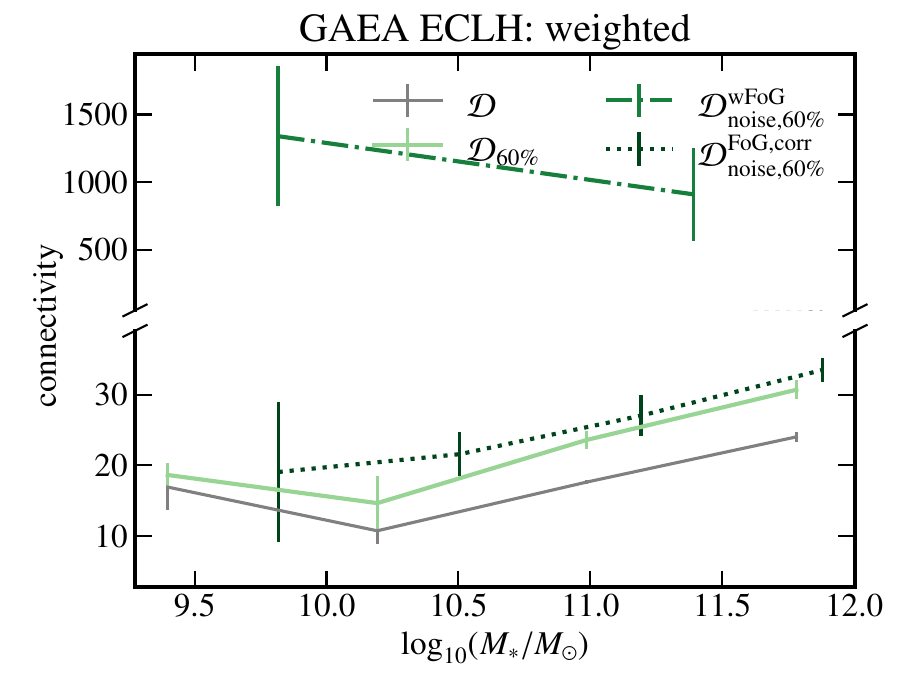}
\centering\includegraphics[width=0.49\textwidth]{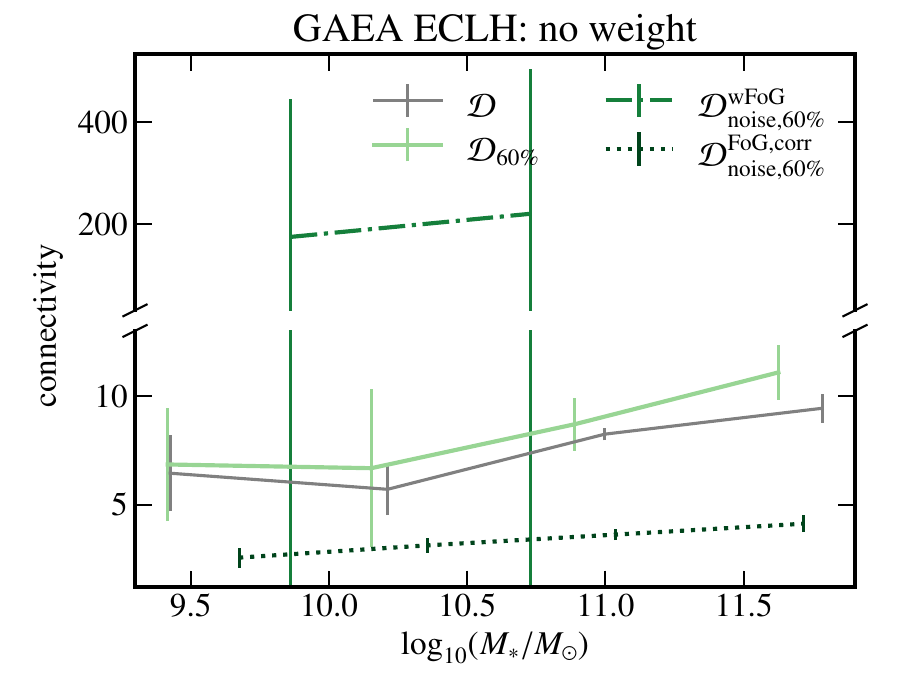}
\caption{Connectivity of central galaxies as a function of their stellar mass in \flag (model m3, top panels) and \gaea (model ECLH, bottom panels) mocks for the fiducial galaxy sample \deepnoisesamp, 
with stellar mass weighting of the skeleton (left) and without weighting (right). As expected, connectivity increases with stellar mass, with weighting of the Delaunay tessellation introducing a bias at lower stellar masses leading to reversed trend. Correction for the FoG effect is needed to recover reasonably well the \mstar-connectivity relation, in particular for weighted tessellations.
}
\label{fig:connect_mass_flag_gaea}
\end{figure*}

\begin{figure*}
\centering\includegraphics[width=0.49\textwidth]{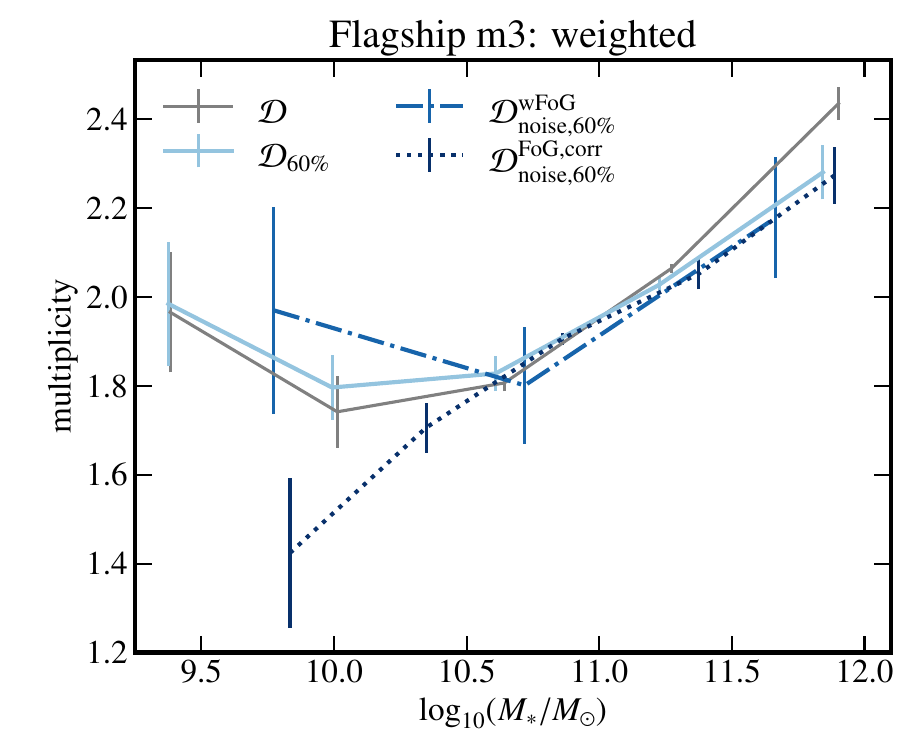}
\centering\includegraphics[width=0.49\textwidth]{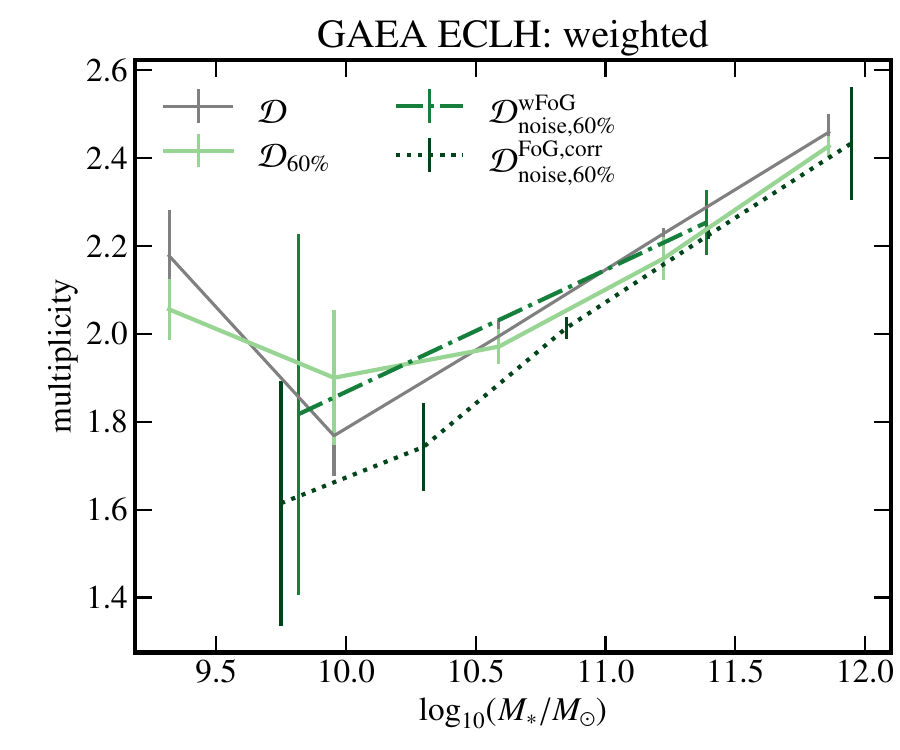}
\caption{Multiplicity of central galaxies as a function of their stellar mass in \flag (model m3, left panel) and \gaea mocks (model ECLH, right panel) for the fiducial sample, with stellar mass weighting of the skeleton. As for connectivity, multiplicity increases with \mstar, with a reversed trend at low \mstar. In contrast to connectivity, the redshift-space distortions do not strongly modify the amplitude of the \mstar-multiplicity relation.
}
\label{fig:multi_mass_flag_gaea}
\end{figure*}

The connectivity and multiplicity of the cosmic web, i.e. the number of filaments respectively globally and locally connected to the nodes of the cosmic web, where massive galaxy groups and clusters predominantly reside, are interesting topological measures. They are expected to depend on the underlying cosmology and to impact the assembly of galaxies and hence their properties. It is therefore important to assess our ability to recover these quantities from the expected configuration for the EDFs.    

Let us start by considering the distribution of connectivity, defined as the number of filaments connected to a given node of the cosmic web. Figure~\ref{fig:PDF_connect} shows the histograms of connectivity of central galaxies\footnote{We focus on central galaxies to limit the complications due to the non-linear nature of the satellite population and to ease the comparison with theoretical predictions \citep[for a more detailed discussion on the effect of satellites on connectivity, see][]{Kraljic2020a}.} in the \ha\ flux-limited sample measured in the \flag and \gaea mocks (models m3 and ECLH, respectively, but qualitatively similar results are obtained for models m1 and ECLQ). The FoG effect has a strong impact on the connectivity of galaxies at the nodes of the cosmic web (\deepwf), modifying both the shape of the distribution, but also its median (indicated by a vertical line), regardless of the sample selection, i.e., including \mstar-limited samples (\deepmasswf; see Fig.~\ref{fig:PDF_connect_mass_lim}). Correction of the FoG effect works very well for the \mstar-limited sample and reasonably well for the \ha\ flux-limited sample in terms of the overall shape of distributions, but also their mean and median values (see Tables~\ref{tab:con_flag} and \ref{tab:con_gaea}).
The weighting of the tessellation by stellar mass for the \ha\ flux-based selection (left panels of Fig.~\ref{fig:PDF_connect}) tends to artificially increase connectivity even in the absence of redshift-space distortions, for example, with a median value of 8 (11) as opposed to a median value of 3 (5) without weighting for \flag m3 (\gaea ECLH). 
For the \mstar-limited selection, the impact of weighting of the tessellation by stellar mass on the connectivity is very weak (see Fig.~\ref{fig:PDF_connect_mass_lim}). Adding redshift errors and incompleteness decreases the quality of the cosmic web reconstruction, in particular for the \ha-based selection (Fig.~\ref{fig:PDF_connect_noise}) due to the reduced ability to correct for FoG. In this configuration, stellar mass weighting of the tessellation improves the cosmic web reconstruction (see Tables~\ref{tab:con_flag} and \ref{tab:con_gaea}).

Let us now explore the multiplicity and our ability to recover this local property of the cosmic web, defined as the connectivity minus the number of bifurcation points \citep[i.e. points where filaments split, even though they are not extrema,][]{Pogosyan2009}, associated with each node.
Figure~\ref{fig:PDF_multi} shows the multiplicity of central galaxies in \flag and \gaea mocks (models m3 and ECLH, respectively, but almost identical results are obtained for the \flag model m1 and the \gaea model ECLQ) for the fiducial sample (\deepnoisesamp). 
Similarly to the connectivity, not correcting for the FoG effect modifies the distribution of multiplicity, but to a much lesser extent, in particular when the Delaunay tessellation is weighted by stellar mass. The mean and median values of the multiplicity agree very well across all mocks for weighted tessellation. For cosmic web reconstruction without weighting the tessellation, the correction of the FoG effect is needed to obtain good agreement between the distributions of multiplicity (Tables~\ref{tab:multi_flag}-\ref{tab:multi_gaea}). 
Qualitatively similar conclusions apply to the \mstar-limited samples. The multiplicity of the cosmic web therefore appears to be a more robust topological property compared to connectivity given its weak sensitivity to the FoG effect, redshift error, sample completeness, and its selection. 

Beyond the statistical measurements of connectivity and multiplicity, it is of interest to explore these quantities as a function of different galaxy properties. In this work, we focus on  stellar mass. Figure~\ref{fig:connect_mass_flag_gaea} shows the connectivity of central galaxies as a function of their stellar mass for the \flag and \gaea mocks (models m3 and ECLH, respectively, with qualitatively similar conclusions for m1 and ECLQ) for the fiducial sample (\deepnoisesamp).
Regardless of the sample selection, i.e. \ha- or \mstar-limited (see Fig.~\ref{fig:connect_mass_flag_gaea_full_mass_lim}), central galaxies in the reference sample follow the expected trend, where the connectivity increases with increasing stellar mass,
for the probed stellar mass range and when no weighting is applied prior to the cosmic web reconstruction (right panels of Fig.~\ref{fig:connect_mass_flag_gaea}). Weighting of the Delaunay tessellation introduces a bias at lower stellar masses, in particular for the \ha-based galaxy selection, leading to an increase of the connectivity with decreasing stellar mass (left panels of Fig.~\ref{fig:connect_mass_flag_gaea}). As already seen from the global distributions, the redshift-space distortions have a strong impact on the connectivity of the cosmic web. This leaves a clear signature in the form of a shift of the \mstar-connectivity relation toward higher values of connectivity, for mocks with FoG compared to the reference sample, regardless of the tessellation weighting. The applied correction of the FoG effect allows us to recover reasonably well the \mstar-connectivity relation of the reference samples, in particular for weighted tessellations. Weighting the tessellation by stellar mass is crucial for recovering the \mstar-connectivity trend if the sample is \ha\ flux-limited, in the presence of the redshift errors, and for reduced sampling, albeit with an introduced bias at lower masses.

Figure~\ref{fig:multi_mass_flag_gaea} shows the multiplicity of central galaxies as a function of their stellar mass for the \flag and \gaea mocks (models m3 and ECLH, respectively, with qualitatively similar conclusions for m1 and ECLQ) for the fiducial sample (\deepnoisesamp).
As in the case of connectivity, the multiplicity of central galaxies increases with increasing stellar mass for galaxies in all reference catalogues (\deep) regardless of sample selection (see Fig.~\ref{fig:multi_mass_flag_gaea_full_mass_lim} for the \mstar-limited sample; \deepmass) and without tessellation weighting (see Fig.~\ref{fig:multi_mass_flag_gaea_appendix}).
At the lowest stellar mass end (below $10^{9.5}$~\msun), accessible only in the \ha\ flux-selected samples, multiplicity tends to increase when stellar mass-weighted tessellation is used for the cosmic web reconstruction.
Contrary to connectivity, the redshift-space distortions do not strongly modify the amplitude of the \mstar-multiplicity relation. The impact of FoG on this relation is overall very limited for the reconstruction with the stellar mass-weighted tessellation, but critically reduces our ability to recover the trend when the sample is \ha\ flux-limited (\deepwf, \deepnoisesampwf) and no weighting is applied. The applied FoG correction significantly improves our ability to recover the \mstar-multiplicity relation in all mocks, but as for connectivity, the quality of this correction decreases with added redshift errors.

In summary, for \ha\ flux-limited samples, as in the case of EDF, the multiplicity appears to be a more robust quantity, compared to connectivity. Indeed, even without weighting the tessellation, and after applying a correction for the FoG effect, we are able to recover the \mstar-multiplicity relation in the presence of redshift error and reduced sampling. 
However, given that the range of multiplicity values is quite restrained, we anticipate that it might still be difficult to retrieve correlations between the multiplicity of galaxies located at nodes of the cosmic web and their properties beyond stellar mass (e.g., morphology or star-formation activity).

\begin{figure*}
\centering\includegraphics[width=\textwidth]{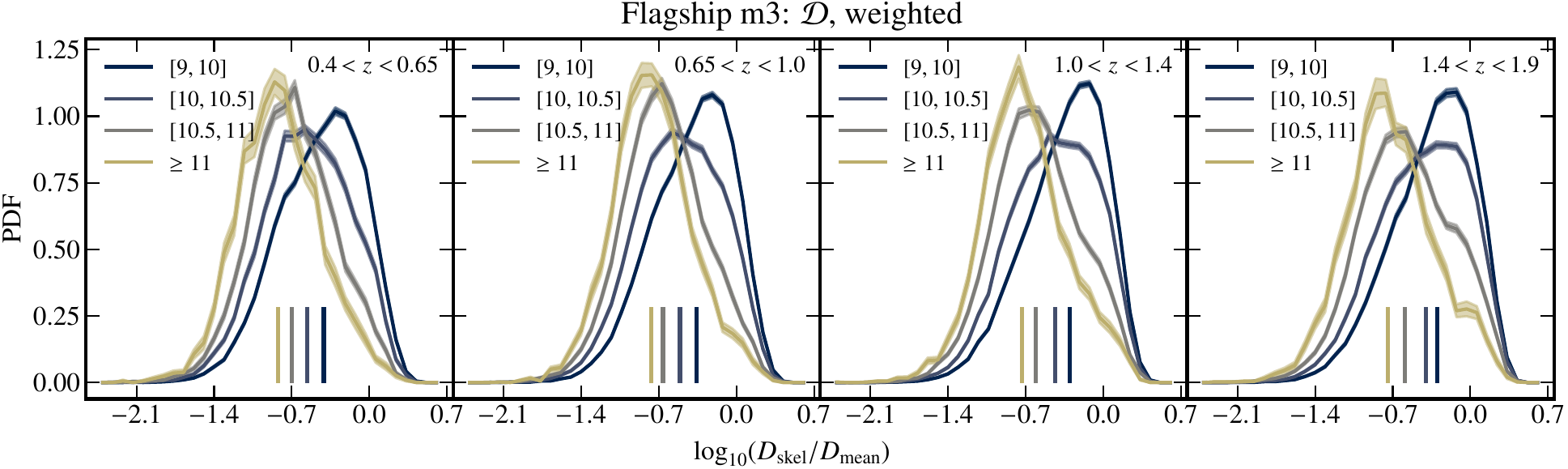}\vspace{2pt}
\centering\includegraphics[width=\textwidth]{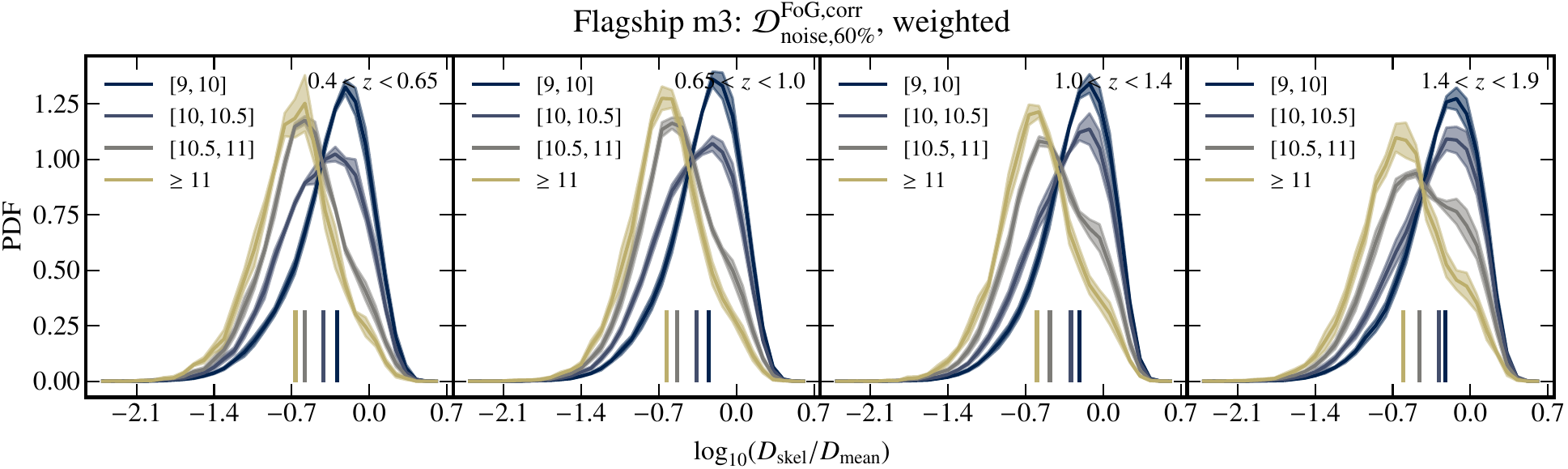}
\caption{PDFs of the closest filament's distance, normalised by the redshift dependent mean intergalactic separation, for galaxies in the \flag mock (model m3) in different stellar mass bins (coloured lines; with numbers indicating $\log_{10}(\mstar/\msun)$ values) and the cosmic web reconstructed using the stellar mass-weighted Delaunay tessellation. Stellar-mass gradients are shown as a function of redshift from the lowest redshift bin (left) to the highest one (right). Stellar-mass gradients, with more massive galaxies located closer to the filaments than their lower mass counterparts, present in the reference \ha\ flux-limited sample (\deep, first row), are recovered when the FoG correction is performed (\deepnoisesampwof, second row) for the sample with reduced completeness and added redshift error.
Vertical lines indicate medians, and the shaded regions are the bootstrap errors for the \deep catalogue and the standard deviation across five mocks for the remaining ones.
}
\label{fig:PDF_dskel_flag_m3_weight}
\end{figure*}

\subsection{Stellar-mass gradients}
\label{sec:grad}

As explained by the theory of biased clustering \citep[e.g.,][]{Kaiser1984,Efstathiou1988}, 
the mass of galaxies is expected to depend on their large-scale environment, such that more massive galaxies are preferentially formed in over-dense regions \citep[e.g.,][]{BondMyers1996,Pogosyan1996,Malavasi2017,Kraljic2018,Laigle2018}.

Let us now explore the stellar-mass gradients with respect to filaments of the cosmic web and our ability to recover this trend with the EDS. 

Figure~\ref{fig:PDF_dskel_flag_m3_weight} shows the PDFs of the distances of galaxies to their closest filament for the \flag m3 model and for the reconstruction of the cosmic web with weighted tessellation as a function of stellar mass and redshift. The equivalent PDFs for the reconstruction without weighting are presented in Fig.~\ref{fig:PDF_dskel_flag_m3_noweight}.
Similar results are obtained for the \flag m1 model and both \gaea models. The distances are normalised by the mean intergalactic separation to take into account the effect of decreasing the density of galaxies with increasing redshift. After accounting for these redshift variations, stellar-mass gradients, where more massive galaxies are preferentially located closer to filaments compared to their lower-mass counterparts, exhibit only minimal evolution in the redshift range $0.4 < z < 1.8$, with galaxies typically found closer to the filaments of the cosmic web at lower redshifts.

Stellar-mass gradients towards filaments of the cosmic web are recovered in the reference sample (\deep, first rows of Figs.~\ref{fig:PDF_dskel_flag_m3_weight} and \ref{fig:PDF_dskel_flag_m3_noweight}) regardless of whether the weighting of Delaunay tessellation is applied or not. We note that the bimodality seen for the distributions without weights is due to our choice for a relatively high value of the persistence threshold $N_{\sigma}$ applied to select filaments. This is caused by the removal of topologically less significant filaments for higher $N_{\sigma}$, which leaves some galaxies associated with filaments further away.

Reducing sample completeness weakens this signal (\deepsamp, Figs.~\ref{fig:PDF_dskel_flag_m3_weight_appendix} and \ref{fig:PDF_dskel_flag_m3_noweight}), especially when the cosmic web is reconstructed without stellar mass-weighted tessellation. Redshift errors further reduce our ability to detect the stellar-mass gradients across the entire mass range (\deepnoisesampwf, Fig.~\ref{fig:PDF_dskel_flag_m3_weight_appendix}), in particular for reconstructions without tessellation weighting. Correcting for the FoG effect does not significantly improve the strength of the signal (\deepnoisesampwof, second rows of Figs.~\ref{fig:PDF_dskel_flag_m3_weight} and \ref{fig:PDF_dskel_flag_m3_noweight}).

\section{Discussion}
\label{sec:discussion}

\subsection{Guidelines for cosmic web reconstruction with the spectroscopic Euclid Deep dataset}

One of the major difficulties identified in this work hampering the reconstruction of the cosmic web in three dimensions is the galaxy selection function. An \ha\ flux-limited, rather than a stellar mass-limited selection of galaxies, is suboptimal for the reconstruction of the cosmic web in three dimensions. On top of that, the redshift errors expected for the EDS further reduce the quality of the cosmic web reconstruction. In this section, we summarise and discuss possible choices to deal with these issues together with their implications. 

Any cosmic web reconstruction in three dimensions employing galaxy redshift-based distances needs to account for the redshift-space distortions. 
Widely used methods to correct for such an effect on small scales depend on a reliable detection of galaxy groups and clusters.
The optimisation of a group finder is typically performed on galaxy mocks designed to capture the specificities of a given survey. We have at our disposal two sets of mock catalogues, the \flag and \gaea simulations, each providing two models of \ha\ flux of galaxies. 
As all these models reasonably represent galaxy populations at $0.4 < z < 1.8$, based on comparisons with observational datasets of the galaxies' main sequence, stellar mass, and \ha\ luminosity functions, we used them to optimise the anisotropic group finder.

However, due to inherent differences in the modelling of \ha\ flux, \flag and \gaea show some fundamental differences, for example in the clustering of galaxies on small scales (below $\sim 1\, \hMpc$; see Fig.~\ref{fig:2pcf}), or number density of galaxies across the entire redshift range (see Fig.~\ref{fig:mean_dist}).
All of these non-trivial differences make it difficult to homogenise the cosmic web reconstruction. Consequently, we decided to treat each model as a plausible representation of our Universe and optimised the reconstruction process individually. The resulting optimal linking lengths are reported in Table~\ref{tab:linking_lengths}. We note that these values are different for the \flag and \gaea mocks, but they are reasonably close, not only to each other but also to the values adopted for the FoF algorithm-based group reconstruction method concentrating on lower redshift spectroscopic surveys \citep[e.g.,][]{Robotham2011,DuarteMamon2014,Treyer2018}. This means that the exact choice of the linking lengths for the reconstruction of galaxy groups in the EDS will have to be made once the data are available on the basis of the comparison with the two mocks.

When reconstructing the cosmic web from discrete galaxy samples that are \ha\ flux- rather than stellar mass-limited, weighting the Delaunay tessellation by stellar mass of galaxies might improve the quality of the reconstruction, in particular in the presence of redshift errors with an RMS of $0.001(1+z)$. However, this only holds for certain quantities. Among those explored in this work, geometrical cosmic web quantities, i.e. the length of filaments, the area of walls, and the volume of voids, are better recovered when stellar mass weighting is applied. 

For topological cosmic web measurements, such as connectivity and multiplicity, the recommendation to apply weighing is less obvious. The main reason is that the stellar mass-weighted tessellation modifies the connectivity of the reconstructed filamentary network by typically shifting it towards higher values. Hence, the comparison with theoretical predictions and existing measurements from hydrodynamical simulations \citep[e.g.][]{Kraljic2020a,GE2024} and low-redshift mass- or volume-limited samples \citep[e.g.][]{Kraljic2020a} becomes less obvious; this should be properly taken into account when interpreting any such connectivity measurement. However, stellar mass weighting helps to recover the \mstar-connectivity correlation. Interestingly, the multiplicity of the cosmic web seems to be very weakly affected by weighting, allowing us to reconstruct the correlation between the stellar mass of central galaxies and the number of locally connected filaments, as well as the global measure of the multiplicity such as its PDF. We therefore recommend using multiplicity rather than connectivity. 

Finally, the segregation of galaxies by stellar mass with respect to their distance to the filaments of the cosmic web, as explored in this work, is found to be strongly impacted by the redshift error expected for the EDS. The stellar-mass gradients are recovered even without weighting by the stellar mass of galaxies. However, this signal is quite weak, and we anticipate that measuring gradients of quantities beyond stellar mass of galaxies, for example, sSFR gradients, would be very challenging. A significant improvement can be obtained by weighting the tessellation by stellar mass. Weighting the tessellation by stellar mass may seem to lead to stellar-mass gradients somewhat by construction. However, the stellar-mass gradients with respect to the filaments are naturally present and recovered in the reference sample without any weighting for the cosmic web reconstruction. 
In addition, the tests we conducted revealed that weighting by stellar masses randomly assigned to galaxies from the same catalogue significantly reduces the presence of stellar-mass gradients. We therefore conclude that while the stellar mass weighting of the tessellation clearly biases to some extent the cosmic web reconstruction (by construction more massive galaxies will follow the cosmic web more closely), the stellar-mass gradients are inherently present in the distribution of galaxies on large scales.

\subsection{Science cases with Euclid Deep Survey}

The current study focuses on the technical aspects of cosmic web reconstruction within EDS, with the aim of providing basic guidelines for future analysis. Validation has been performed by analysing some of the more fundamental properties of the cosmic web, with correlations involving filaments on the one hand and stellar mass of galaxies on the other. \Euclid data provide the foundation to extend this analysis to other properties beyond their stellar mass, potentially allowing us to study the co-evolution of galaxies and the cosmic web at epochs close to cosmic noon.

It will be of particular interest to assess the impact of the cosmic web on the star formation of galaxies. Through measurements of the \ha\ flux of galaxies, \Euclid will provide access to their star-formation rate (SFR) for the spectroscopic sample within the EDS. 
It will therefore be possible to perform the analysis of the gradients of the specific star-formation rate (sSFR) of galaxies with respect to the distance to filaments and walls of the cosmic web. At low redshifts, galaxies with lower sSFR (or redder colours) are found to be located closer to filaments and walls, compared to galaxies with higher sSFR (or bluer colours) at fixed \mstar. Qualitatively similar trends are found in cosmological hydrodynamical simulations, with such a behaviour persisting up to at least redshift $z \sim 1$, and weakening or vanishing at $z\sim2$ \citep[e.g.][]{Xu2020,Hasan2023,Bulichi2024}, with a hint of a reversal with increasing sSFR in the vicinity of filaments
\citep[][]{Bulichi2024}.
\Euclid will allow us to test these predictions, and extend to higher redshifts our knowledge on the impact of the cosmic web on the star-formation activity of galaxies.

Low-redshift studies, both observational and numerical, also highlight the important role of connectivity (and multiplicity) of the cosmic web in shaping galaxy properties beyond their stellar mass. At fixed \mstar, galaxies at the nodes of the cosmic web with higher number of connected filaments tend to have lower sSFR and to be less rotationally supported compared to galaxies with lower connectivity \citep[][]{Kraljic2020a}. As connectivity is a practical observational proxy for past and present accretion of galaxies and therefore controls their assembly history, it is of great importance to extend this kind of study to higher redshifts. \Euclid will allow us to infer the cosmic evolution of connectivity (and multiplicity) of the cosmic web and its impact on various galaxy properties, such as sSFR, morphology, but also metallicity, to name a few. 
In particular, detailed morphological information will be obtained from photometry \citep[with various methods: profile-fitting, non-parametric estimates, deep learning;][]{Bretonniere-EP26} provided by the  VIS instrument \citep{EuclidSkyVIS}. The Euclid pipeline includes a model-fitting algorithm evaluated through the Euclid Morphology Challenge using simulated datasets. \Euclid is expected to deliver robust structural parameters for more than 400 million galaxies, with a less than 10\% scatter for single Sérsic profiles down to $\IE=23$ and double Sérsic profiles down to $\IE=21$. For complex morphologies, performance can be improved using the Zoobot CNN and larger crowd-sourced datasets. Zoobot's adaptability to new morphological labels was demonstrated with peculiar galaxies, highlighting its utility for detailed Euclid catalogues \citep{EP-Aussel}.

Besides morphology measurements, the VIS instrument will provide precise estimates of the position angle of galaxies that jointly with access to large-scale structure will enable us to investigate the alignment of angular momentum (or spin) of galaxies
with respect to the filaments and walls of the cosmic web. At low redshifts ($z \lesssim 0.2$), statistical studies using data from the SDSS, SDSS-IV MaNGA and SAMI surveys revealed that rotationally supported, disk-dominated low-mass galaxies tend to have their spin aligned with their neighbouring filaments and walls, while bulge-dominated (S0-type) massive galaxies have their spin preferentially in the perpendicular direction \citep[e.g.,][]{Tempel2013,TempelLibeskind2013,Welker2020,Kraljic2021}. 
As suggested by hydrodynamical cosmological simulations, such a spin alignment signal is expected to be stronger at $z\gtrsim1$ \citep[e.g.,][]{Dubois2014,Codis2018,Wang2018,Kraljic2020b}, making the EDS an ideal data set for measuring the 3D orientation of galaxies' angular momentum with respect to large-scale filaments and walls at high redshift.

Understanding the co-evolution of galaxies and large-scale structure is of interest not only in the context of galaxy formation and evolution, but it is also of paramount importance for cosmology. An important example is the spin alignment of galaxies with regard to large-scale structure, as it is a known source of contamination for weak-lensing-based dark energy surveys \citep[e.g.,][]{Chisari_2017}. 
Another example is cosmic connectivity and its evolution that is expected to depend on cosmology \citep[e.g.,][]{Codis2018b,WST_2024,Boldrini2025} and therefore may be an interesting probe of cosmological models beyond $\Lambda$CDM.
Beyond connectivity, other cosmic web properties, such as exclusion zones present in the cross-correlations of critical points of the density field that were found to behave as standard rulers, represent a promising tool for constraining cosmological parameters \citep[][]{Shim_2024}.

\subsection{Synergies with other surveys}

Connecting the evolution of galaxies and the large-scale cosmic web requires spectroscopic surveys with high spatial sampling across a wide area in order to properly sample a wide range of cosmic environments. 
The EDS will in this regard be complementary to the PFS Galaxy Evolution survey \citep[PFS hereafter;][]{PFS_GE2022}.
The PFS main continuum-selected sample of $\sim 3\times 10^5$ galaxies with $J<22.8$ in the redshift range $0.7 \lesssim z \lesssim 1.7$ will allow one to capture $\sim 90\%$ of the population with \mstar $\gtrsim 10^{10}$ \msun. With 70\% average completeness, PFS will enable the observation of multiple galaxies in groups with halo masses down to $\sim 10^{13}\, \msun$. This makes it ideal not only for studying the assembly history of galaxy groups, but also for the cosmic web-related analysis, thanks to an improved reconstruction of the large-scale structure (Kraljic et al. in prep.). 

The EDS is also complementary to MOONRISE \citep[the main Guaranteed Time Observation MOONS extra-galactic survey at the VLT;][]{MOONRISE2020}, an upcoming massively multiplexed spectroscopic survey that enables the study of galaxy properties with rest-frame optical line coverage up to $z \sim 2$. With its unique observing capabilities and strategy, MOONS is expected to capture the environment of galaxies across four orders of magnitude in over-density, from large under-dense void regions to high-density groups and clusters at $z \sim 1$--2. This will, in turn, allow for the reconstruction of the cosmic web, particularly the filaments connected to galaxy clusters, and facilitate the study of their impact on galaxy properties at the peak of the cosmic star-formation rate in the Universe.

In the comparable redshift range $z \sim 1$--2, additional complementarity will be provided by the planned Nancy Grace Roman Space Telescope \citep[Roman;][]{Akeson_2019}. Galaxy samples from Roman's deeper observations with the High-Latitude Wide-Area Survey \citep[HLWAS;][]{Wang_2022}, in particular its ``Deep'' and ``Ultra-Deep'' tiers covering respectively $\sim19$ and $\sim 5$ deg$^2$ areas, are expected to allow for a reconstruction of the cosmic web with a mean galaxy number density comparable to that of EDS \citep[][]{Hasan_2025}.
However, thanks to the larger wavelength coverage of Roman grisms, the emission lines [OIII] will be detectable beyond $z\sim 1.9$, extending the Roman redshift coverage to $z\sim2.8$ \citep[][]{Zhai_2021}. This represents a true complementarity to Euclid, pushing potentially the cosmic web reconstruction in 3D to even higher redshifts.

Another potential complementarity with the EDS will be offered, at lower redshifts, by DESI, which is already collecting spectra \citep[e.g.,][]{DESI_DR1_2024,DESI_DR1_2025}.
In particular, the DESI magnitude-limited Bright Galaxy Survey \citep[BGS;][]{DESI-BGS2023}, comprising approximately ten million galaxies over the redshift range $0<z<0.6$, will provide a galaxy sample up to two magnitudes fainter than the SDSS Main Galaxy Sample \citep[][]{SDSS_MGS}, with doubled median redshift ($z \approx 0.2$). With significantly higher number density than any previous survey in this redshift range, 
spectra and photometry for ten million galaxies, 
BGS will enable measurements of their physical properties and the reconstruction of the cosmic web in three dimensions. This will in turn enable studies of the connection between the large-scale environment and galaxies, by focusing on their properties and statistics, e.g. the mass-metallicity relation, the star-forming main sequence, stellar mass and luminosity functions at lower redshifts.

Finally, the Spectro-Photometer for the History of the Universe, Epoch of Reionization and Ices Explorer \citep[SPHEREx,][]{bock2025} and the Vera C. Rubin Observatory's Legacy Survey of Space and Time \citep[LSST,][]{2019ApJ...873..111I} will  both provide photometric data on wavelength range complementary to the Euclid ground-based and NIR photometric filters. On the one hand, SPHEREx, launched on 11 March 2025, will provide an all-sky low-resolution spectro-photometric survey in 102 narrow-bands from 0.75 to 5.0 $\mu$m, at a depth comparable to the \Euclid spectroscopic survey but with much lower spatial (6\arcsec.2 pixels) and spectral ($\lambda/\Delta \lambda= 35 - 130$) resolutions. The EDFs and SPHEREx deep fields are planned to overlap \citep{dore2018}, making SPHEREx a tool of choice to mitigate the fraction of catastrophic outliers in the \Euclid spectroscopic survey, by providing additional constraints for line identification on a larger wavelength range. On the other hand, LSST, will provide a $18000 \deg^2$ coverage of the south hemisphere, hence overlapping with EDFS and EDFF. In these fields, LSST will provide deep optical photometry in six filters $ugrizy$ covering the wavelength range 0.32 to 1.05 $\mu$m (in a range where \Euclid has only one broad-band). This  photometry could also be used to help tuning the spectroscopic solution of galaxies in the spectroscopic EDS. Beyond constraining the redshifts, these photometric data will be valuable to improve the stellar mass measurements of the \Euclid galaxies, which in turn is pivotal to weight the density field at the stage of skeleton extraction.

\section{Conclusions}
\label{sec:conclusion}

In this study, we have investigated the quality of the reconstruction of the cosmic web with the expected spectroscopic dataset of galaxies in the EDFs in the redshift range $0.4<z<1.8$, and with an \ha\ flux limit of $6\times 10^{-17}{\rm erg}\,{\rm cm}^{-2}\,{\rm s}^{-1}$. This analysis was carried out using the \flag and \gaea mock galaxy catalogues. 

Our main findings are the following:

\begin{itemize}
\item Geometry of the cosmic web:
The small-scale redshift-space distortions
strongly impact the cosmic web reconstruction, its filamentary network in particular, regardless of the sample selection (\ha\ flux- or \mstar-limited) and regardless of what weighting of the tessellation by the stellar mass of galaxies is chosen. The correction applied for
the FoG effect works better for stellar mass-limited samples. In addition to sample selection, the redshift uncertainties hinder our
ability to correct for the FoG effect.
\item Connectivity of the cosmic web:
For the overall properties of the PDF of the connectivity, the correction of the FoG effect works very well for the \mstar-limited sample and reasonably well for the \ha\ flux-limited sample.  Weighting the tessellation by stellar mass for the \ha\ flux-based selection artificially enhances the connectivity, even in the absence of redshift-space distortions.
Adding noise and incompleteness reduces the quality of the cosmic web reconstruction for both \ha-based and \mstar-based galaxy selections, due to the diminished ability to correct for FoG effects. In this configuration, stellar mass weighting of the tessellation significantly improves the cosmic web reconstruction, particularly for an \ha-based galaxy selection.
Weighting is similarly crucial to recover the
\mstar-connectivity relation when the sample is \ha\ flux-limited, in the presence of redshift errors, and for reduced sampling.
\item Multiplicity of the cosmic web:
For \ha\ flux-limited samples, rather than \mstar-limited samples, the multiplicity appears to be a more robust quantity compared to the connectivity. Even without weighting the tessellation, the applied correction for the FoG effect enables us to recover the \mstar-multiplicity relation. This holds true even in the presence of redshift errors and reduced sampling.
\item Stellar-mass gradients:
Stellar-mass gradients, where more massive galaxies are preferentially located closer to filaments compared to their lower-mass counterparts, are recovered regardless of whether the tessellation is weighted by stellar mass. However, the redshift error and limited sampling significantly weaken the strength of the measured signal for the \ha\ flux-limited sample in the absence of tessellation weighting.
\end{itemize}

In this work, we performed the cosmic web reconstruction in three dimensions, using simulated spectroscopic redshifts in the range $0.4 < z < 1.8$, expected in EDFs. 
The reconstruction methodology will therefore be applicable once the EDS data will be available and the performance of spectroscopic measurements fully assessed in this redshift range. The first results using the \Euclid Quick Data Release (Q1) of the Spectroscopic Processing Functions of the \Euclid pipeline \citep[][]{Q1-TP007} in the redshift range $0.9 < z < 1.8$ are promising. It was shown that a success rate above 80\%, and excellent redshift precision ($\sim 10^{-3}$) and accuracy (better than $3\times10^{-5}$) can be reached in only one visit. The final EDS will be made of about 40 visits and with much better performance.

The analysis of the cosmic web in two dimensions using photometric redshifts from the Q1 release, focused on the galaxy morphology and shape alignments at $0.5< z <0.9$, is presented in \cite{Q1-SP028}, while in \cite{Q1-SP005}, the focus is on cosmic connectivity of galaxy clusters at $0.2< z <0.7$.

This work was focused on the technical aspects of cosmic web reconstruction, with the goal of providing guidelines for future applications to real data. For this reason, we specifically only considered trends with the stellar mass of galaxies, leaving the exploration of the impact of the cosmic web on other galaxy properties for future studies. 
\Euclid will enable such an investigation for the first time across cosmic time, up to $z\approx1.8$.

\begin{acknowledgements} 
\AckEC \AckCosmoHub 
This work is partially supported by the grant \href{https://www.secular-evolution.org}{\emph{SEGAL}} ANR-19-CE31-0017
of the French Agence Nationale de la Recherche. This research has made use of computing facilities operated by CeSAM data centre at LAM, Marseille, France. We thank St\'ephane Rouberol for the smooth running of the Infinity cluster, where part of the computations was performed.
\end{acknowledgements}

\bibliographystyle{aa.bst}
\bibliography{biblio}

\newpage
\appendix
\onecolumn

\clearpage
\onecolumn

\section{The cosmic web statistics}
\label{appendix:stats}

\begin{figure*}
\centering\includegraphics[width=0.45\textwidth]{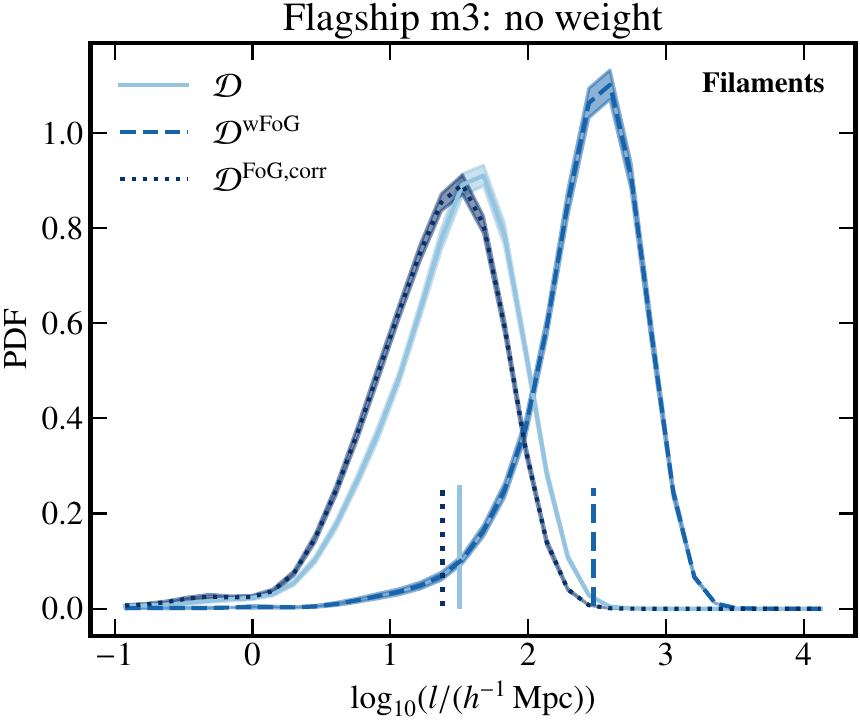}
\centering\includegraphics[width=0.45\textwidth]{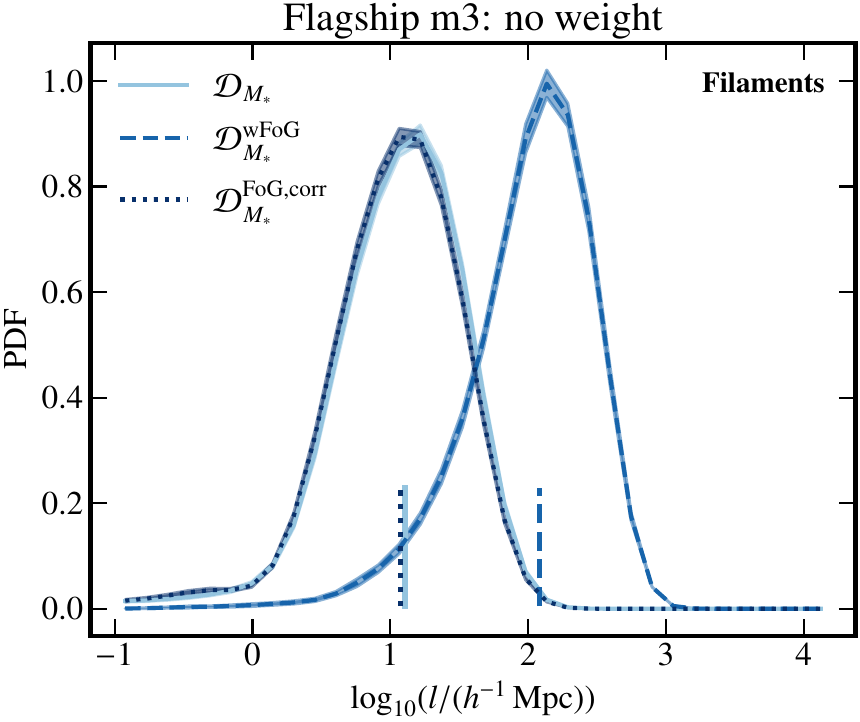}\\[5pt]
\centering\includegraphics[width=0.45\textwidth]{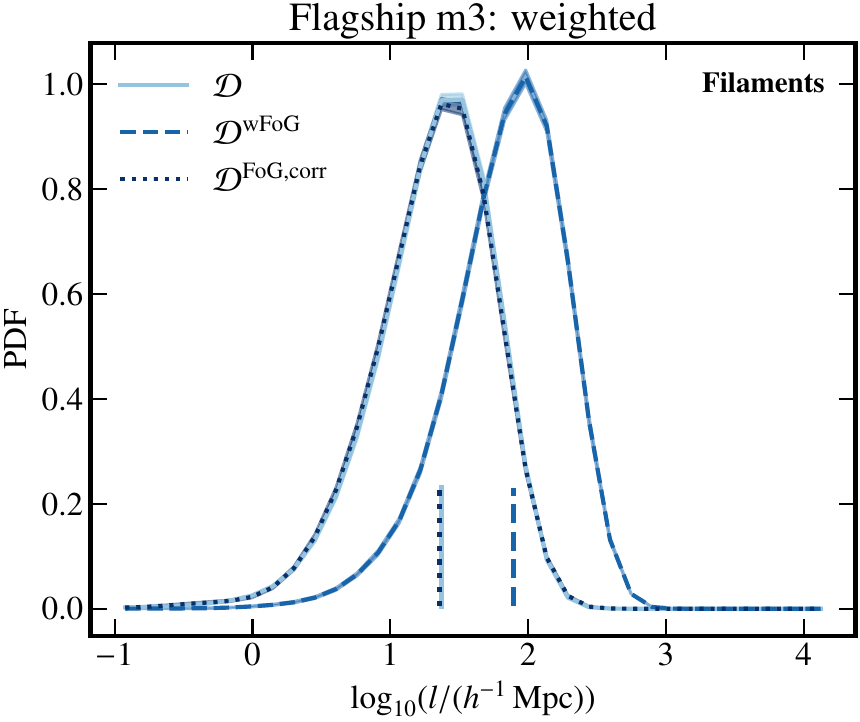}
\centering\includegraphics[width=0.45\textwidth]{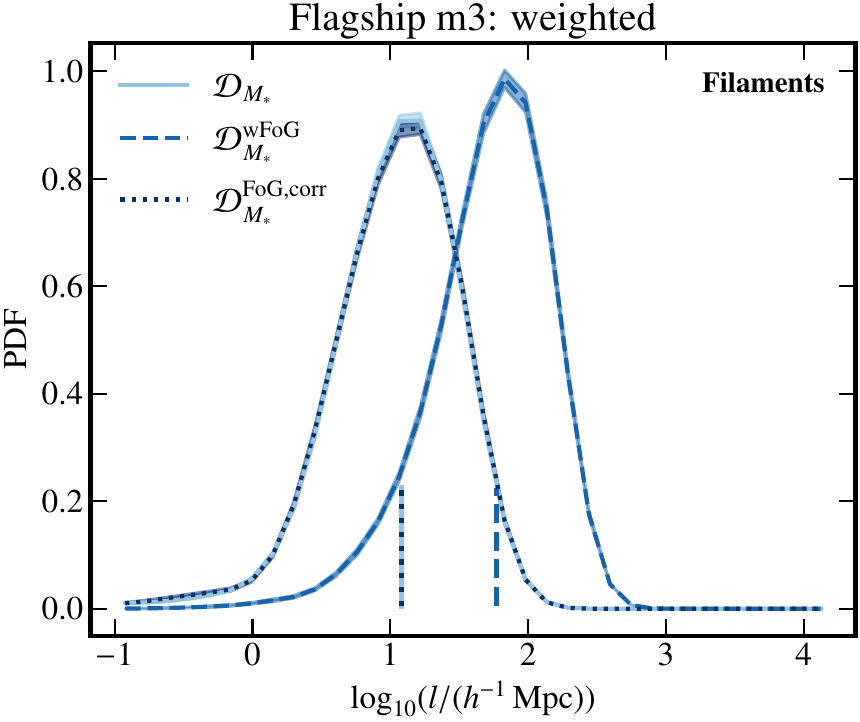}\\[5pt]
\caption{PDF of filament lengths for \ha- (left) and stellar mass-limited (right) samples. Top and bottom panels compare non-weighted and weighted tessellations, respectively, for the \flag m3 model. 
Vertical lines indicate the medians of distributions. Shaded regions correspond to the bootstrap error bars. For the stellar mass-limited galaxy sample, weighting of the Delaunay tessellation for the cosmic web reconstruction is not required to obtain a good agreement for the distribution of filaments' length.
}
\label{fig:PDF_stats_mass_vs_halpha}
\end{figure*}

\begin{figure*}
\centering\includegraphics[width=0.45\textwidth]{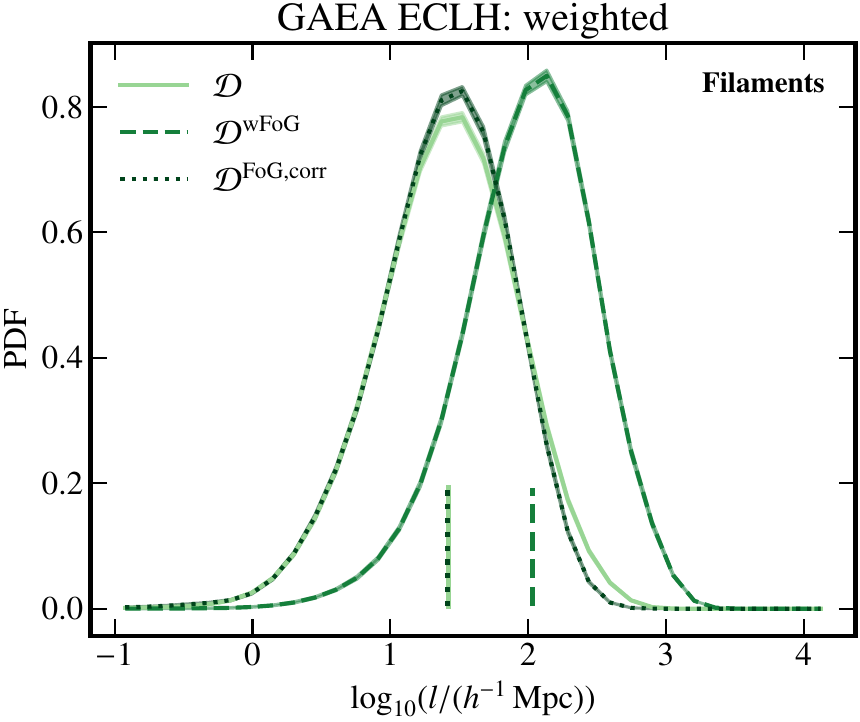}
\centering\includegraphics[width=0.45\textwidth]{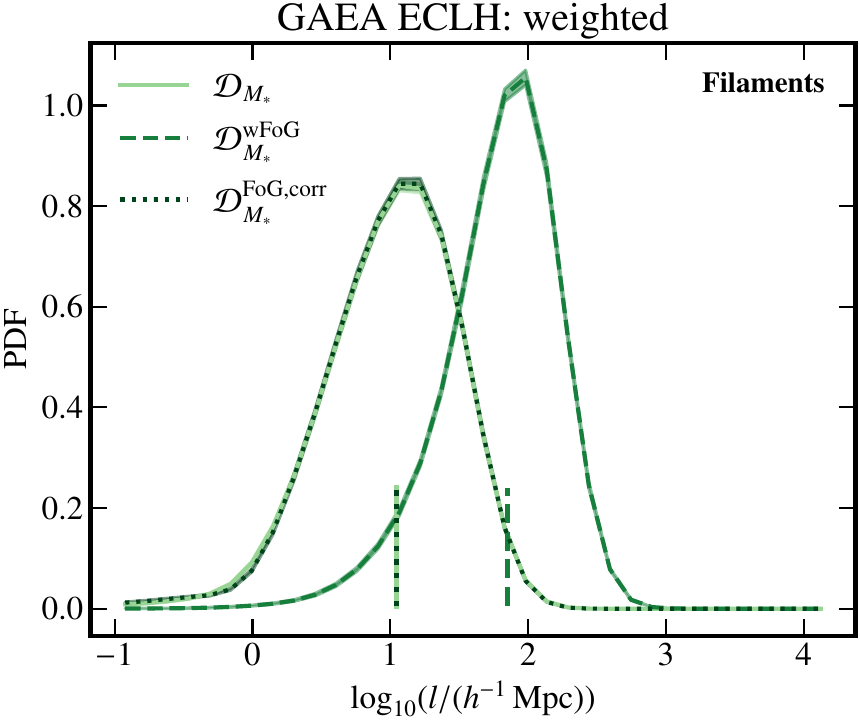}
\caption{As in Fig.~\ref{fig:PDF_stats_mass_vs_halpha}, bottom panels, i.e. using the stellar mass-weighted tessellation for the cosmic web reconstruction, but for the \gaea ECLH model.
}
\label{fig:PDF_stats_mass_vs_halpha_gaea}
\end{figure*}

Figure~\ref{fig:PDF_stats_mass_vs_halpha} illustrates the effect of the selection function of galaxies and the mass weighting of the Delaunay tessellation on the distribution of filament lengths in the \flag m3 model (see Fig.~\ref{fig:PDF_stats_mass_vs_halpha_gaea} for the \gaea ECLH model with weighting). For the \ha-limited samples (left panels), weighting of the tessellation by stellar mass is necessary to correct for the FoG effect (compare the top left and bottom left panels). For a \mstar-limited sample (right panels), weighting is not needed to obtain a good agreement for the distribution of filaments' length after correcting for the FoG effect (compare top right and middle right panels). This behaviour is recovered in all models used in this work.

\begin{figure}
\centering\includegraphics[width=0.45\textwidth]{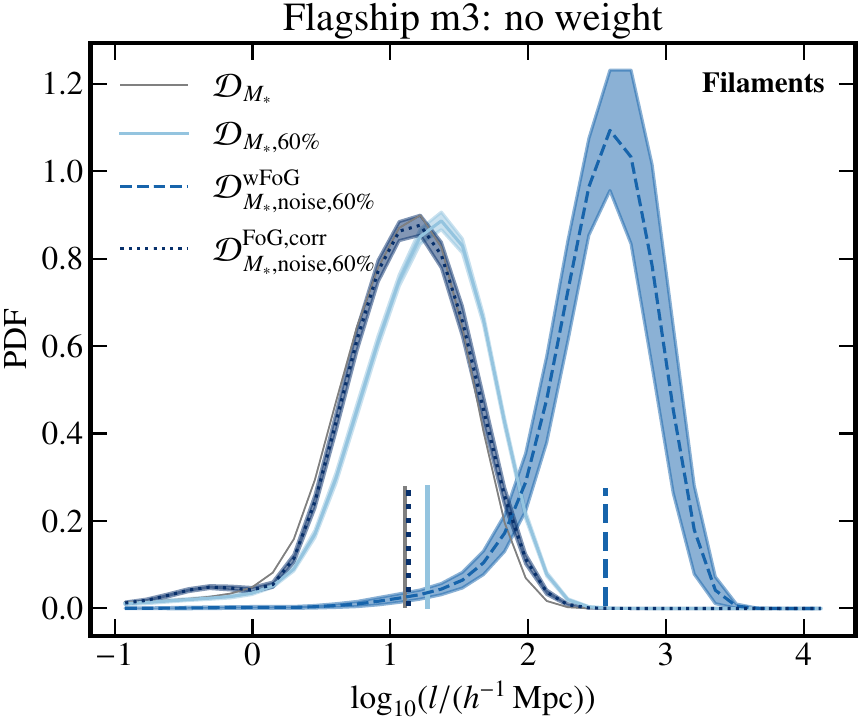}
\centering\includegraphics[width=0.45\textwidth]{./figs/PDF_fil_len_fof_l0.07_R25.0_flagship2_1_10b_deep_m3_full_noise_all_K3_S5}
\caption{PDF of filaments' length for stellar mass- (left) and \ha-limited (right) samples with 60\% sampling.
}
\label{fig:PDF_stats_mass_noise}
\end{figure}

Figure~\ref{fig:PDF_stats_mass_noise} highlights the effect of the reduced sampling on the distribution of filament lengths. As expected, reduced sampling (60\%, as estimated for the EDF) shifts the distribution of lengths of filaments to higher values, regardless of sample selection (\ha- and \mstar-limited sample, left and right panels, respectively).

\section{Connectivity and multiplicity}
\label{appendix:connect}

\begin{figure*}
\centering\includegraphics[width=0.49\textwidth]{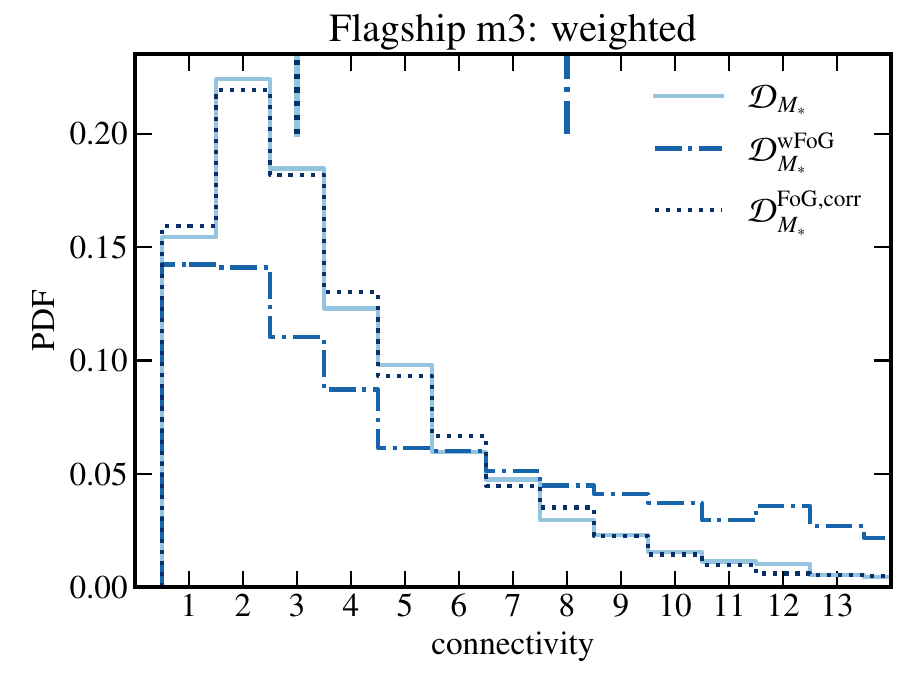}
\centering\includegraphics[width=0.49\textwidth]{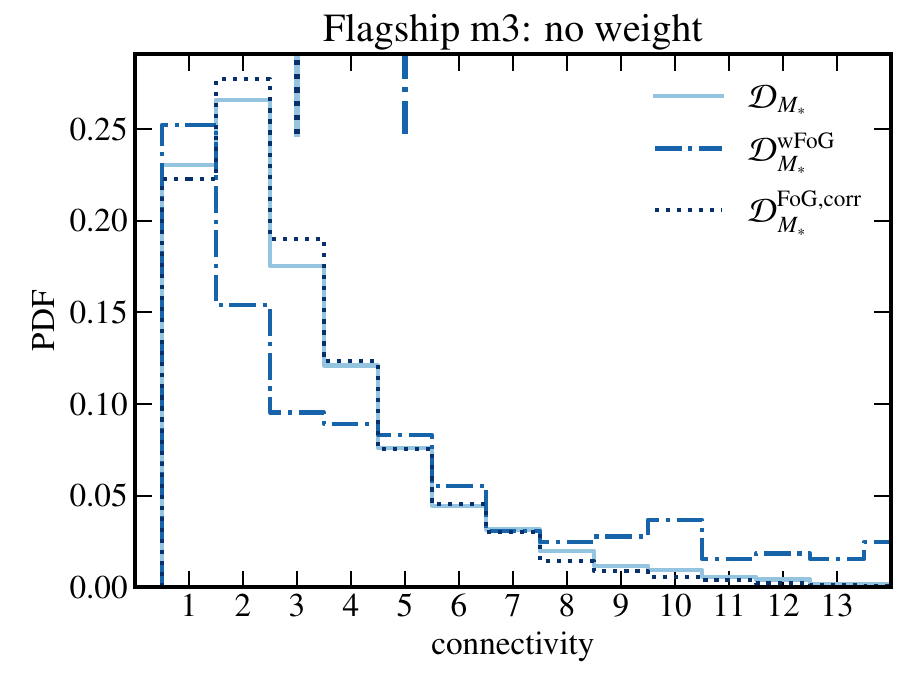}\\
\centering\includegraphics[width=0.49\textwidth]{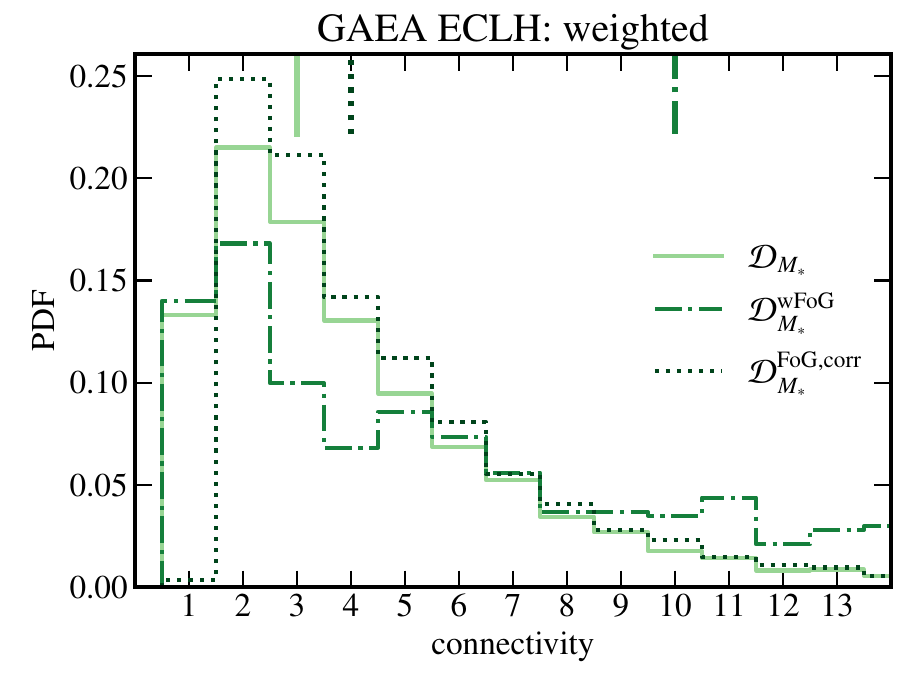}
\centering\includegraphics[width=0.49\textwidth]{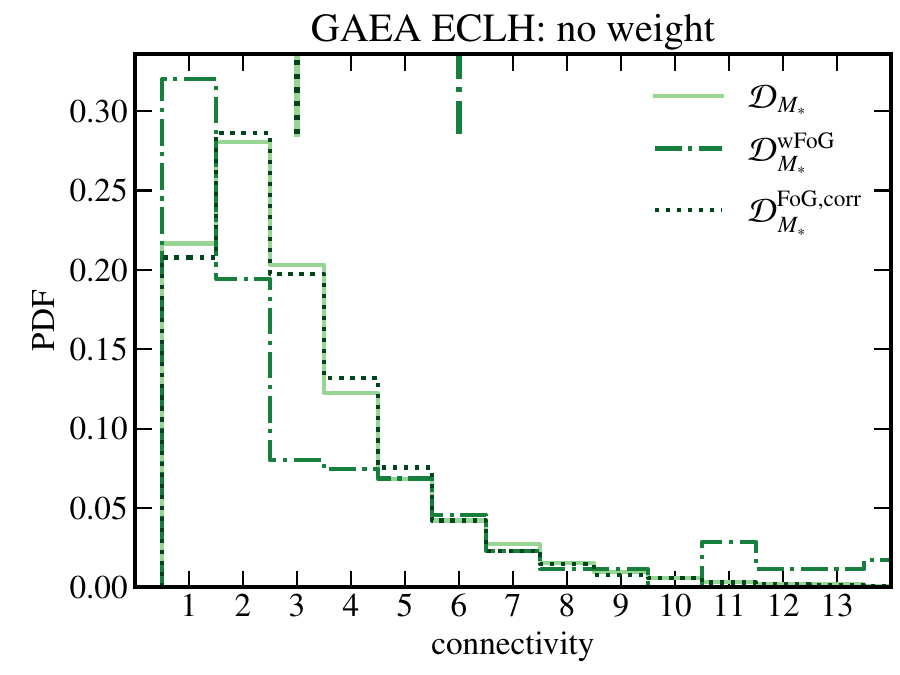}
\caption{As in Fig.~\ref{fig:PDF_connect}, but for the \mstar-limited galaxy selection. The mean and median values for all distributions are reported in Tables~\ref{tab:con_flag} and \ref{tab:con_gaea}.
}
\label{fig:PDF_connect_mass_lim}
\end{figure*}

Figure~\ref{fig:PDF_connect_mass_lim} shows the PDFs of the connectivity of central galaxies for \mstar- limited selection of galaxies in the full sample without redshift error (\deep) for \flag (model m3, top panels) and \gaea (model ECLH, bottom panels), with stellar mass weighting of the Delaunay tessellation (left) and without weighting (right). The stellar mass weighting impacts the PDF of the connectivity to a much lesser degree when the galaxy sample is \mstar-limited compared to an \ha-limited selection. After correcting for the FoG effect, the shape of the distributions and their medians (vertical lines) are in very good agreement with the measurements in the absence of redshift-space distortions
without tessellation weighting for both \flag and \gaea. Stellar mass weighting introduces some spurious effects, preventing us from completely correcting for the FoG effect in \gaea. However, the distribution of the connectivity and its median are substantially improved. The mean and median values of connectivity for all mocks and selections are reported in Tables~\ref{tab:con_flag} and \ref{tab:con_gaea}.

Regarding the multiplicity, in general, PDFs of the multiplicity and their median values are only weakly impacted by the FoG effect when stellar mass-weighted tessellation is used for cosmic web extraction. Without weighting, the applied FoG compression efficiently corrects for the redshift-space distortion for the distribution of the multiplicity and its median value. These conclusions apply to all tested models (i.e., including \mstar-limited samples). 
The mean and median values of multiplicity for all mocks and selections are reported in Tables~\ref{tab:multi_flag} and \ref{tab:multi_gaea}.

\begin{figure*}
\centering\includegraphics[width=0.49\textwidth]{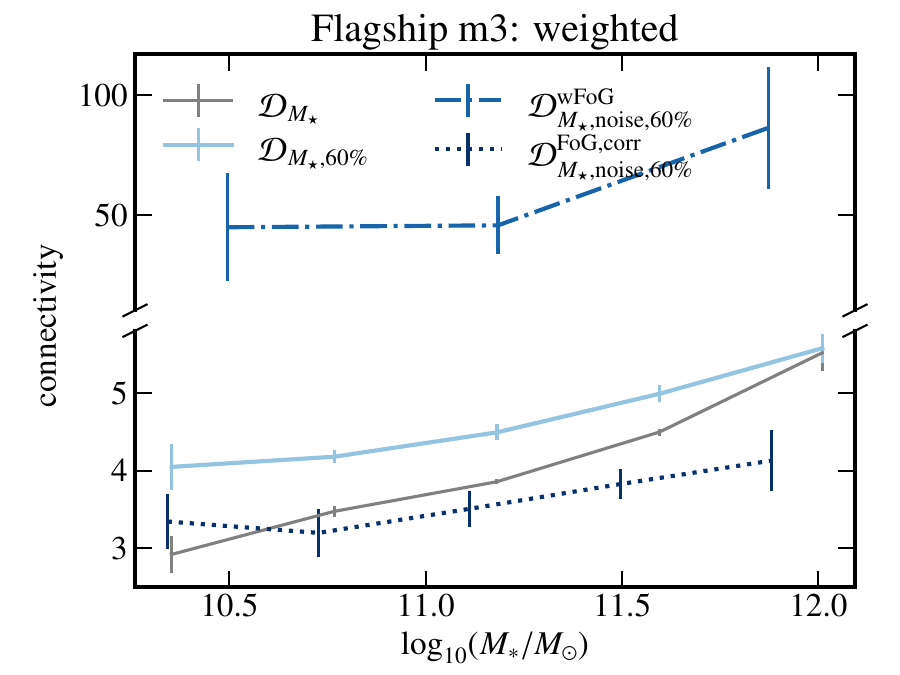}
\centering\includegraphics[width=0.49\textwidth]{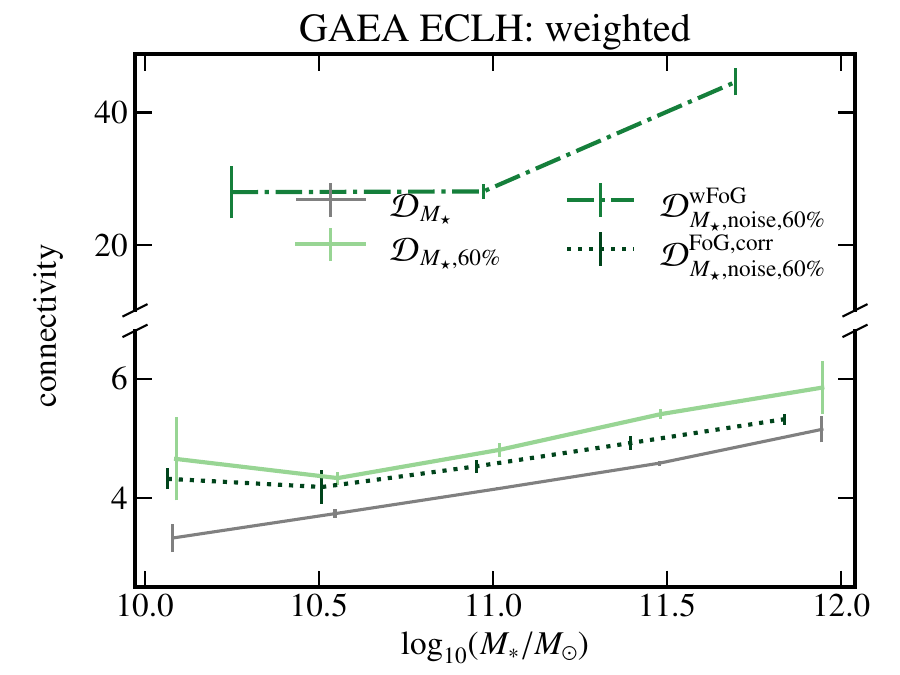}
\caption{Connectivity of central galaxies as a function of their stellar mass in the \flag (model m3) and \gaea (model ECLH) mocks, right and left panels, respectively, for \mstar-limited samples (\deepmassnoisesamp), with \mstar-weighted tessellation (qualitatively similar results are obtained without weighting).
}
\label{fig:connect_mass_flag_gaea_full_mass_lim}
\end{figure*}

Figure~\ref{fig:connect_mass_flag_gaea_full_mass_lim} illustrates the mean connectivity of central galaxies as a function of their stellar mass in the model m3 of the \flag (left) and model ECLH of \gaea (right) for \mstar-limited samples with reduced sampling and redshift uncertainties (\deepnoisesamp) 
with tessellation weighting, but qualitatively similar results are obtained without weighting.
Regardless of weighting, the FoG effect increases the connectivity by an order of magnitude at all stellar masses. The method used to correct for the small-scale redshift-space distortions helps bring the connectivity close to the values obtained for the sample without FoG. Weighting the tessellation for the \mstar-limited sample does not seem to introduce the bias seen for the selection of galaxies based on their \ha\ flux, that is the increasing connectivity at low stellar masses. Similar conclusions apply to all other models.

\begin{figure*}
\centering\includegraphics[width=0.49\textwidth]{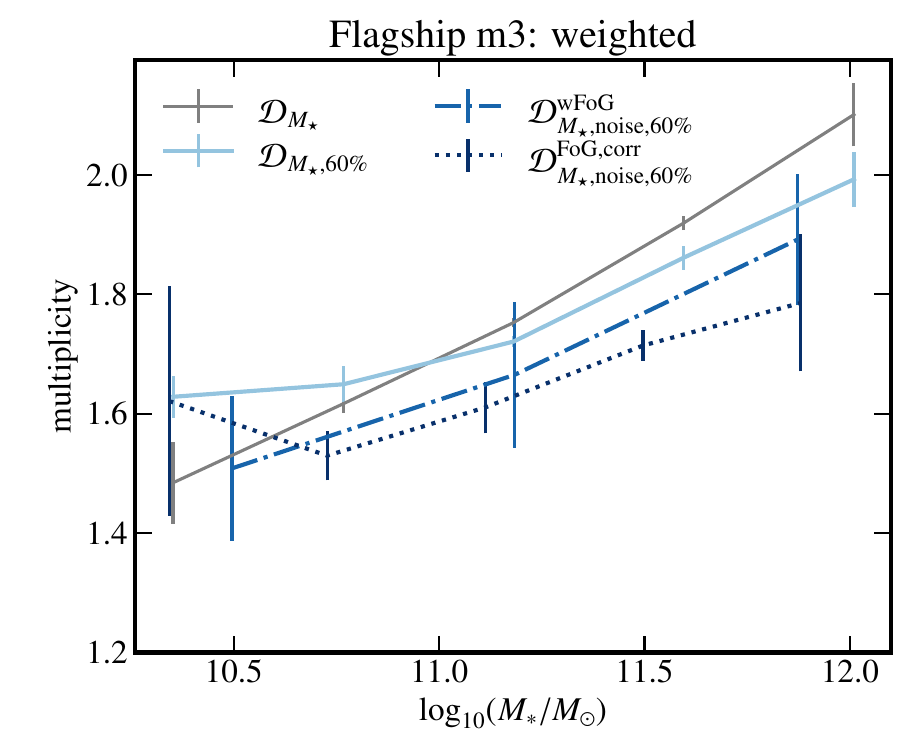}
\centering\includegraphics[width=0.49\textwidth]{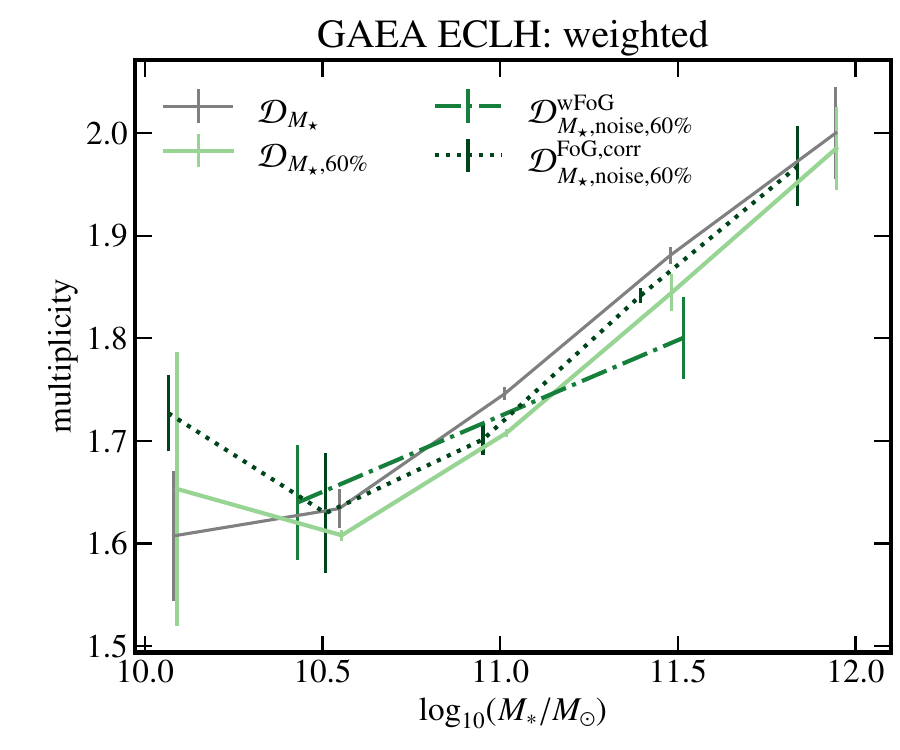}
\caption{Multiplicity of central galaxies as a function of their stellar mass in \flag (model m3) and \gaea (model ECLH) mocks, left and right panels, respectively, for \mstar-limited samples
(\deepmassnoisesamp), with \mstar-weighted tessellation (qualitatively similar results are obtained without weighting).
}
\label{fig:multi_mass_flag_gaea_full_mass_lim}
\end{figure*}

Figure~\ref{fig:multi_mass_flag_gaea_full_mass_lim} shows the mean multiplicity of central galaxies as a function of their stellar mass in the model m3 of \flag (left) and model ECLH of \gaea (right) for \mstar-limited samples with reduced sampling and redshift uncertainties (\deepnoisesamp) with 
tessellation weighting, but qualitatively similar results are found without weighting. 
The multiplicity appears to be a more robust quantity than the connectivity, given that the FoG effect has an overall much weaker impact on its values, in particular when the tessellation is weighted by stellar mass.

\begin{figure*}
\centering\includegraphics[width=0.49\textwidth]{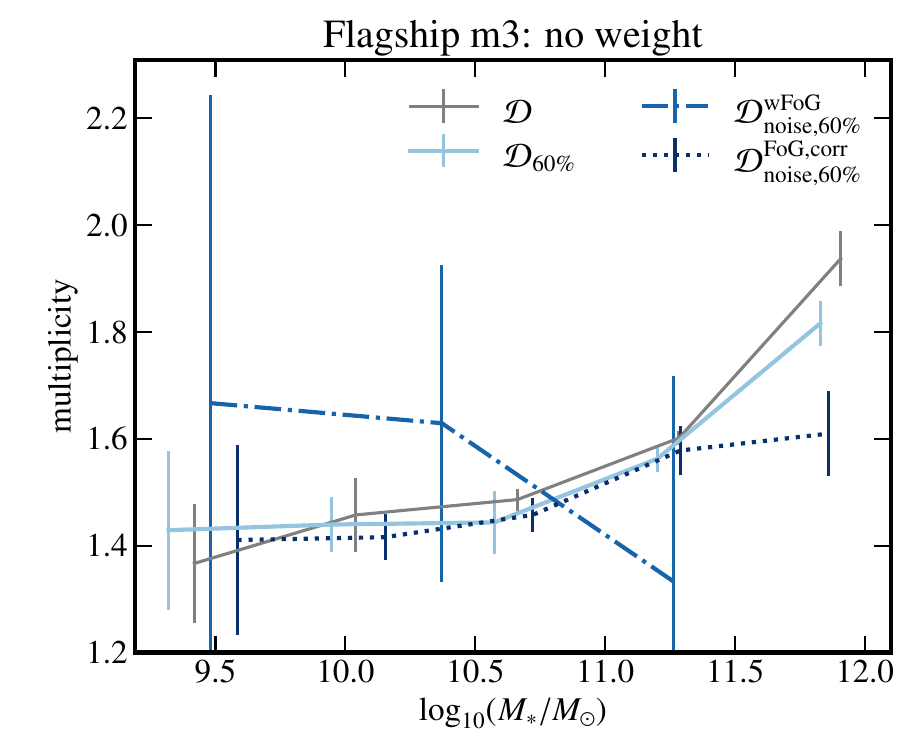}
\centering\includegraphics[width=0.49\textwidth]{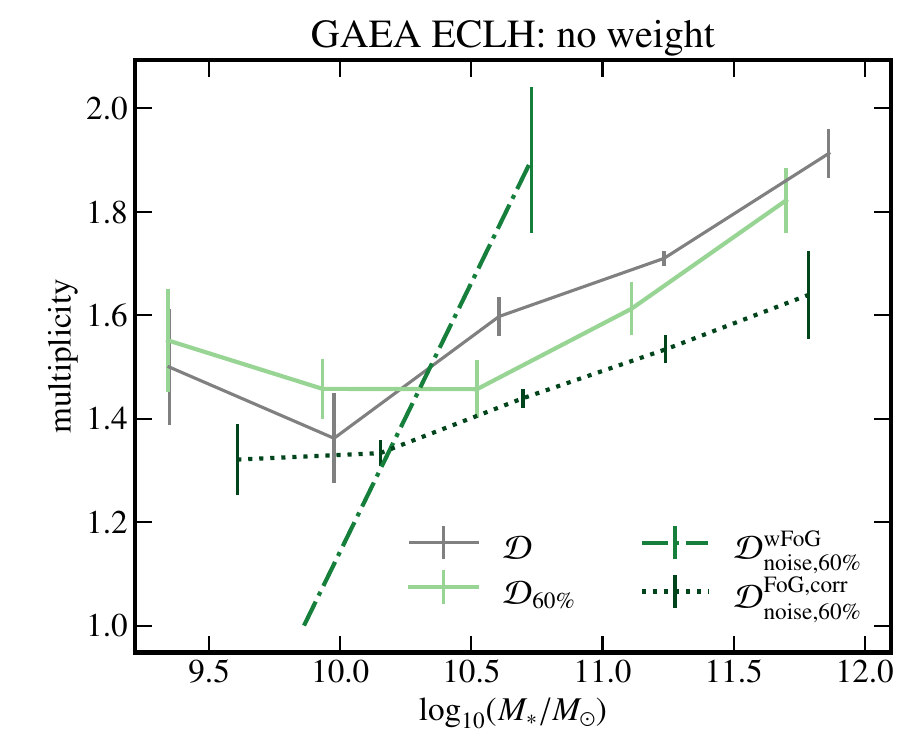}
\caption{As in Fig.~\ref{fig:multi_mass_flag_gaea} but without stellar mass weighting of the skeleton. 
}
\label{fig:multi_mass_flag_gaea_appendix}
\end{figure*}

Figure~\ref{fig:multi_mass_flag_gaea_appendix} shows the multiplicity of central galaxies as a function of their stellar mass for the \flag and \gaea mocks (models m3 and ECLH, respectively, with qualitatively similar conclusions for m1 and ECLQ) for the fiducial sample (\deepnoisesamp) without stellar mass-weighted tessellation.
The multiplicity of central galaxies increases with increasing stellar mass for galaxies in all reference catalogues (\deep), catalogues with reduced sampling and redshift uncertainties (\deepnoisesamp). The applied FoG correction significantly improves our ability to recover the \mstar-multiplicity relation.

\begin{table}
\centering
\caption{Median and mean connectivity for the \flag m3 and \flag m1 mocks, for tessellations with and without stellar mass weighting.} 
\label{tab:con_flag}
\setlength{\tabcolsep}{4.3pt}
\begin{tabular*}{\textwidth}{@{\extracolsep{\fill}}l|cc|cc|cc|cc}
\hline
 & \multicolumn{4}{c|}{} & \multicolumn{4}{c}{} \\[-9pt]
 & \multicolumn{4}{c|}{\flag m3} & \multicolumn{4}{c}{\flag m1}\\
 \cline{2-9}
 &  \multicolumn{2}{c|}{median} & \multicolumn{2}{c|}{mean} &  \multicolumn{2}{c|}{median} & \multicolumn{2}{c}{mean}\\
 \cline{2-9}
 & & & & & & & & \\[-9pt]
 &  no weight & weight & no weight & weight &  no weight & weight & no weight & weight\\
 \hline
 & & & & & & & & \\[-9pt]
\deep & 3 & 8 & $4.95\pm0.11$ & $10.94\pm0.17$& 3 & 7 & $4.52\pm{0.06}$ & $10.07\pm{0.11}$\\
\deepwf & 6 & 19 & $50.60\pm9.31$ & $58.54\pm3.8$ & 6.5 & 17 & $48.83\pm{7.51}$ & $53.12\pm{2.47}$\\
\deepwof & 3 & 7 & $4.15\pm0.06$ & $10.95\pm0.18$ & 3 & 7 &  $3.79\pm{0.04}$&$10.55\pm{0.12}$ \\
\hline
& & & & & & & & \\[-9pt]
\deepmass & 3 & 3 & $3.18\pm0.03$ & $3.99\pm0.03$&  3&3  &$3.19\pm{0.02}$  &$4.24\pm{0.03}$ \\
\deepmasswf & 5 & 8 & $17.12\pm1.49$ & $16.64\pm0.72$& 5 &8  & $18.19\pm{1.45}$ &$18.51\pm{0.67}$ \\
\deepmasswof & 3 & 3 & $3.04\pm0.02$ & $3.93\pm0.03$&  2& 3 & $2.96\pm{0.02}$ &$4.21\pm{0.03}$ \\
\hline
& & & & & & & & \\[-9pt]
\deepsamp & 4 & 9 & $5.51\pm{0.45}$& $14.56\pm{0.87}$ & 3&8 &$5.03\pm{0.29}$ &$12.96\pm{0.53}$ \\
\deepnoisesampwf & 32 & 99 & $160.71\pm{247.66}$& $282.84\pm{156.16}$ & 24&86 & $153.18\pm{270.71}$& $275.60\pm{139.61}$ \\
\deepnoisesampwof & 3 & 8 & $3.25\pm{0.14}$& $13.87\pm{0.81}$ &2 &8 & $2.95\pm{0.09}$& $13.26\pm{0.57}$ \\
\hline
& & & & & & & & \\[-9pt]
\deepmasssamp & 3 & 4 & $3.40\pm{0.11}$& $4.62\pm{0.14}$ &3 & 4&$3.38\pm{0.09}$ &$5.02\pm{0.12}$ \\
\deepmassnoisesampwf & 10.5 & 23.5 & $46.30\pm{27.75}$&$56.03\pm{23.76}$ & 13& 28&$62.97\pm{50.37}$ &$80.22\pm{26.05}$\\
\deepmassnoisesampwof & 2 & 3 & $2.68\pm{0.15}$&$3.64\pm{0.22}$ &2 & 3&$2.59\pm{0.11}$ &$4.15\pm{0.19}$\\
\hline
\end{tabular*}
\end{table}

\begin{table}
\centering
\caption{Median and mean connectivity for the GAEA ECLH and GAEA ECLQ mocks, for tessellations with and without stellar mass weighting.} 
\label{tab:con_gaea}
\setlength{\tabcolsep}{3.5pt}
\begin{tabular*}{\textwidth}{@{\extracolsep{\fill}}l|cc|cc|cc|cc}
\hline
& \multicolumn{4}{c|}{} & \multicolumn{4}{c}{} \\[-9pt]
 & \multicolumn{4}{c|}{\gaea ECLH} & \multicolumn{4}{c}{\gaea ECLQ}\\
 \cline{2-9}
 &  \multicolumn{2}{c|}{median} & \multicolumn{2}{c|}{mean} &  \multicolumn{2}{c|}{median} & \multicolumn{2}{c}{mean}\\
 \cline{2-9}
 & & & & & & & & \\[-9pt]
 &  no weight & weight & no weight & weight &  no weight & weight & no weight & weight\\
 \hline
 & & & & & & & & \\[-9pt]
\deep & 5 & 11 & $8.20\pm0.38$ & $19.09\pm0.44$&6  &16  &$14.97\pm{1.37}$  &$34.07\pm{1.45}$ \\
\deepwf & 3 & 18 & $116.33\pm32.70$ & $129.37\pm13.09$ & 3 & 18 & $128.77\pm{50.96}$  &$164.69\pm{20.19}$ \\
\deepwof & 4 & 11 & $5.43\pm0.12$ & $19.24\pm0.40$ & 5 & 19 &$7.75\pm{0.26}$  &$37.27\pm{1.19}$ \\
\hline
& & & & & & & & \\[-9pt]
\deepmass & 3 & 3 & $3.05\pm0.02$ & $4.26\pm0.03$& 3 & 4 & $3.13\pm{0.02}$  &$5.21\pm{0.04}$ \\
\deepmasswf & 6 & 10 & $30.66\pm4.06$ & $27.13\pm1.41$& 8 & 11 & $33.85\pm{4.36}$ &$35.67\pm{2.26}$ \\
\deepmasswof & 3 & 4 & $3.04\pm0.02$ & $4.73\pm0.03$& 3 & 4 &  $3.08\pm{0.02}$&$5.18\pm{0.04}$ \\
\hline
& & & & & & & & \\[-9pt]
\deepsamp &5 & 13& $9.31\pm{1.61}$&$25.25\pm{2.50}$ &7 & 19& $17.84\pm{6.21}$&$46.69\pm{7.49}$ \\
\deepnoisesampwf & 145& 253.5& $495.88\pm{486.04}$&$852.72\pm{944.33}$ &33 &126 & $616.36\pm{}1460.80$&$1340.72\pm{1704.38}$ \\
\deepnoisesampwof & 3& 14& $3.61\pm{0.21}$&$26.30\pm{1.94}$ & 3&20 & $4.18\pm{0.32}$&$43.05\pm{4.45}$ \\
\hline
& & & & & & & & \\[-9pt]
\deepmasssamp & 3& 4& $3.28\pm{0.10}$& $5.02\pm{0.15}$& 3&5 & $3.41\pm{0.12}$&$6.55\pm{0.24}$ \\
\deepmassnoisesampwf & 7&12 & $35.57\pm{12.73}$&$33.81\pm{5.95}$ & 7& 12& $36.59\pm{14.23}$& $42.94\pm{9.00}$\\
\deepmassnoisesampwof & 3&4 &$3.14\pm{0.10}$ & $4.71\pm{0.16}$&3 &4 & $3.20\pm{0.12}$&$5.90\pm{0.25}$ \\
\hline
\end{tabular*}
\end{table}

\begin{table}
\centering
\caption{Median and mean multiplicity for the \flag m3 and \flag m1 mocks, for tessellations with and without stellar mass weighting.} 
\label{tab:multi_flag}
\setlength{\tabcolsep}{9pt}
\begin{tabular*}{\textwidth}{@{\extracolsep{\fill}}l|cc|cc|cc|cc}
\hline
& \multicolumn{4}{c|}{} & \multicolumn{4}{c}{} \\[-9pt]
 & \multicolumn{4}{c|}{\flag m3} & \multicolumn{4}{c}{\flag m1}\\
 \cline{2-9}
 &  \multicolumn{2}{c|}{median} & \multicolumn{2}{c|}{mean} &  \multicolumn{2}{c|}{median} & \multicolumn{2}{c}{mean}\\
 \cline{2-9}
  & & & & & & & & \\[-9pt]
 &  no weight & weight & no weight & weight &  no weight & weight & no weight & weight\\
 \hline
 & & & & & & & & \\[-9pt]
\deep &2 &2 &$1.58\pm{0.01}$ &$2.03\pm{0.01}$ &2 &2 &$1.62\pm{0.01}$ &$2.07\pm{0.01}$  \\
\deepwf &1 &2 &$1.48\pm{0.06}$ &$2.07\pm{0.03}$ &1 &2 &$1.48\pm{0.04}$ &$2.13\pm{0.02}$\\
\deepwof &2 &2 &$1.62\pm{0.01}$ &$2.04\pm{0.01}$ &2 &2 &$1.62\pm{0.01}$ &$2.07\pm{0.01}$ \\
\hline
& & & & & & & & \\[-9pt]
\deepmass &2 &2 &$1.62\pm{0.01}$ &$1.78\pm{0.01}$ &2 &2 &$1.63\pm{0.01}$ &$1.81\pm{0.01}$ \\
\deepmasswf &1 &2 &$1.55\pm{0.03}$ &$1.77\pm{0.02}$ &1 &2 &$1.52\pm{0.03}$ &$1.83\pm{0.02}$ \\
\deepmasswof &2 &2 &$1.65\pm{0.01}$ &$1.79\pm{0.01}$ &2 &2 &$1.63\pm{0.01}$ &$1.83\pm{0.01}$ \\
\hline
& & & & & & & & \\[-9pt]
\deepsamp &1 &2 & $1.56\pm{0.05}$&$2.02\pm{0.04}$ & 1&2 & $1.58\pm{0.04}$&$2.06\pm{0.03}$ \\
\deepnoisesampwf &1 &2 &$1.57\pm{0.64}$ &$2.07\pm{0.25}$ & 1&2 &$1.29\pm{0.35}$ &$2.19\pm{0.18}$ \\
\deepnoisesampwof &1 &2 &$1.52\pm{0.04}$ &$1.97\pm{0.04}$ & 1&2 &$1.51\pm{0.03}$ &$2.04\pm{0.03}$ \\
\hline
& & & & & & & & \\[-9pt]
\deepmasssamp &2 &2 &$1.59\pm{0.03}$ &$1.76\pm{0.03}$ &2 & 2&$1.59\pm{0.02}$ & $1.80\pm{0.02}$\\
\deepmassnoisesampwf &1 &2 &$1.53\pm{0.27}$ &$1.74\pm{0.16}$ &1.5 & 2&$1.55\pm{0.27}$ &$1.91\pm{0.14}$ \\
\deepmassnoisesampwof &1 &2 &$1.56\pm{0.06}$ &$1.66\pm{0.05}$ &1 & 2&$1.53\pm{0.04}$ & $1.71\pm{0.04}$\\
\hline
\end{tabular*}
\end{table}

\begin{table}
\centering
\caption{Median and mean multiplicity for the \gaea ECLH and \gaea ECLQ mocks, for tessellations with and without stellar mass weighting.} 
\label{tab:multi_gaea}
\setlength{\tabcolsep}{9pt}
\begin{tabular*}{\textwidth}{@{\extracolsep{\fill}}l|cc|cc|cc|cc}
\hline
& \multicolumn{4}{c|}{} & \multicolumn{4}{c}{} \\[-9pt]
 & \multicolumn{4}{c|}{\gaea ECLH} & \multicolumn{4}{c}{\gaea ECLQ}\\
 \cline{2-9}
 &  \multicolumn{2}{c|}{median} & \multicolumn{2}{c|}{mean} &  \multicolumn{2}{c|}{median} & \multicolumn{2}{c}{mean}\\
 \cline{2-9}
 & & & & & & & & \\[-9pt]
 &  no weight & weight & no weight & weight &  no weight & weight & no weight & weight\\
 \hline
 & & & & & & & & \\[-9pt]
\deep & 2 &2  &$1.69\pm{0.02}$ &$2.21\pm{0.01}$ &2  &2  &$1.61\pm{0.03}$ &$2.22\pm{0.02}$  \\
\deepwf & 1 &2 &$1.36\pm{0.07}$ &$2.07\pm{0.04}$ &1 &2 &$1.28\pm{0.07}$ &$2.08\pm{0.04}$ \\
\deepwof & 2 &2 &$1.61\pm{0.01}$ &$2.14\pm{0.01}$ &2 &2 &$1.62\pm{0.02}$ &$2.19\pm{0.02}$ \\
\hline
& & & & & & & & \\[-9pt]
\deepmass &2  &2  &$1.66\pm{0.01}$ & $1.78\pm{0.01}$&2 &2 &$1.65\pm{0.01}$ &$1.80\pm{0.01}$ \\
\deepmasswf &1 &2 &$1.46\pm{0.04}$ &$1.78\pm{0.02}$ &1 &2 &$1.54\pm{0.04}$ &$1.79\pm{0.03}$ \\
\deepmasswof &2 &2 &$1.67\pm{0.01}$ &$1.80\pm{0.01}$ &2 &2 &$1.67\pm{0.01}$ &$1.85\pm{0.01}$ \\
\hline
& & & & & & & & \\[-9pt]
\deepsamp &2 &2 &$1.64\pm{0.07}$ &$2.17\pm{0.05}$ &2 &2 &$1.59\pm{0.11}$ &$2.15\pm{0.07}$ \\
\deepnoisesampwf &1 &2 &$1.50\pm{0.81}$ &$2.18\pm{0.33}$ &1 &2 & $1.28\pm{0.63}$&$2.15\pm{0.46}$ \\
\deepmassnoisesampwof & 1 &2 &$1.49\pm{0.04}$ &$2.10\pm{0.05}$ &1  &2  &$1.47\pm{0.05}$ &$2.05\pm{0.06}$ \\
\hline
& & & & & & & & \\[-9pt]
\deepmasssamp &2 &2 &$1.60\pm{0.03}$ &$1.76\pm{0.02}$ & 2 & 2& $1.60\pm{0.03}$ & $1.80\pm{0.03}$ \\
\deepmassnoisesampwf & 1&2 &$1.43\pm{0.11}$ &$1.76\pm{0.08}$ & 1&2 & $1.50\pm{0.13}$&$1.81\pm{0.09}$ \\
\deepmassnoisesampwof & 2 &2  &$1.63\pm{0.03}$  &$1.77\pm{0.03}$ & 2& 2& $1.62\pm{0.04}$&$1.81\pm{0.04}$ \\
\hline
\end{tabular*}
\end{table}

\section{Stellar-mass gradients}
\label{appendix:grads}

\begin{figure*}
\centering\includegraphics[width=\textwidth]{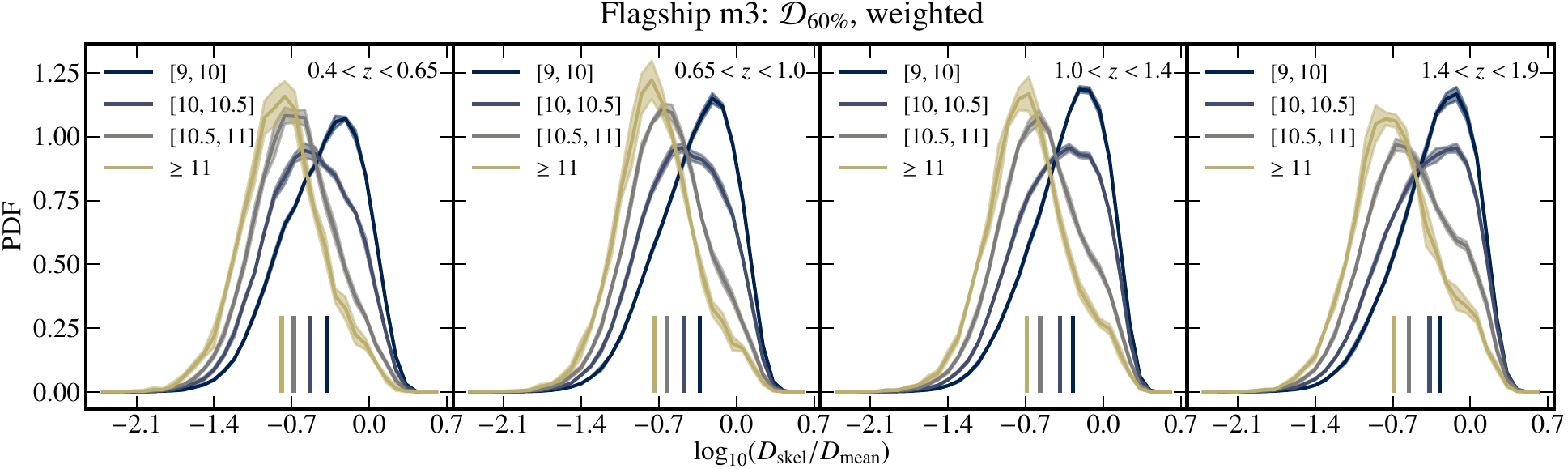}\vspace{2pt}
\centering\includegraphics[width=\textwidth]{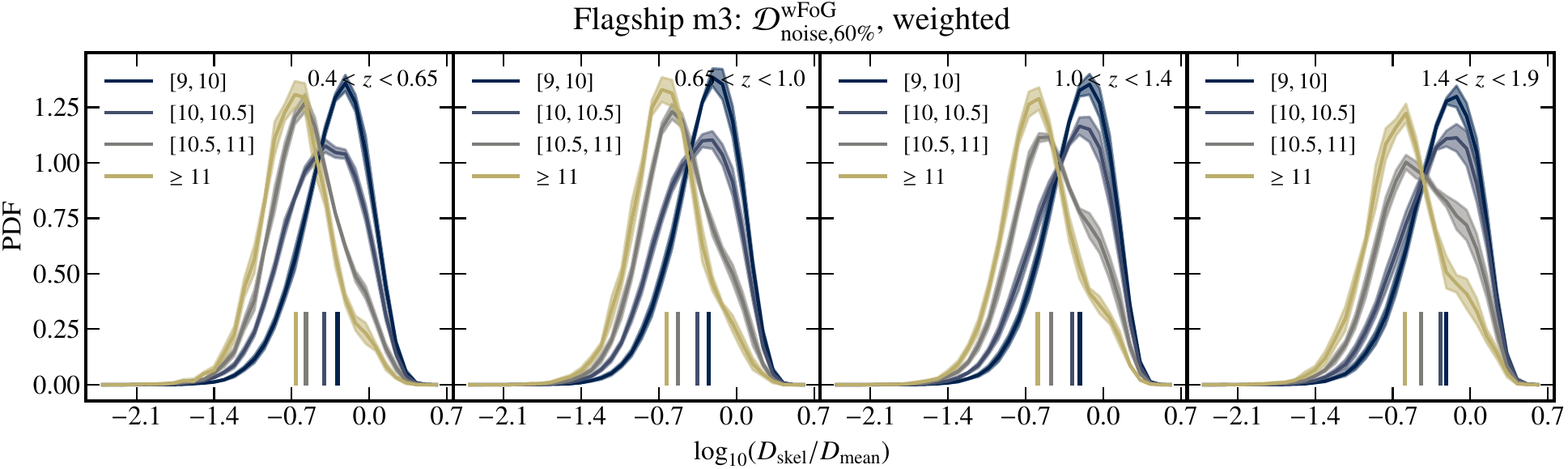}\caption{As in Fig.~\ref{fig:PDF_dskel_flag_m3_weight}, but for a catalogue with reduced sampling (\deepsamp, first row) and with added redshift error and FoG (\deepnoisesampwf, second row). Stellar-mass gradients are recovered in all configurations.}
\label{fig:PDF_dskel_flag_m3_weight_appendix}
\end{figure*}

Figure~\ref{fig:PDF_dskel_flag_m3_weight_appendix} shows the PDFs of the distances of galaxies to their closest filament for the \flag m1 model (the results for the \flag m3 and \gaea models are comparable) and for the reconstruction of the cosmic web with weighting the tessellation as a function of stellar mass and redshift, for \ha\ flux-limited sample with reduced sampling (\deepsamp, first row) and when redshift error and FoG are added (\deepnoisesampwf, second row).
Stellar-mass gradients, with more massive galaxies located closer to the filaments than their lower mass counterparts, present in the reference \ha\ flux-limited sample are recovered.

\begin{figure*}
\centering\includegraphics[width=\textwidth]{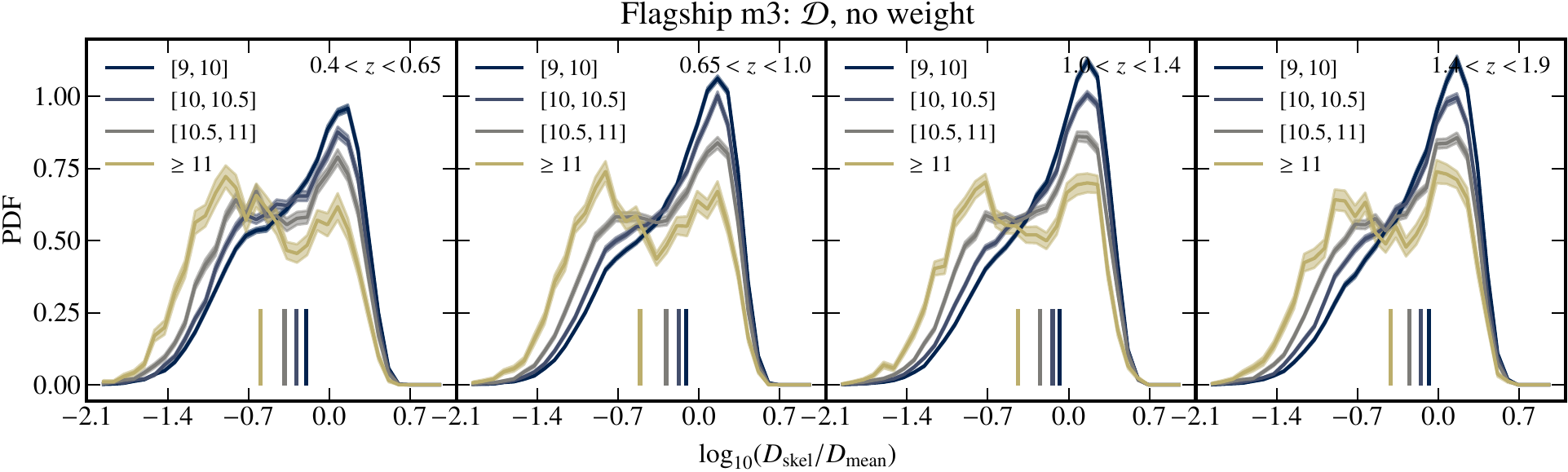}\vspace{5pt}
\centering\includegraphics[width=\textwidth]{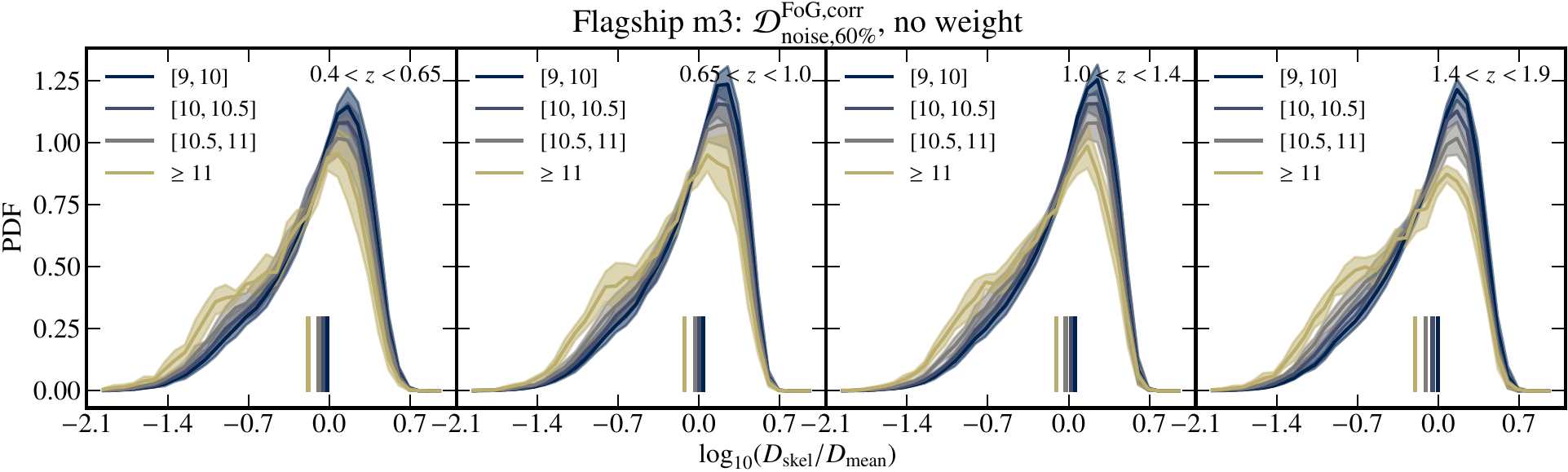}
\caption{As in Fig.~\ref{fig:PDF_dskel_flag_m3_weight}, but for the reconstruction without stellar mass-weighted Delaunay tessellation. 
The error on redshift in combination with the redshift-space distortions strongly reduce the stellar-mass gradients, that are only slightly improved after the correction of the FoG effect.
}
\label{fig:PDF_dskel_flag_m3_noweight}
\end{figure*}

Figure~\ref{fig:PDF_dskel_flag_m3_noweight} shows the PDFs of the distances of galaxies to their closest filament for the \flag m1 model (the results for the \flag m3 and \gaea models are comparable) and for the reconstruction of the cosmic web without weighting the tessellation as a function of stellar mass and redshift. 
The distances are normalised by the mean intergalactic separation to take into account the effect of decreasing the density of galaxies with increasing redshift.

The first row shows the PDFs for the reference sample (\deep), while the second row shows the PDFs after correction of the FoG effect (\deepnoisesampwof). When the cosmic web is reconstructed without weighting the Delaunay tessellation, reduced galaxy sampling tends to decrease the stellar mass gradient signal (not shown), as in the case of the reconstruction including the weighting. Redshift errors have the strongest impact, by significantly reducing our ability to recover the stellar-mass gradients towards filaments of the cosmic web (not shown). Correcting for the FoG effect only mildly improves the signal.

\end{document}